# Toward Refactoring of DMARF and GIPSY Case Studies – A Team XI SOEN6471-S14 Project Report

Zinia Das, Mohammad Iftekharul Hoque, Renuka Milkoori, Jithin Nair, Rohan Nayak, Swamy Yogya Reddy, Dhana Shree Sankini, Arslan Zaffar

Concordia University

Montreal, QC, Canada

{zi_das, mo_hoqu, r_milkoo, j_nair, r_nayak, sw_reddy, d_sankin, a_zaffa}@encs.concordia.ca

**Abstract**

*This literature focuses on improving the internal structure of the Distributed Modular Audio recognition Framework (DMARF) and the General Intensional Programming System (GIPSY) case studies without affecting their original behavior. At first, the general principles, and the working of DMARF and GIPSY are understood by mainly stressing on the architecture of the systems by looking at their frameworks and running them in the Eclipse environment. To improve the quality of the structure of the code, a furtherance of understanding of the architecture of the case studies and this is achieved by analyzing the design patterns present in the code. The improvement is done by the identification and removal of code smells in the code of the case studies. Code smells are identified by analyzing the source code by using Logiscope and JDeodorant. Some refactoring techniques are suggested, out of which the best suited ones are implemented to improve the code. Finally, Test cases are implemented to check if the behavior of the code has changed or not.*

**Keywords: Distributed** Modular Audio Recognition Framework (MARF), Generic Intentional Programming Language (GIPSY), Logiscope, JDeodorant, Design Patterns, Refactoring

## 1. Introduction

In this survey, the frameworks of DMARF and GIPSY are studied with the main interest in the architecture and the design of the systems. The primary focus is to understand and extract the needs, requirements and relationships between all the components of the case studies and to improve the architecture. The design requirements include different stake holders, use cases for both the systems. For further understanding the architecture of both the case studies domain models are designed which helped in understanding the commonalities in the architecture for DMARF and GIPSY. A survey of the software structures of both the systems is done by looking at the design patterns used to build the systems which helped in identifying the problematic classes. In addition to this, Refactoring is performed on the problematic classes. Moreover in this survey the changes are made to the original code by first identifying code smells within the problematic classes are collected.

## 1. Background

MARF and GIPSY are the two Open Source Software (OSS) case studies used in this study where we attempt to gain an understanding about their needs, requirements, goals, domains, High-level Architectures and the requirements along with their working and applications.

### 1.1 DMARF

**DMARF**, which stands for **D**istributed **M**odular **A**udio **R**ecognition **F**ramework which is a distributed extension of MARF, an open-source and extensible framework written in Java [1]. MARF is used as a research platform for programmers and software architects to test and compare new algorithms with the existing collection of pattern recognition, signal processing and natural language processing (NLP) algorithms. MARF's working is mainly based on the pattern recognition pipeline which consists of various stages as shown in figure 1 [2]. The goal of DMARF is to use the various stages shown in the pipeline as a distributed system [2]. It has been observed that DMARF can be autonomic in nature combined with independent self-protecting capabilities by using Autonomic System Specification Language (ASSL) [3].

#### 1.1.1 Domain

DMARF is biologically inspired and hence finds its wide use in application domains such as biometric forensic applications and processing audio, imagery and text and many other scientific applications in a distributed environment such as a web-service [4]. It employs pattern recognition, signal processing and NLP algorithms to help process the aforementioned data. DMARF can be used in many other fields by making it autonomic with the help of ASSL [1]. It can be used in conference recordings, phone conversations for forensic analysis and subject identification [3] [4]. Distributed MARF offers a number of service types such as [5] [6] [7] -

- Application Services
- General MARF Pipeline Services
- Sample Loading Services
- Preprocessing Services
- Feature Extraction Services
- Training and Classification Services

It can be used to defy security attacks as well which is the goal of the Java Data Security Framework (JDSF) which is still a work in progress. The JDSF employs the distributed properties of DMARF to counter security threats [8].

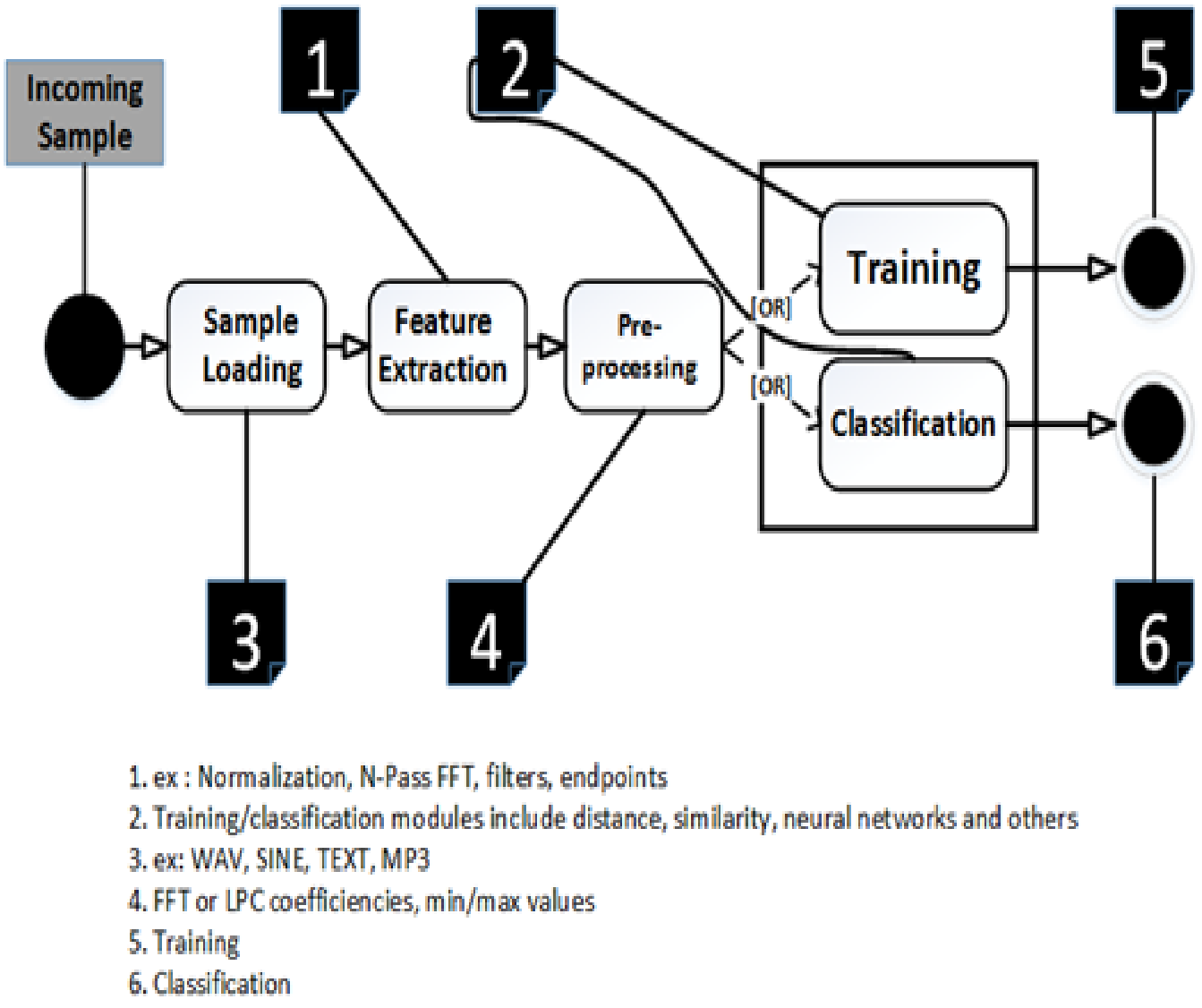

**Figure 1: MARF Pattern Recognition Pipeline [2]**

### 1.1.2 High Level Requirements

DMARF is based on the pipeline architecture of MARF which are run as distributed nodes. The applications that are created using DMARF employ these capabilities of MARF. An efficient communication protocol is used to exchange data between layers over Java RMI [9], CORBA [10], and XML-RPC WebServices [11].

***Autonomic System Specification Language (ASSL)***

Autonomic System Specification Language (ASSL) is used to develop autonomic properties for DMARF. Autonomic DMARF can be written as ADMARF. These autonomic properties provide DMARF with an autonomic middleware that helps enhance the management of MARF's pattern recognition pipeline. ASSL was used to generate prototype models for NASA and the Voyager missions [1]. These models employed the autonomic properties of DMARF for the space exploration missions. The three autonomic properties of ADMARF were –

- *Self-Healing* – Ability to recover itself through replication to keep at least one route of the pattern-recognition pipeline open.
- *Self-Optimization* – Communication between nodes to exchange data by selecting the most efficient communication protocol.
- *Self-Protection* – Protection against malicious alteration or denial of service in the global environment.

***Simple Network Management Protocol (SNMP)***

Simple Network Management Protocol (SNMP) is the most commonly used protocol for network management. SNMP works by collecting and configuring information from network devices on Internet Protocol (IP) network. There are three major components of SNMP [2] -

- **Manager**: Administrative computer that manages or monitors a group of hosts [12].
- **Agent**: A software component on managed devices which sends information to manager via SNMP [12].
- **Management Information Base (MIB):** Database this is used for managing the entities in a communications network [12].

DMARF's components do not understand SNMP as they are standalone components capable of listening only the RMI, XML-RPC, CORBA and TCP connections [5].

Therefore, each managed service is required to have a proxy SNMP-aware agent for management tasks and delegate instrumentation proxy in order to communicate according to service specifications [5].

### 1.1.3 Architecture

DMARF makes use of these various stages of the pipeline and allows them to run as distributed nodes [1] as shown in figure 2 [3] [16].

The figure describes the architecture of the DMARF pipeline which shows the various front-end and back-end modules. The front-end modules are invoked by client applications and services invoking other services through the pipeline during execution of the pipeline mode. The back-ends are in charge of providing servant implementations like replication, monitoring and disaster recovery modules [2].

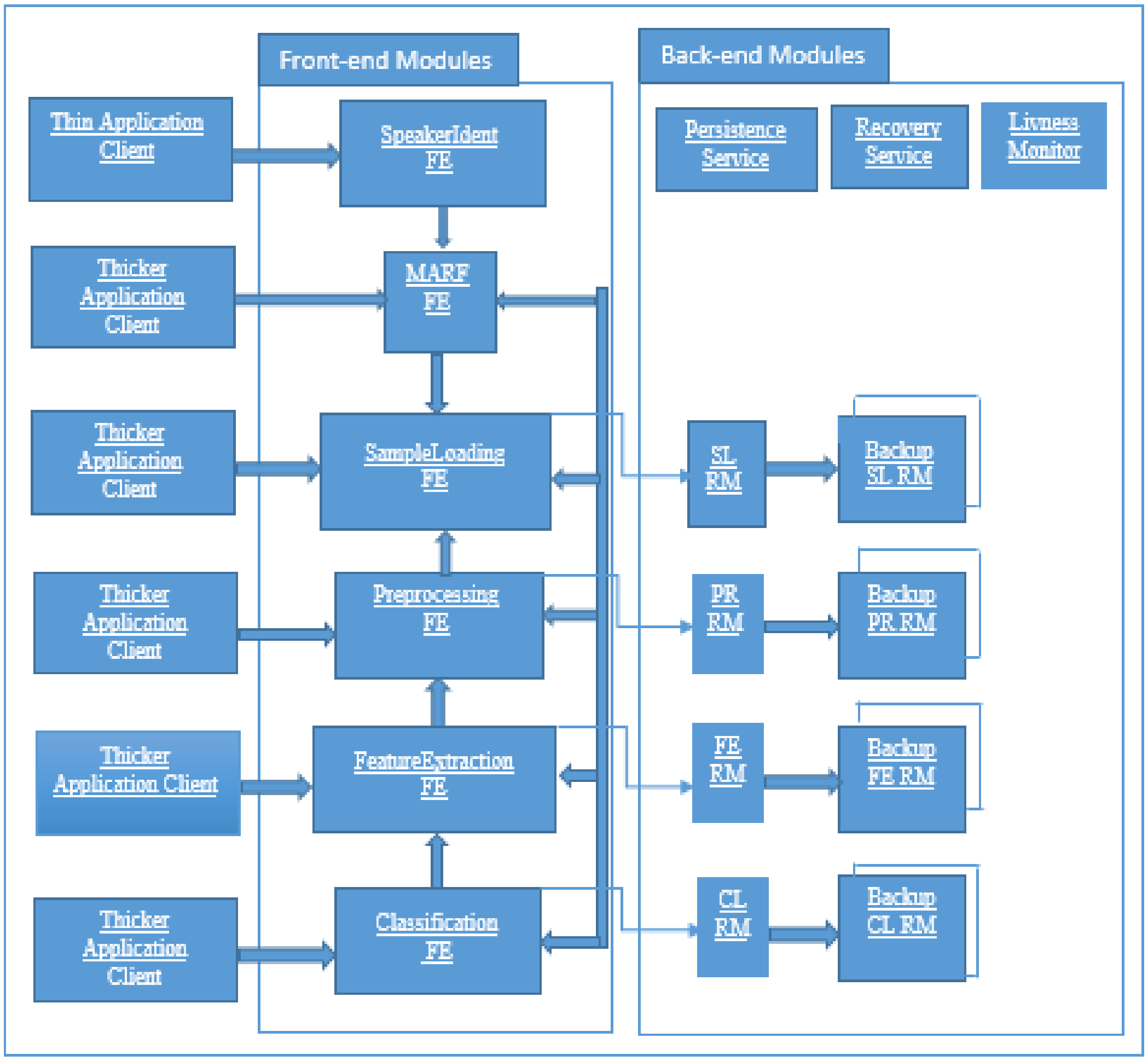

**Figure 2: The Distributed MARF Pipeline [13] [16]**

There are several distributed services in DMARF which needs to inter-communicate in order to produce a reliable system, such as -

- *General MARF* – It introduces the MARF pipeline to clients and other services and communicate with other nodes present in the pipeline.
- *Sample Loading Services* – Loads the file and convert it accordingly for pre-processing.
- *Pre-processing Services* – Accepts the input file and does the processing such as, filtering and normalization.
- *Feature Extraction Service* – It takes the pre-processed data and extracts certain features out of it given a requested algorithm. MinMax, FFT, LPC are some of the algorithms used for feature extraction.
- *Classification and Training* – It accepts the feature vectors and either updates the database with the new training sets or simply performs classification from the existing training sets [2].The distributed architecture of DMARF allows synchronization, recovery and replication which are described later.
- *Synchronization* – Synchronization has a great significance in an application where multiple clients can access a shared resource in order to ensure data integrity. At server side it is important when the clients want to access the database. It is achieved if critical paths of the objects are locked when it is accessed by multiple clients. Also, multiple servers maintains their own set of data structure thus making the system concurrent.
- *Recovery and Replication* - During system crash, in order to recover the data, log is maintained using Write-Ahead Logging (WAL). Recovery of the data is done by capturing the before and after snapshots of the object. WAL can hold a fixed number of committed transactions. Garbage collection and memory back up are performed in a periodic manner. Replication is performed by transferring WAL to another host or by using lazy update. Delegates broadcast requests before computing anything, if no response is received the delegate starts its computations or else a transfer is initiated from another delegate.

## 1.2 GIPSY

GIPSY is expanded as General Intensional Programming System which is primarily written in Java [13]. It provides an integrated framework to compile the programs which are written in the LUCID language [15] or any other Intensional language. It is an ongoing project that helps in providing a flexible platform for the analysis of the results generated by Intensional programming languages. LUCID is a context-aware language which is designed in such a way that the programs are written in terms of real-world semantics [14]. GIPSY has a distributed nature which is depicted in the logo in figure 3.

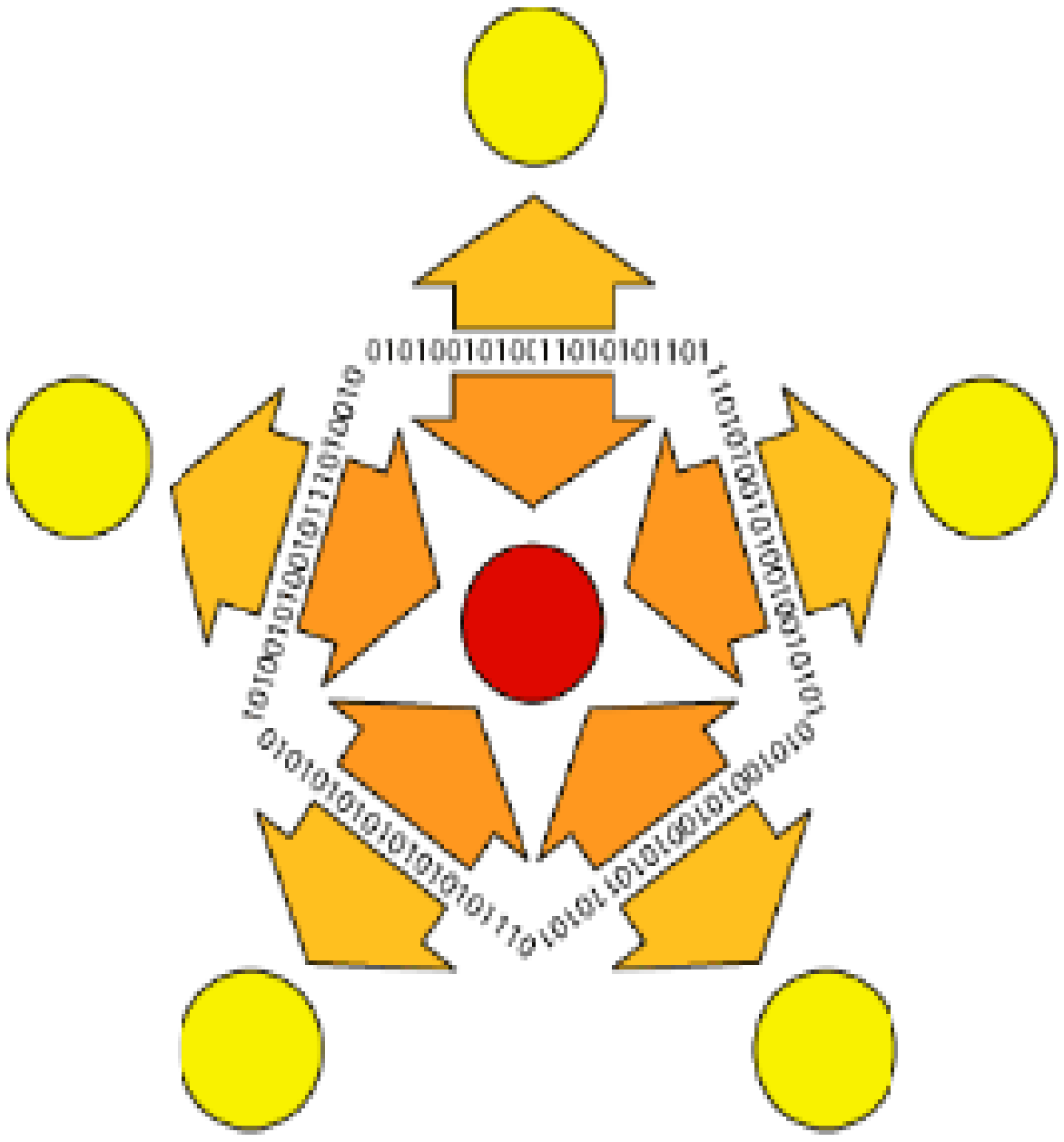

**Figure 3: The GIPSY logo showing the distributed nature of GIPSY [14]**

### 1.2.1 Domain

GIPSY (General Intensional Programming System) is a tool that provides a flexible platform for investigation of Intensional programming whose design reflects three main goals which are adaptability, generality and efficiency. They support for hybrid languages including Intensional and imperative languages for various needs. GIPSY provides an integrated framework for executing programs written in all variants of Lucid. GIPSY finds its domain in Hybrid Intensional Programming, with Lucid, JLucid, Objective Lucid and other Generic Intensional Programming Languages [13]. Security Frameworks and Intensional Cyberforensics and Self-Forensics [14] [28] also form a part of GIPSY's domain. The properties of GIPSY also find their use in the evaluation of Higher-Order Intensional Logic Expressions (HOIL) [30

### 1.2.2 High-Level Requirements

Nowadays Intensional programming is evolving very fast and thus there is a necessity of designing new tool which has better generality, adaptability and efficiency. The General Intensional Programming System (GIPSY) is developed keeping in mind of those qualities [20]. GIPSY has been built with respect to the LUCID family of programming languages mainly Intensional in nature. These languages depend on Higher Order Intensional Logic (HOIL) to provide a multidimensional reasoning platform for the analysis and execution of Intensional expressions and statements. Its framework is implemented in Java and its compilers provide a platform to execute LUCID family of programs that run in a distributed environment with the help of its1distributed environment with the help of its own demand-driven eductive engine [14].

### 1.2.3 Design and Implementation Goals

The main purpose of GIPSY is to set contexts as first-class values and usage of context calculus operators, which will lead into the construction and manipulation of contexts [17].

### 1.2.4 Architecture

GIPSY has a multi-tier architecture and the declared expressions in the Intensional programming are evaluated in a multi-dimensional context space. It aims at providing a flexible platform for investigating Intensional and Hybrid Intensional Imperative Programming [16]. The High-Level architecture is described in figure 2. Figure 3 describes a simplified version of the different layers. It is divided into four tasks assigned to four layers shown in table 1.

| | Layer Type | Purpose |
|---|---|---|
| **1.** | Demand Generator Tier (DGT) | Encapsulate demands [18] |
| **2.** | Demand Store Tier(DST) | Middleware for other tier [18] |
| **3.** | Demand Worker Tier (DWT) | Process Demands [18] |
| **4.** | GIPSY Manager Tier (GMT) | Register GIPSY nodes [18] |

**Table 1: GIPSY Multi-tier Layers [18]**

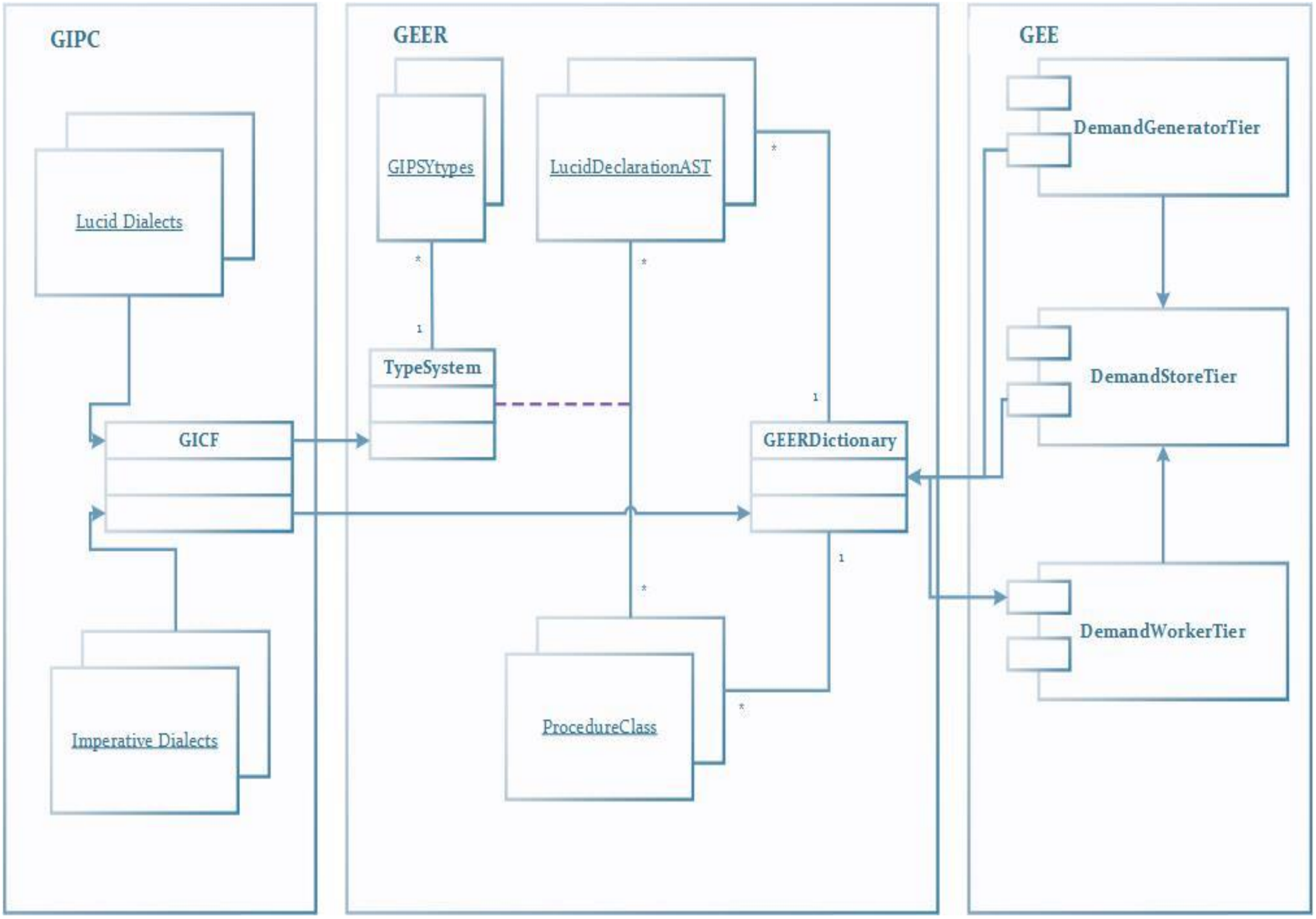

**Figure 2: High-Level Architecture of GIPSY [14]**

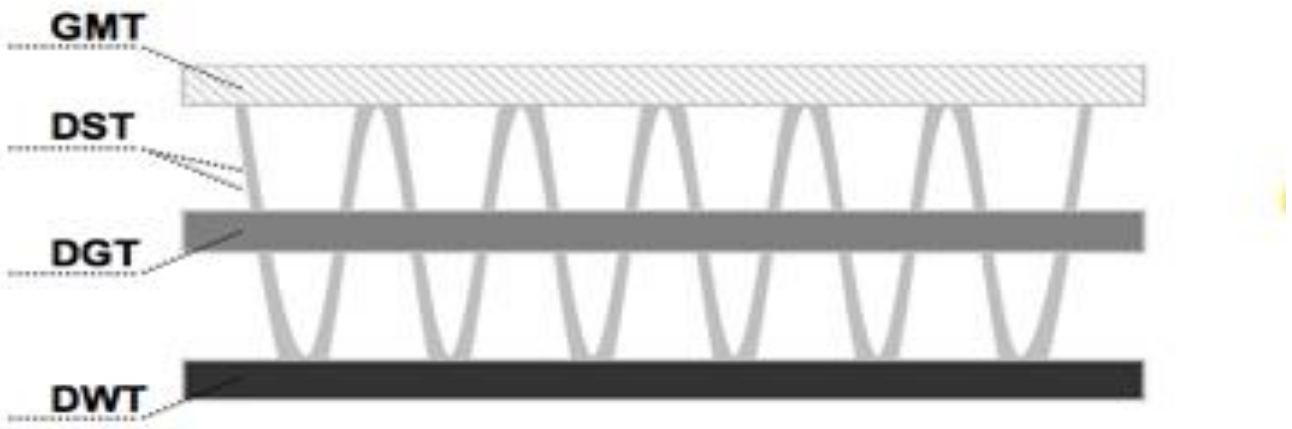

**Figure 3: GIPSY Multi-Tier Architecture [23, 24]**

GIPSY tiers can interact with each other using demands. Here, based on initial demand, the Demand Generator Tier (DGT) creates Intensional and procedural demands. For migration of the demands between various tiers the Demand Store Tier (DST) is used. It is designed like a peer-to-peer architecture. Here, the Demand Worker Tier (DWT) is responsible to process the procedural demands. The GIPSY Instance Manager (GIM) communicates with DGT, DST, DWT controllers to decide if any new tiers or nodes are required and based on the situation it registers them [23] [24] [25].

### 1.2.4.1 General Intensional Programming Language Compiler (GIPC)

The GIPSY software architecture is shown in Figure 4. Here, GIPSY programs are first translated into C and then compiled. Source code is composed of the Lucid part which explains the dependencies among variables and the sequential part which illustrates the computational units. After compilation Intensional Data Dependency Structure (IDS) defines the dependencies among variables in the Lucid part. The General Eduction Engine (GEE) which is based on demand propagation technique interprets IDS at execution time. The GIPC produced data communication procedures based on the data structure which yield the Intensional Communication Procedures (ICP). Using C compiler syntax the sequential part of the GIPSY program are converted to C program which yield C Sequential Threads (CST) [20].

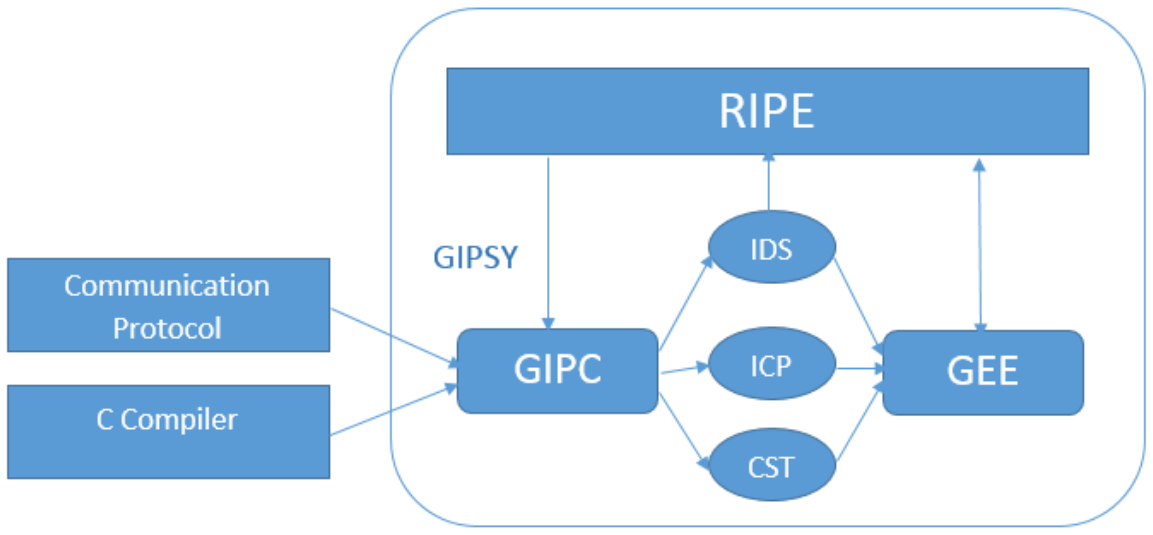

**Figure 4: The GIPSY Software Architecture [20]**

### 1.2.4.2 General Eduction Engine (GEE)

In GIPSY a computation starts if there is a demand for it. Procedure calls are generated locally or remotely by these demands and then stored in the warehouse. The values which are already computed are removed from the warehouse. The GIPSY generator-worker execution architecture is shown in Figure 5. The generator interprets the IDS produced by the GIPC. Here, the generator is comprised of the Intensional Demand Propagator (IDP) and the Intensional Value Warehouse (IVW). The technique of demand generation and propagation is implemented by the IDP and the IVW is responsible for the warehouse. The worker is composed of Ripe Functional Executor (RFE) which is in charge of calculating the ripe sequential threads. Remote workers can upload or download these sequential threads using shared network system [20].

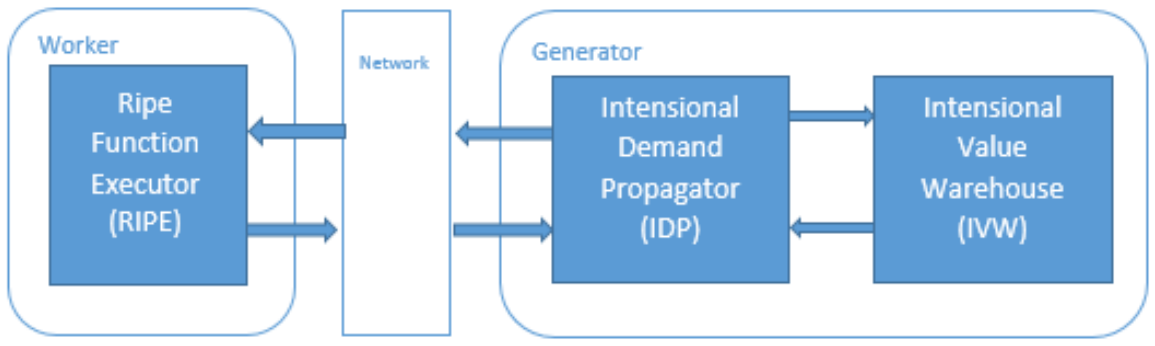

**Figure 5: Generator-Worker Execution Architecture [20]**

### 1.2.4.3 Run-Time Interactive Programming Environment (RIPE)

At runtime the RIPE which is a programming environment helps to envision the dataflow diagram of the Lucid program. Here users are also permitted to communicate with the RIPE by inspecting the IVW, changing the channels of the code, again compiling the threads which are sequential, altering the rules of communication or adjusting components of the GIPSY at run-time [20].

### 1.2.5 Multi-Tier Architecture for a runtime system

Multi-Tier Architecture system is adopted for the GIPSY runtime system, where the entire system is divided into four kinds of GIPSY tiers. Each kind of GIPSY tier consists of any number of differently implemented tier instances, and each instance is a separate process that runs within a computer and also communicates with other tiers via demands [21].

The detailed descriptions of the four kinds of GIPSY tiers are as follows -

**1) Demand Store Tier (DST):** A GIPSY tier whose instance serves as the middleware and cache to provide demand delivery and storage services for other tiers [21].

**2) Demand Generator Tier (DGT):** A GIPSY tier whose instance generates various types of demands by traversing the abstract syntax tree contained in a GEER. A GEER is a dictionary that provides an information about runtime resources generated by the compiler GIPC from a hybrid Intensional program [21].

**3) Demand Worker Tier (DWT):** Its instance processes procedural demands by executing the procedural function calls defined in GEERs contained in its Procedural Class Pool [21].

**4) General Manager Tier (GMT):** An instance of GMT manages the GIPSY node registration and the allocation and de allocation of DGT, DST and DWT instances. The GIPSY node registration is the process of registering a computer into the GMT so that the GMT is able to allocate GIPSY tiers in that computer when necessary [21].

### 1.2.6 Advances in the Design and Implementation of A Multi-Tier Architecture in the Gipsy Environment

GIPSY provides a platform for investigation on Intensional and hybrid Intensional-imperative programming. Intensional programming is a declarative programming language which is based on denotational semantics. GIPSY Compiler is based on the Generic Intensional Programming Language (GIPL) which is said to be the core run time language where all the Intensional programming languages can be translated [13][22][23][24][25][26].

### 1.2.6.1 Node & Tier Properties

In a GIPSY program, Nodes and tiers may be effected at run time i.e. a specific node or tier may be computed for different programs. Any tier or node can fail without the system is fatally affected. The tiers and nodes can be added or removed during the time of computation [16].

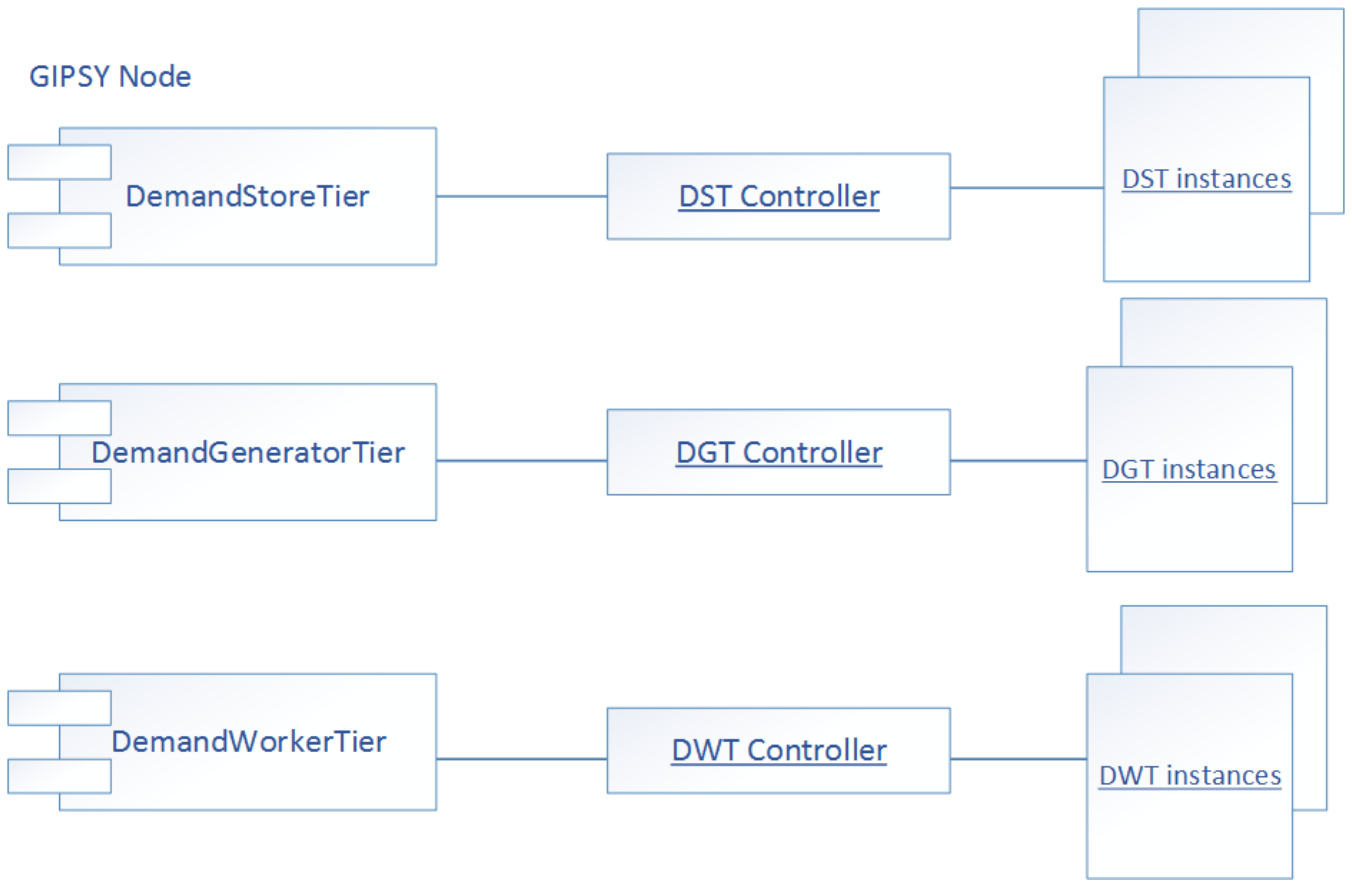

**Figure 6: Design of the GIPSY Node [16]**

### 1.2.6.2 Demand Driven Computation

It is mainly focused on the notion of generation, propagation and computation of demands along with their resulting values. Where the demands may be of several types like Intensional, procedural, resource and system [16].

- Intensional demands
- Procedural demands
- Resource demands
- System Demands
- Demand Identifiers

Two types of demand identifiers.

- Universally unique identifier
- Demand Signature identifier

to be used as experimental results, thus providing a strong basis for the use of the GIPSY as an experimentation platform. This feature has already been put into experimental action for the testing of the DST implementation [19].

Using the General Intensional Programming System (GIPSY) for Evaluation of Higher-Order Intensional Logic (HOIL) Expressions

- *Eductive Model of Computation*: The first operational model designed for computing the LUCID programs with a concept of executing the notion of generation and consumption of resulting values.
- *Intensional Logic and Programming*: This programming language is based on the notion of declarative language in an inherent multidimensional context space [27].

### 1.2.6.3 Eductive Execution of Hybrid Intensional Programs

Additionally GIPSY provides an integrated framework for executing programs written in all variants of Lucid. Architectural framework for the run-time system that enables the distributed execution of such programs following the eductive model of computation is been discussed below [19]. The multi-tier architecture adopted by the solution permits the addition or suppression (or failure) of new nodes and tiers as programs are being executed. It also permits the allocation of new programs to be executed by GIPSY Instances already executing other program. It is observed that the solution reduces the communication by relying on a scalable DST, where DST instances can be spawned upon need (e.g. when the available DST instances are overloaded), thus distributing the communication load [19].

There is a peer-to-peer solution that is been suggested by allowing the tiers to interrogate a single DST transparently to its potentially distributed aspect. GIPSY program can be executed in different execution topologies, which can be set prior to the starting of the program's execution, or even as the program is being executed. This was one of the greatest features of GLU that has been retained in the solution discussed [19].

***Solution Set***

The solution designed is in a highly modular manner, where each component interacts through a simple interface, mostly based on exchanging demands. These demands are all stored in the DST and embed lifetime statistics and can thus be further queried and analyzed during run-time for run-time optimizations or postmortem

***Applications***

Self-Forensics is a concept that is introduced to be included in complex, autonomic hardware and software systems. It is a module which keeps track of the events and logging for automatic forensic processing, deduction and event re-construction in case of incidents [2].

### 1.2.6.4 ASSL Toolset

ASSL performs a checking of syntax and semantics of the specification of properties of autonomous system and on success, it creates a collection of Java classes and interfaces corresponding to that of the specifications.

The ASSL framework includes the autonomic multi-tier system architecture to specify service-level objectives (SLOs), self-CHOP properties such as, self-configuration, self-healing, self-optimization, self-protection [28].

The study of self-forensics includes the concept of forensic lucid, a functional-Intensional forensic case programming language. It has undergone extensive design and development including its syntax, semantics, the corresponding compiler, run-time environment, and interactive development environments provided by the General Intensional Programming System (GIPSY) [29].

### 1.2.7 Autonomic GIPSY

The aim of AGIPSY is to improve and manage the workload of complex GIPSY by enabling self-monitoring [1].

Generally the execution of GIPSY programs is divided into four tasks and assigned to each tier since GIPSY is a multi-tier architecture. The tiers create their own processes and communicate using demands with other processes with same or through tiers [1].

***AGIPSY Architecture and Behavior***

It consists of autonomous GNs and Node managers (NM) to guide them in execution. And the architecture is multi-agent distributed system with decentralized control and data allocation which is loosely connected. The architecture consists of GNs and GIPSY Managers (GM). GN plays major role in behavior and communication with other nodes in the architecture to accomplish tasks and some of the GNs which run the GMT instances are called global autonomic managers. These manage the whole system and are called GMs [1].

The self-management features like Fault tolerance and recovery, Self-Maintenance, Self-Configuration, Self-optimization, Self-Healing and Self-Protection are responsible for the autonomic features of AGIPSY which are provided by NM [1].

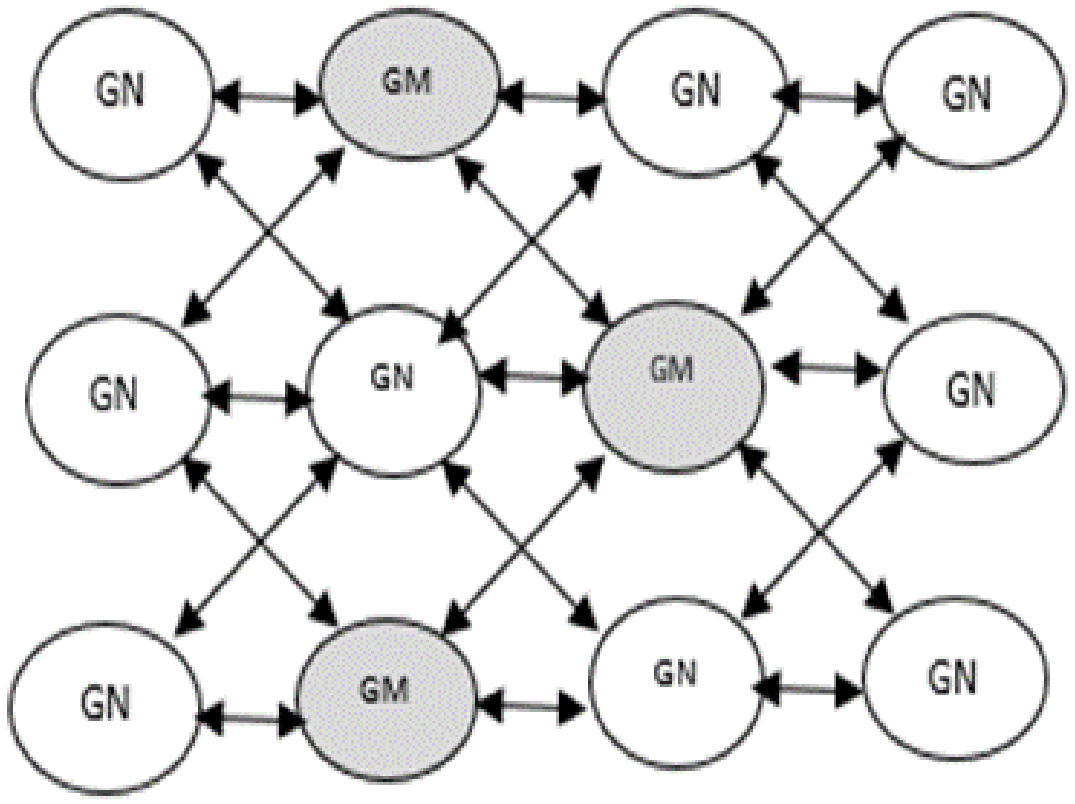

**Figure 7: AGIPSY Architecture [1]**

**GIPSY AE**

Gipsy AE consists of NM and GIPSY tier controllers which communicate through GIPSY tier controller interface. And it has four control loop components Monitor, simulator, decision maker and executor and also have two controllers channel controller and sensor controller. They all work in combination to provide control loop functionality by sharing data with ASSL [1].

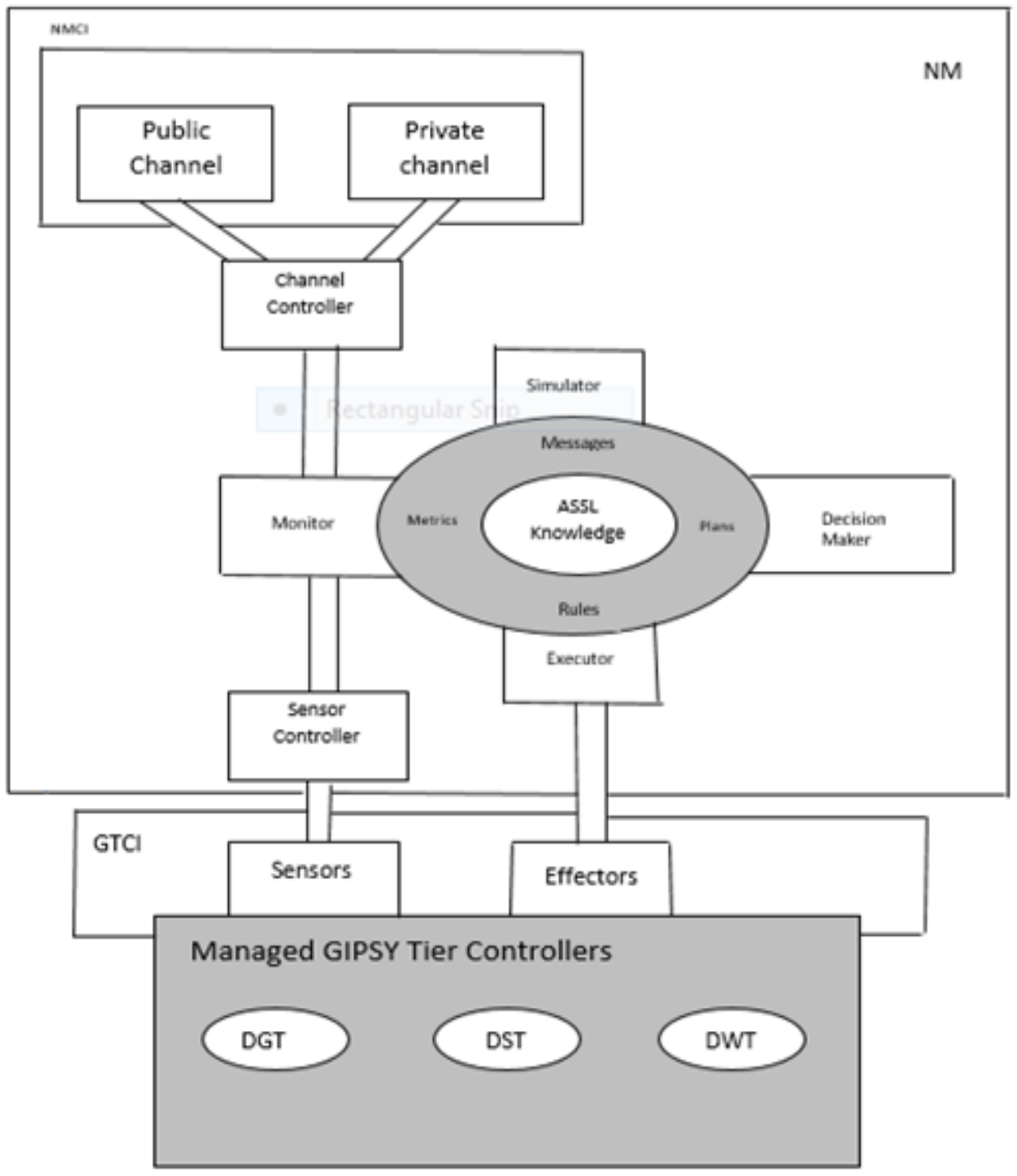

**Figure 8: GIPSY AE Architecture [1]**

### 1.3 DMARF Vs GIPSY

In this section we do a comparison in between DMARF and GIPSY and discuss the similarities and differences between the two case studies.

The GIPSY has a demand-driven eductive execution model: its General Eduction Engine (GEE) evaluates Intensional (Lucid dialects) or imperative expressions (e.g. Java methods, C++ functions, etc.) in presence of which a procedural demand is generated and delivered to a networked demand store (data value warehouse), from where it can be picked up from by an observing worker on some other node for evaluation. The described type of distributed asynchronous communication is managed by the DMS [18, 19].On the other hand DMARF due its nature of application, implements a pipelined or chained way of connecting some of its distributed nodes following the pattern recognition pipeline stages, thereby making it look synchronous.

Both systems are written in the Java language and share some common utility base. They also overlap in the use of the available Java communication technologies, such as the RMI. DMARF has Web Services and CORBA implementation, while GIPSY has JINI and JMS [18, 19].

**JDSF architecture**

JDSF [21, 22] is one of the proposed frameworks to allow security researches working with several types of data storage or databases in Java to evaluate different security algorithms and methodologies in a consistent environment. The JDSF design mainly aims at the following aspects of data security: confidentiality (data is private), integrity (data is not altered in transit), origin authentication (data coming from a trusted source), and SQL randomization (for relational databases only).The main disadvantage of JDSF is that, it can't deal with DDoS and cannot detect malicious code or any type of network protocol [21, 22].

***Threat Model, Security Requirements***

**A. Confidentiality**

Confidentiality refers to limited access and disclosure of information to authorized users only. It is noticed that in GIPSY, the confidentiality requirement is generally less applicable than in DMARF except the aforementioned cyber forensic case. In DMARF, depending on the application front-end, confidentiality, and more specifically privacy may be more desired when some subject's identity data is not required to produce results such as accuracy or performance statistics. Application of the JDSF's confidentiality sub framework would resolve the confidentiality concerns in the two systems. Its integration into the security layer would be minimal because all the systems share the common implementing language, Java [21, 22].

**B. Integrity**

Integrity refers to the trustworthiness of information resources. Integrity of the data in transit is achieved by the JDSF's integrity sub framework for the similar reasons and mechanisms similar to the confidentiality aspect. Integrity of the SQL based data can also be maintained by randomizing the SQL statements [17].

The JDSF's authentication sub framework can be used here in part similarly in the way it is done for the confidentiality and the integrity aspects. The authentication can be achieved similarly to the DNSsec methodology and can be hierarchical [17].

Therefore, JDSF does cover a wide array of these aspects, but it is observed that it cannot be an all-in-one solution, and the availability aspect along the lines with the malicious code detection is the most difficult to tackle; thus, suggesting to look at other similar frameworks and enhance the secure coding practices engineered back into the studied system [17].

**1.4 Summary**

**I. DMARF**

DMARF allows MARF to run as a distributed framework and it is biologically inspired to run various applications by using the MARF pattern recognition pipeline. It takes data identification and processing to the next level and provides and enhanced framework for the working of the biometric applications [1]. It is mainly used for speech and text identification and as a test platform for scientist and researchers. The architecture of DMARF is a distributed version of the classical MARF pattern recognition pipeline. The Clint applications and the services of the systems communicate and process information between each other with the help of the front-end and back-end modules. The SNMP allows efficient communication between the devices of the network. With the help of the ASSL, the distributed architecture of DMARF can allow it to have autonomic properties that allows it recover automatically, communicate accurately and effectively and also protect itself from malicious intent.

**II. GIPSY**

The necessity and the motivation behind the multi-tier architecture of the GIPSY framework have been identified and the core components namely, GIPC (General Intensional Programming Compiler), GEE (General Eduction Engine) and RIPE (Run time Interactive Programming Environment) have been reviewed. The GIPC and GEE frameworks have been discussed in terms of the goals of flexibility which help in understanding the system's functionalities in a wider array of entities such as data types, languages and replaceable components. GIPSY finds a wide-array of application domain and is used mostly for Intensional Programming Languages. It also finds its domain in the evaluation of Higher-Order Intensional Logic Expressions. Design of security frameworks are helped with the help of GIPSY framework.

## III. Code Analysis

Estimations of the case studies are performed by using different tools and plugins such as McCabe [31], Logiscope [32], CodePro AnalytiX [33], CLOC 1.6.0 [34], SLOCCount [35], and InCode [36]. These plugins are installed in eclipse. The results of the metrics are achieved after compiling the project. SLOCCount and CLOC are used to achieve the number of programming languages. SLOCCount is used in Linux platform. To compute the number of lines of text, first the projects were compiled using the Eclipse IDE. The projects were then run in McCabe, Logiscope and SLOC Count. The results are compared in the tables. Linux platform is used to calculate the number of Java files in the project. The following command is used to perform this task.

find marf/src –type f – name "*.java"| wc –l

| | **DMARF** | **GIPSY** | |
|---|---|---|---|
| | **Values Achieved** | **Values Achieved** | **Support Used** |
| **Java Files** | 1024 | 599 | CLOC [34] |
| | 1024 | 602 | LINUX |
| **Classes** | 1058; (public) 1045; (private) 5; (protected) 2; (package) 6 | 626; (public) 540; (private) 21; (protected) 0; (package) 65 | CodePro[33] |
| | 1058 | 702 | InCode [36] |
| **Methods** | 7554 | 6468 | InCode |
| | 6305; (public) 5132; (private) 350; (protected) 463; (package) 360 | 5680; (public) 2993; (private) 62; (protected) 336; (package) 2289 | CodePro |
| **Lines of Code** | 77297 | 104073 | CodePro |
| | 52343 | 83976 | McCabe[31] |
| | 77221 | 104083 | Logiscope[32] |
| | 77166 | 104024 | CLOC |
| | 68068 | 98478 | SLOCCount[35] |

**Table 2: Summary of Code Analysis of DMARF and GIPSY**

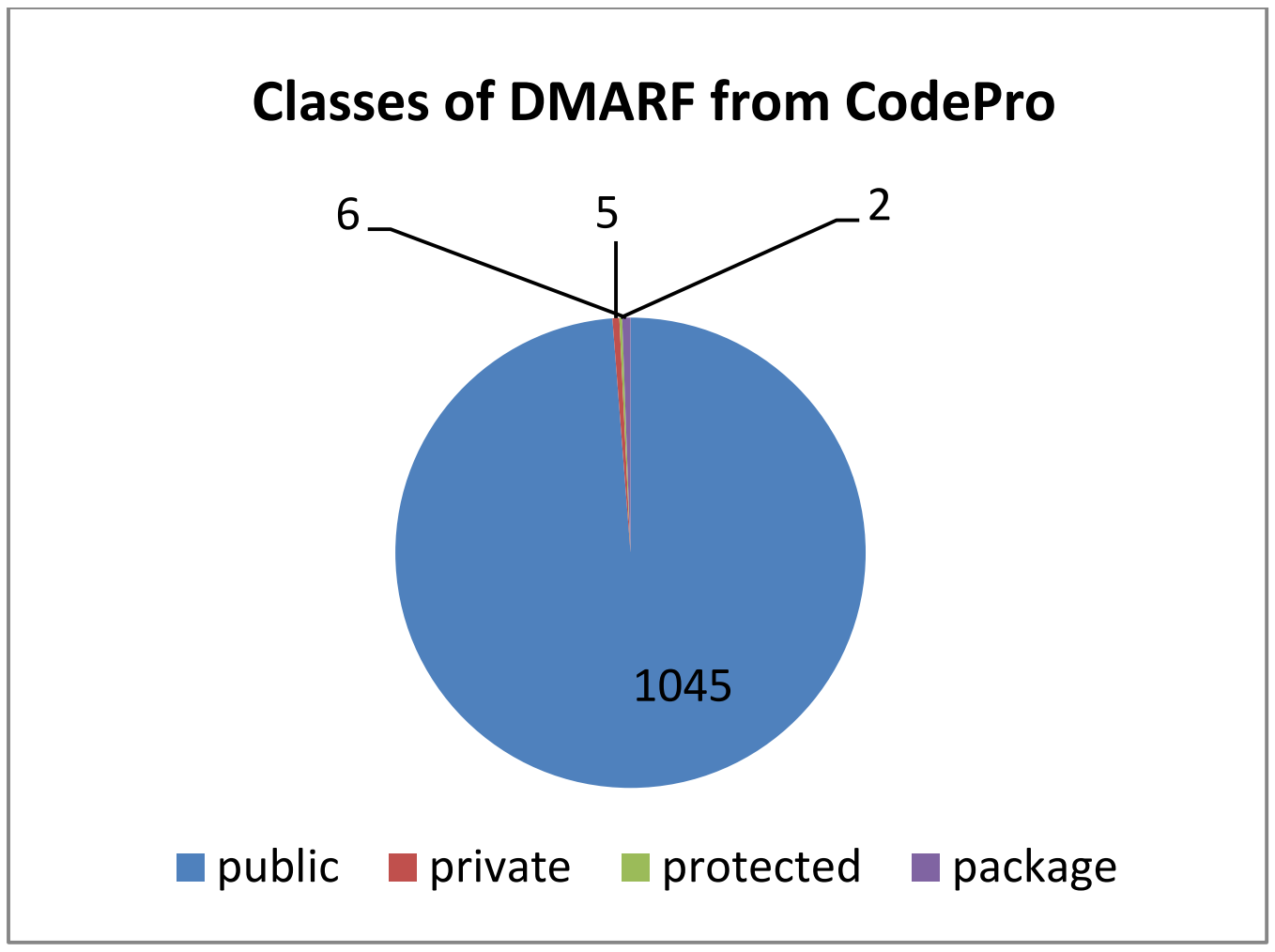

**Figure 9: Breakdown of classes of DMARF using CodePro**

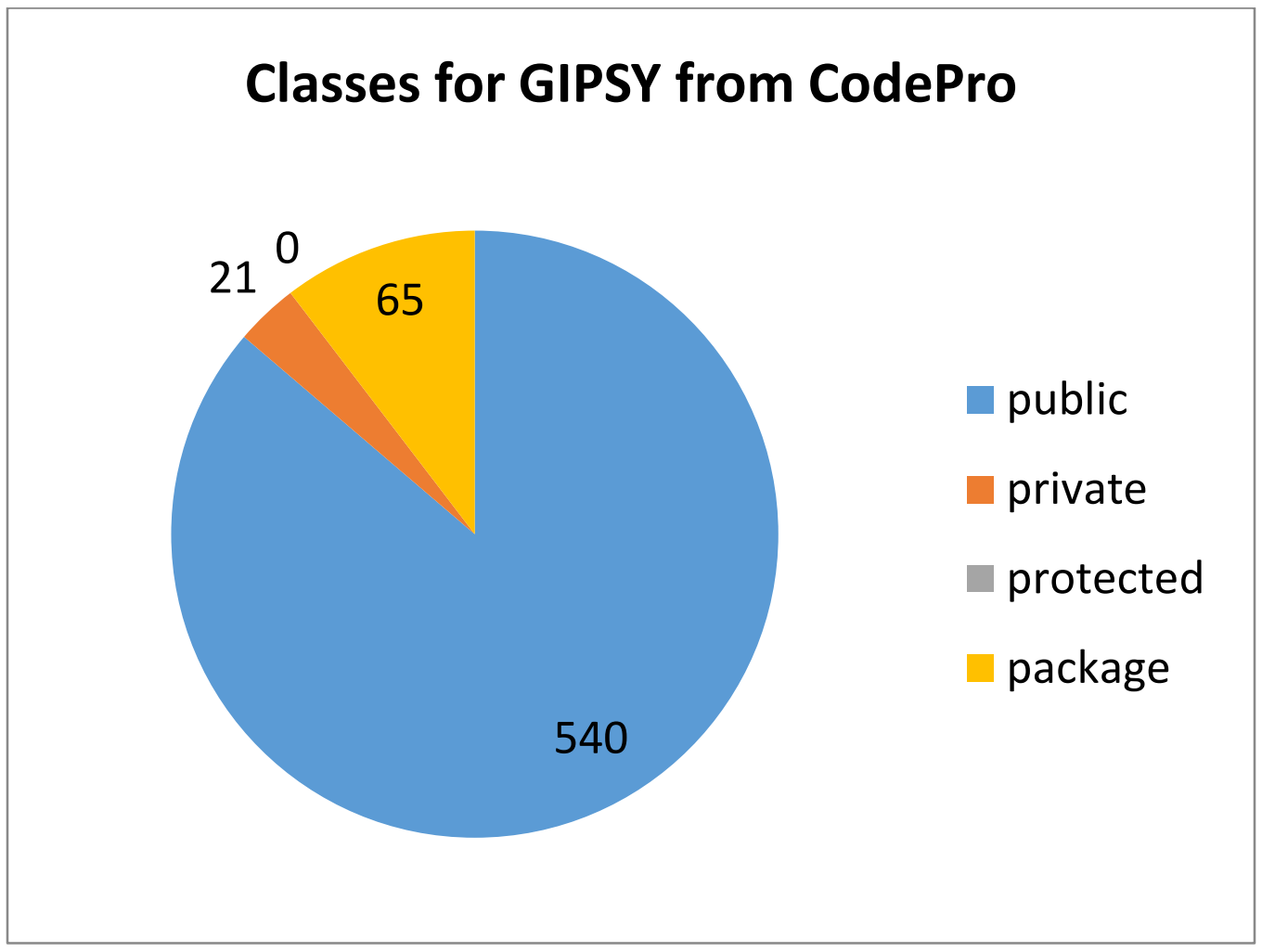

**Figure 10: Breakdown of classes of GIPSY using CodePro**

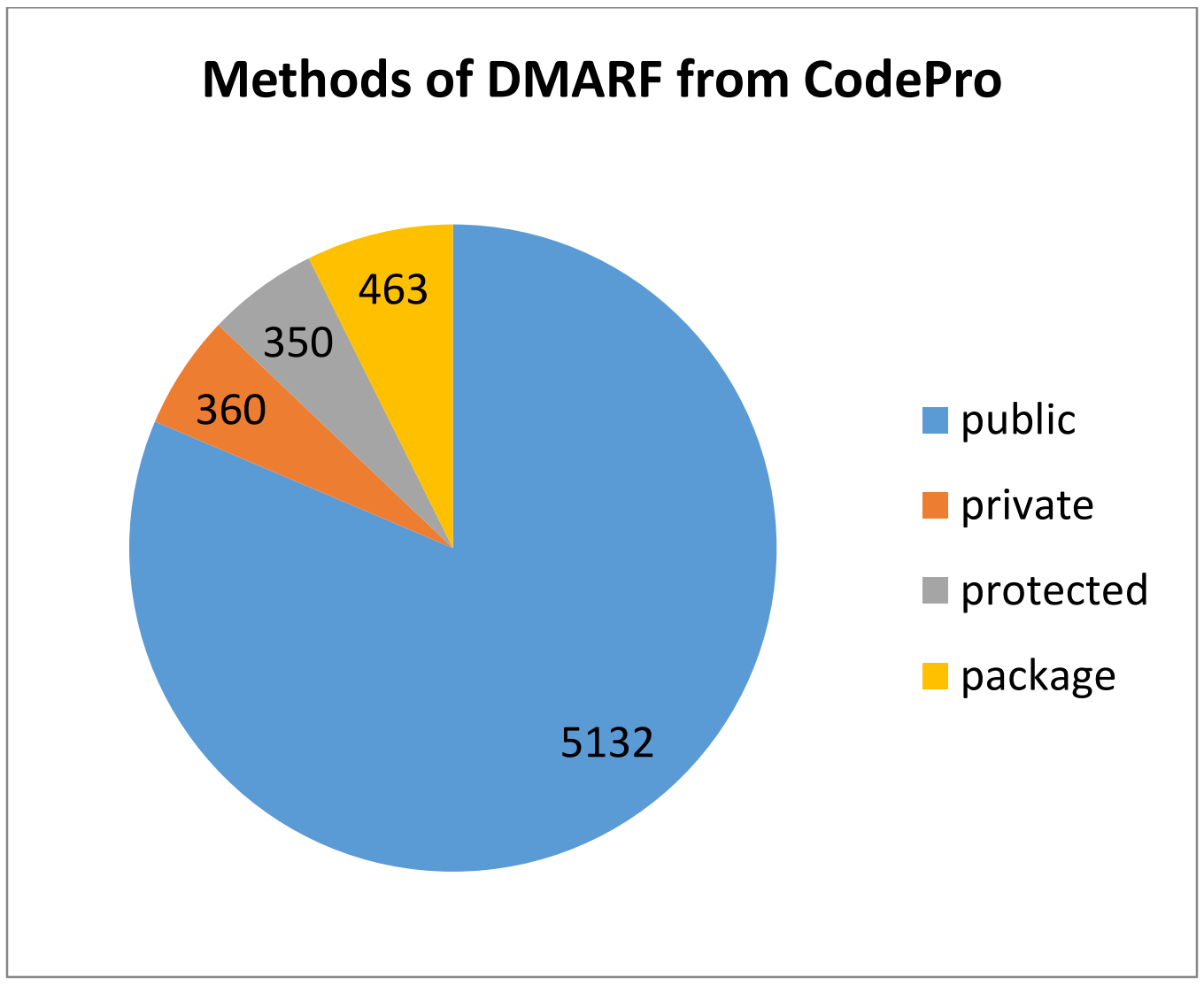

**Figure 11: Breakdown of methods of DMARF using CodePro**

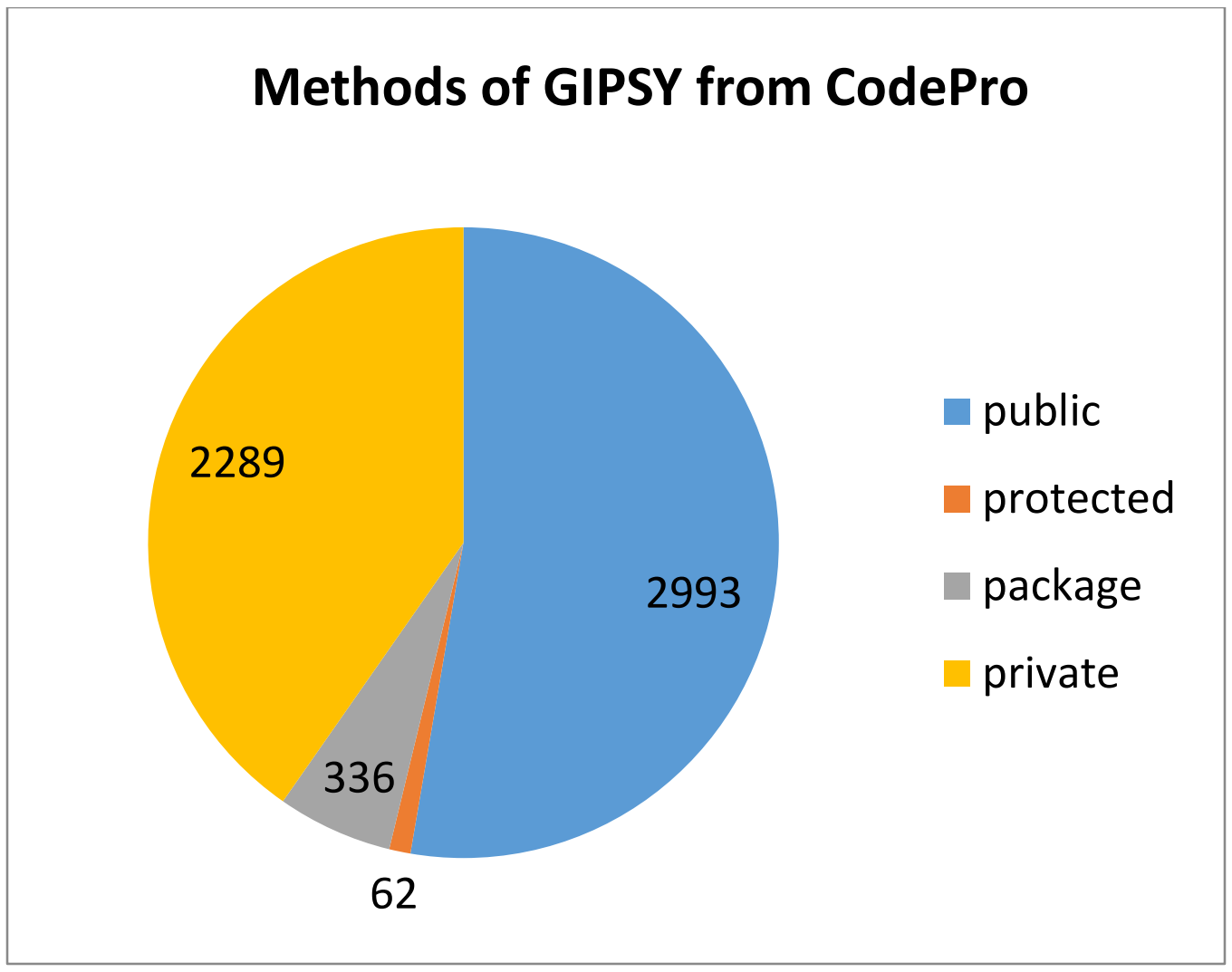

**Figure 12: Breakdown of methods of GIPSY using CodePro**

## 2. Requirements and Design Specifications

In this section, we gain an understanding behind the architecture of the case studies by conceptually looking at the system using the UML and the roles of the actors and the stakeholders that are involved in the projects. Generally the domain model is considered as the non-technical stakeholders point of view for the project. Domain model is the visual dictionary of stakeholders view of concepts and it also reduces the representational gap between the domain model and design model like using the similar class names and also for the attributes and the relationship between the classes are taken from real world representation of conceptual classes. Sample scenarios in terms of use cases for each of the case studies are also examined.

### 2.1 Actors and Stakeholders and Personas

Personas, actors and stakeholders help give a broader understanding about the requirements that are needed to build the system as well as the working of the system in different scenarios. Personas have been generated with respect to the intended users of each system. Actors are the people and other non-living entities that are responsible in the working of the system. Stakeholders focus on the users who use the system. Stakeholders can also be the users who are affected or influenced by the computer system. The personas, users and stakeholders are listed out for both systems and are mostly based on the application domains of both the systems.

#### 2.1.1 DMARF

Personas, Actors and Stakeholders are identified by researching papers and understanding the architecture and the application domains of the DMARF system.

**Users and Stakeholders of DMARF**

**Researchers:** Researchers define an open source and extensible framework for the research platform. They can also design case studies to understand how the system will perform and provide in different scenarios. Each case study describes a different scenario which can be used to understand the functionality of the system in terms of maintainability, usability and reliability. A student can be a researcher. The role of the student is to learn the various applications in DMARF under the guidance of the professor. Students can ask queries on the working principles of these applications and can suggest some new ideas for better implementation like refactoring or finding out the reusable components etc.

**Application Developers:** Application developers define the design of the application like Speaker Identification, biometric identification and Cyber forensics. These developers describe the specified algorithms like pattern recognition, signal processing and NLP used for various usage in the DMARF architecture. They use the principles and properties of DMARF to develop applications in speech, audio, video and text recognition. Applications relating to security (ex. CyberForensics) are also dealt with using the distributed nature of DMARF.

**Architects**: Architects are responsible for the implementation and development of the architecture and also in applications in the DMARF framework. One such framework is Java Data Security Framework (JDSF) which comprises of Java classes and methods for encountering the security threats. A professor can be an architect in case of DMARF. The role of the professor is to understand and execute the application under some valid conditions. One such application is Speaker Identification App where the voice samples are stored with their properties to identify and compare with the other voice samples.

**Maintainers:**

Maintainers can help in improving the system and identifying bugs and smells. Identifying of these smells and bugs further helps to improve the system. They are responsible for creating diffs and commits for the system.

**Persona:**

| Name | George Peter |
|---|---|
| Age | 40 |
| Job Title | Research in Cybercrime Forensic at McGill University |
| Experience | Expert in the fields of pattern recognition and signal processing in the DMARF application |
| Goal | Implementing Forensic Lucid programming for security attacks in DMARF application using the Java Data Security Framework |
| Description | George is highly passionate in learning the Forensic Lucid programming language which encounters security threats. Java Data Security framework is mainly used to identify the security attacks with the help of distributed properties of DMARF. He can even lend aid to law enforcement agencies as well as any kind of private firm involved in data security and cyber-crime to analyze information retrieved from computers and data storage devices. He also has the skill sets to test the security of any other private firm who might require his services. Finally, he writes up technical reports detailing all his work and publishes them in order to allow other researchers to help him improve his work. |

**Table 3: Persona for DMARF**

### 2.1.2 GIPSY

Personas, Actors and Stakeholders are identified by researching papers and understanding the architecture and the application domains of the GIPSY system.

**Users and Stakeholders of GIPSY**

**Researchers:** Researchers defines the declarative and explanatory analysis of an event with the underlying principles. Each case study is considered as a research strategy. They involve in decision making, concrete data techniques and methodology paradigms. In GIPSY architecture, the case study developers defines a flexible platform to identify the communication between application systems and external systems. A student can play the role of the researcher. The role of the student is to learn the application under the guidance of the professor. She/he can ask queries like execution and maintenance of the application.

**Application Developers:** Application developers define an integrated environment for supporting hybrid programming languages like Intensional and Imperative languages. This platform can be used for executing the programs written in various hybrid Intensional Programming Languages such as LUCID, JLUCID, and Objective LUCID etc.

**Architects**: General Developers are the developers who are responsible for writing the code of the application involved in the GIPSY framework. One such language is JLUCID programming language which comprises of Java classes and methods for implementation. As most of the research has been done with respect to research papers published by different professors from universities, a professor can play as an Architect. The role of the professor is to understand and execute the application under some valid conditions. She/he needs to go through the entire documentation of the designed framework in order to meet the specified requirements.

**Maintainers:** Maintainers can help in improving the system and identifying bugs and smells to further improve the system. As mentioned earlier, a student can be a maintainer and perform refactoring tasks on the GIPSY system by testing it with the application of different Hybrid Intensional Programming languages such as LUCID, JLUCID and OBJECTIVE LUCID.

**Persona:**

| Name | Maria Williams |
|---|---|
| Age | 38 |
| Job title | Application Developer at Oracle Corporation |
| Experience | Expert in the field of Data processing and Algorithm designs |
| Goal | Implementing the Higher Order Intensional logic expressions with the Lucid programming language |
| Description | Maria is interested in developing a logical expression language which can provide context oriented multidimensional reasoning of various Intensional expressions built around the GIPSY architecture. Using her experience with the evaluation higher Order Logical Expressions and Hybrid Intensional Programming, she develops applications which can be used in Security Frameworks like and Intensional Cyberforensics and Self-Forensics. |

**Table 4: Persona for GIPSY**

### 2.2 Use Cases

#### 2.2.1 DMARF

In this section we look at a scenario (fully dressed use case) with respect to the Application Domain of DMARF.

| | |
|---|---|
| **Use Case** | **Speaker Identification** |
| **Primary Actor** | **Developer** |
| **Stakeholders and Interests** | **Security Analyst**: wants accurate and fast identification of **voice samples**<br>**Forensic Analyst:** analyzes **voice samples** to use it as evidence in criminal proceedings<br>**Student:** wants to learn about the working pattern of the **system**<br>**Professor:** wants to update the **system** in order to produce a better **system** |
| **Pre-conditions** | 1. Clear and legible **voice samples** to be present for experimentation |
| **Post-conditions** | 1. **System** will be trained for the new **samples** provided by the **developer**<br>2. New **voice samples** will be stored in the **database**<br>3. **System** will correctly classify the **voice samples** provided by the **developer** |
| **Main Success Scenario** | 1. The **developer** uploads a **voice sample** into the **system**<br>2. More than one **sample** can be uploaded using the **samples** directory<br>**3.** Using the **SampleLoader** service the **system** converts the **sample** to a corresponding format for **further pre-processing**<br>4. The **system** performs **pre-processing** of the **voice sample** by using various **algorithms**(e.g. **Filtering, normalization**)<br>5. The **system** performs **feature extraction** on the **voice sample** and extracts various features using **algorithms** such as, **FFT, LPC, Minmax** etc<br>6. If the user selects the "**Train**" option, the **system** performs training and updates the **database** according to the new **voice sample**<br>7. If the **user** selects the "**Recognize**" option, the **system** performs **classification** by comparing the provided **voice sample** with the existing **database** |
| **Extensions** | 1. a. The **system** throws an **error** if the uploaded **file** is not an **audio file** |
| **Special Requirements** | **Audio codecs** to identify different formats of **audio files** |
| **Technology and Data variations List** | 1. Microphone for audio recording |
| **Open Issues** | 1. What if the speaker is wrongly identified?<br>2. Will the system classify new samples and overwrite existing samples of the same user correctly?<br>3. What if the audio sample is of a very large duration?<br>4. What if the audio sample is recorded in a very noisy environment? Will it affect the preprocessing phase? |

**Table 5: Fully Dressed Use-Case for DMARF**

### 2.2.2 GIPSY

In this section we look at a scenario (fully dressed use case) with respect to the Application Domain of GIPSY.

| | |
|---|---|
| **Use Case** | Evaluating a **LUCID Program** |
| **Primary Actor** | **Developer** |
| **Stakeholders and Interests** | **Student:** wants to learn about the working pattern of the **system**<br>**Professor:** wants to update the **system** in order to produce a better **system**<br>**Application Developer:** wants to develop an application using **GIPSY system**<br>**Case Study Developer:** wants to understand the **GIPSY system** to generate different **case studies** based on how the **system** performs under different circumstances |
| **Pre-conditions** | 1. The **developer** should specify **data** to be processed<br>2. The transformations to be applied on the computed **data** should be mentioned |
| **Post-conditions** | 1. The **LUCID program** is compiled and **code segments** are generated<br>2. The **code segments** are divided according to specific languages and are defined by **language tags**<br>3. **Demands** are generated from the **GEE (General Eduction Engine)** by **the DGT (Demand Generator Tier)** for the **GEER (General Eduction Engine Resources)**<br>4. The generated demands are stored in the **DST (Demand Store Tier)** |
| **Main Success Scenario** | 1. The **developer** enters the **LUCID program** into the **GIPSY environment**<br>2. The **GIPSY framework** compiles the **program** by using the **GIPC (Generic Intensional Programming Compiler)**<br>3. The **preprocessor** divides the compiled **code** into **code segments**<br>4. Each **code segment** is language specific and is identified by a **language tag**<br>5. The **Code Segments** are stored in the **GEER**<br>6. The **DGT** generates the **demands** according to the declaration and **resources** stored in **GEER**<br>**7.** The **DWT** processes these demands by executing the **methods** defined in **DGT dictionary**. These **demands** can either be **pending demands** or the **newly generated demands**<br>8. The **DST** migrates the **demands** and performs persistent storage of **demands** and resulting values<br>9. The **GMT (General Manager Tier)** remotely controls other **tiers** and exchanges the **system demands** |
| **Extensions** | 4.a The **code segments** can be visualized in terms of an **abstract syntax tree** |
| **Special Requirements** | **Graphical GMT Tool** Support for **the GIPSY Run- Time System** [37] |
| **Technology and Data variations List** | **General Intensional Compiler Framework (GICF)** that consists of a number of compiler interfaces which performs many functions such as preprocessing and generating **abstract syntax trees** |
| **Open Issues** | 1. Is it possible to integrate the **Demand Migration System (DMS)**? [23]<br>2. Is it possible to visualize and **control communication patterns**? [23]<br>3. Is it possible to introduce **functional language compilers**? [23]<br>4. Will this framework work in **security environments** and in dealing with **DDoS**? [23] |

**Table 6: Fully Dressed Use-Case for GIPSY**

### 2.3 Domain models and UML Diagrams

A domain model illustrates meaningful concepts for DMARF and GIPSY and it represents real-world concepts.

It describes various entities, their attributes, roles and relationships including the constraints that govern the problem domain [41].

#### 2.3.1 DMARF

**Domain Description:**

The Domain Diagram for the GIPSY case study can be seen in figure 13. **DMARF** is an extension of the **MARF** case study whose architecture is based on the **pattern recognition** pipeline making MARF one of the applications or in other words a subject of **Pattern Recognition**. There are a number of client applications that implement DMARF such as **Biometrics**, **Speaker Identification** and **Cyber forensics**. **DMARF** is composed of the **Front end** and the **Back end** modules. The **front end** contains the main modules that make up the pattern recognition pipeline. The flow of the process is sequential and the first module is **Sample Loading**, followed by **Preprocessing**, **Feature Extraction**, **Training** and **Classification**. The **Training** and **Classification** modules are followed by the **Feature Extraction** module and are executed based on user request. The responsibilities of each of these modules are explained as follows –

- **Sample loading**: The function of this module is to allow the incoming audio samples of different kinds of file types so that it a request is sent to the next module to process them.
- **Preprocessing**: The audio sample from the sample loader is fed into the preprocessor. This module responds to the sample loadings request and then preprocessing is done by different filters like CFE or FFT or by normalizing the loaded files which are received from sample loading module.
- **Feature Extraction**: This module allows the preprocessed data for further extraction of features by using different algorithms for training and classification. The feature extraction is split into two modules namely, training and classification.
- **Training:** It accepts feature vectors and updates the database with new samples for training sets. There are many approaches to analyze these feature vectors to allow accurate classification in the next phase.
- **Classification:** Classification is done on the current training sets. Additionally it is used to query the feature extraction service for feature vectors.

The **Back end** is responsible for making the system fault tolerant and reliable. **Back end** consists of three main components. Firstly, the **Monitoring** module, which checks live ness of the system. In other words it checks whether the system is properly functioning or not. Secondly, the **disaster recovery** module helps to restore a system to a last best working state. These states are stored on a backup module. The **Backup** module forms an integral part of the **disaster recovery** process and stores the necessary files to restore a system. Thirdly, the **Replication** service creates a copy of the main modules of the front end in order to make a system highly available. It is associated with the **backup** module which contains a **backup** storage of the replicas which can be used to reconstruct the replica using a replication scheme [4].

The modules in the front end are associated with the **replication** module within the **back end** in order to provide a reliable result.

According to the analysis made there are few Stakeholders identified which are shown in the domain model. Relative explanation and description of each stakeholder is given below:

**Maintainers**: The maintainer is responsible to update or make changes to the system (DMARF). A student can be a maintainer who create reports based on the analysis performed on the system (DMARF) and then the reports are submitted to the professor. A professor can review the reports submitted by the students and is responsible to make changes to the system (DMARF) as required (Refactoring)

**Application Developers**: Application developers work on the client applications which implements DMARF. For instance DMARF is used in the domains of Speaker Identification, Cyber Forensics and Biometrics.

**Researchers**: A researcher is mainly responsible to understand the architecture and the working of the system. He can also develop case studies. The researcher can affect the success and failure scenarios. A professor and a student can act as a researcher for the DMARF project as it is an Open source project mainly derived from plenty of research work.

**Architects**: Architects are mainly responsible for designing the system. As this is mainly a research based case study and is mainly based on the research papers written by professors from different Universities, the professor can play the role of an architect. He can assigns students as architects as well to improve and add modules and new features to the system.

**Domain Model (DMARF):**

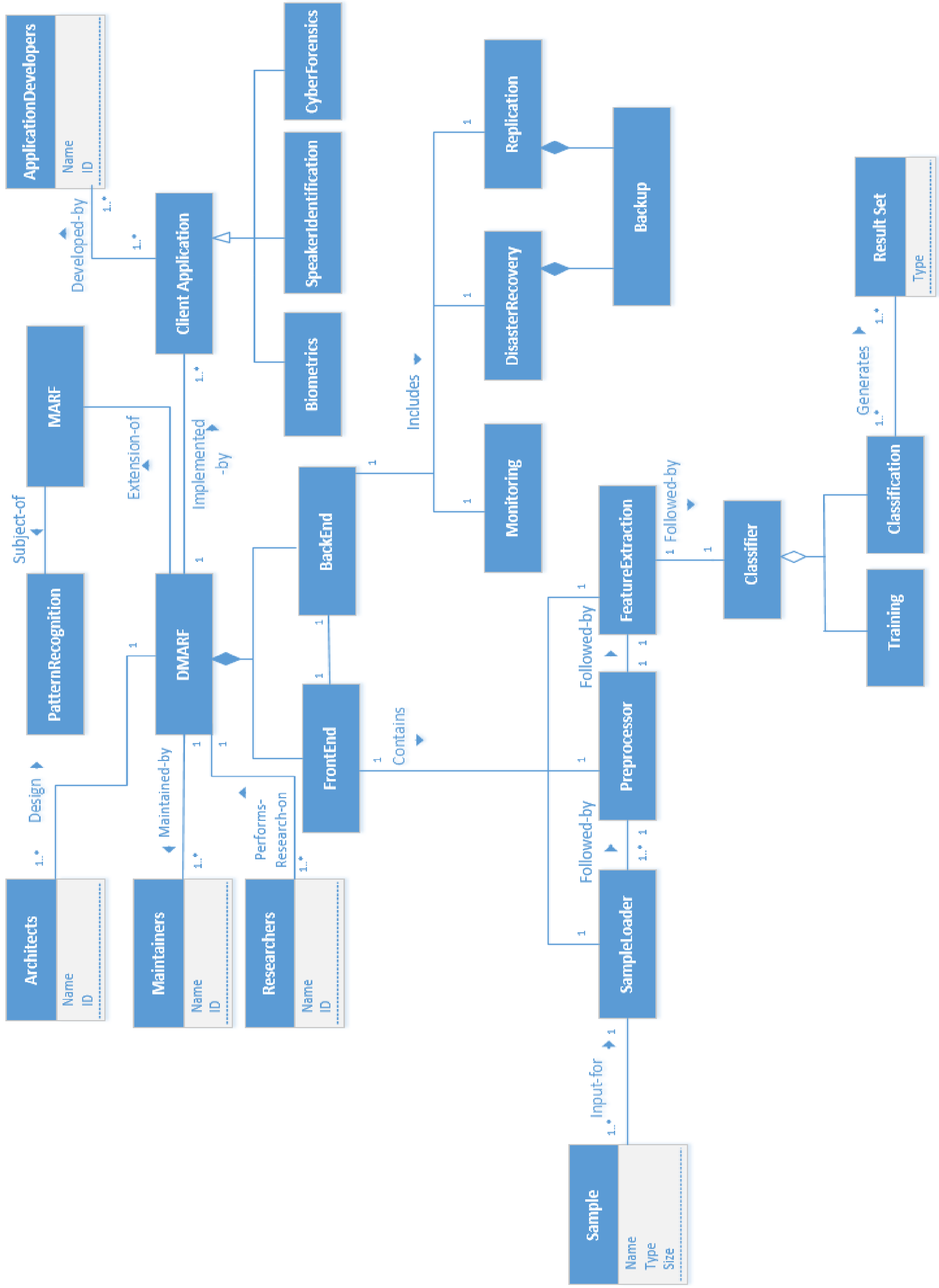

**Figure 13: Domain Model Diagram for DMARF**

### 2.3.2 GIPSY

**Domain Description:**

The Domain Diagram for the GIPSY case study is shown in figure 14. This is a brief description of the domain model based on the components and the relationships between the components.

**GIPSY** is a distributed framework that evaluates **Intensional Programming languages**. **LUCID JLUCID** and **Objective LUCID** form a family of Intensional programming Languages relationship. GIPSY as a system can be divided into three main modules which are **General Intensional Programming Compiler (GIPC)**, **General Eduction Engine (GEE)** and **Runtime Interactive Programming Environment (RIPE)**. Firstly, The General Intensional Programming Compiler is responsible for compiling the code. Each compiler consists of a preprocessor. It consists of a **preprocessor** which is responsible for splitting the program into **Code Segments**. The code segments are divided according to the type of the languages they are written in and are defined by the Language type attribute. The **General Eduction Engine** contains the **GIPSY Node** which acts as a Wrapper class for the **GIPSY Manager Tier (GMT), Demand Generator Tier (DGT), Demand Store Tier (DST)** and the **Demand Worker Tier (DWT)**. The **GIPSY Node** module issues a request to the **GIPSY Manager Tier**. The GIPSY Manager Tier is responsible for managing **–**

- **Demand Store Tier**
- **Demand Generator Tier**
- **Demand Worker Tier**

The **Demand Generated Tier** generates demands and these demands are stored in the **Generic Eduction Engine Resources (GEER)** along with the **code segments**. Code segments refer to chunks of code broken down into simpler forms that are carried forward for processing different requests. **GEER** also acts as a dictionary of resources. It provides the necessary information about the runtime resources that are generated by the compiler **GIPC** from Hybrid Intensional programs. **DGT** is also responsible for generating **Intensional Demands**. The **Demand Store Tier** exposes **Transport Agents**. These **Transport Agents** are used to deliver demands between demand store, demand worker and demand generator. The **Demand Worker Tier** retrieves a request (demand) from the **Demand Store Tier** and processes **Procedural Demands**. **Procedural Demands** are the demands that are generated to execute functions or methods defined in a procedural language and the **Intensional Demands** are the demands that are generated to execute functions or methods defined in an Intensional language. Therefore the **Intensional** and **Procedural demands** are related using association in the domain diagram. **Runtime Interactive Programming Environment (RIPE)** is an essential part of GIPSY, it handles the user interaction aspect of the system. It consists of two editors, **Textual Editor** and **DFG Editor**.

A variety of **Client Applications** use the GIPSY resources as a backbone for their working and execution. **Hybrid Intensional Programs, Intensional Cyberforensics and Security Frameworks** form an example family of Client Applications.

After analyzing GIPSY as a system some of the stakeholders are identified. These are shown in the domain model, a relative explanation of each stakeholder can be described as below:

**Maintainers**: Maintainers are responsible to make changes to the system (GIPSY) as required. These include adding features or removing unnecessary features. Student performs analysis on the system and the corresponding reports are created. The professor reviews the reports that are submitted by the student and changes to the system are made accordingly if required.

**Application Developer**: The application developer makes changes to the applications that implements GIPSY for instance security frameworks, Intensional Cyber Forensics. The application developer is solely responsible to create more applications using GIPSY as a system or make changes to the existing applications. Few of which are shown in the domain diagram in figure 14.

**Researchers**: A Researchers role is to understand and develop case studies. Researchers mainly focus on use cases. These include success scenarios as well as failed use cases. The use case scenario provides detailed description about the system which can be used to understand the system in a better way.

**Architects**: Architects are mainly responsible for designing the system. As this is mainly a research based case study and is mainly based on the research papers written by professors from different Universities, the professor can play the role of an architect.

The stakeholders identified in both the systems have individual responsibilities to play. The reports created by one stakeholder can be used by other stakeholders if necessary to customize the system accordingly. Therefore the stakeholders are interrelated and are shown in the domain diagrams.

**Domain Model (GIPSY):**

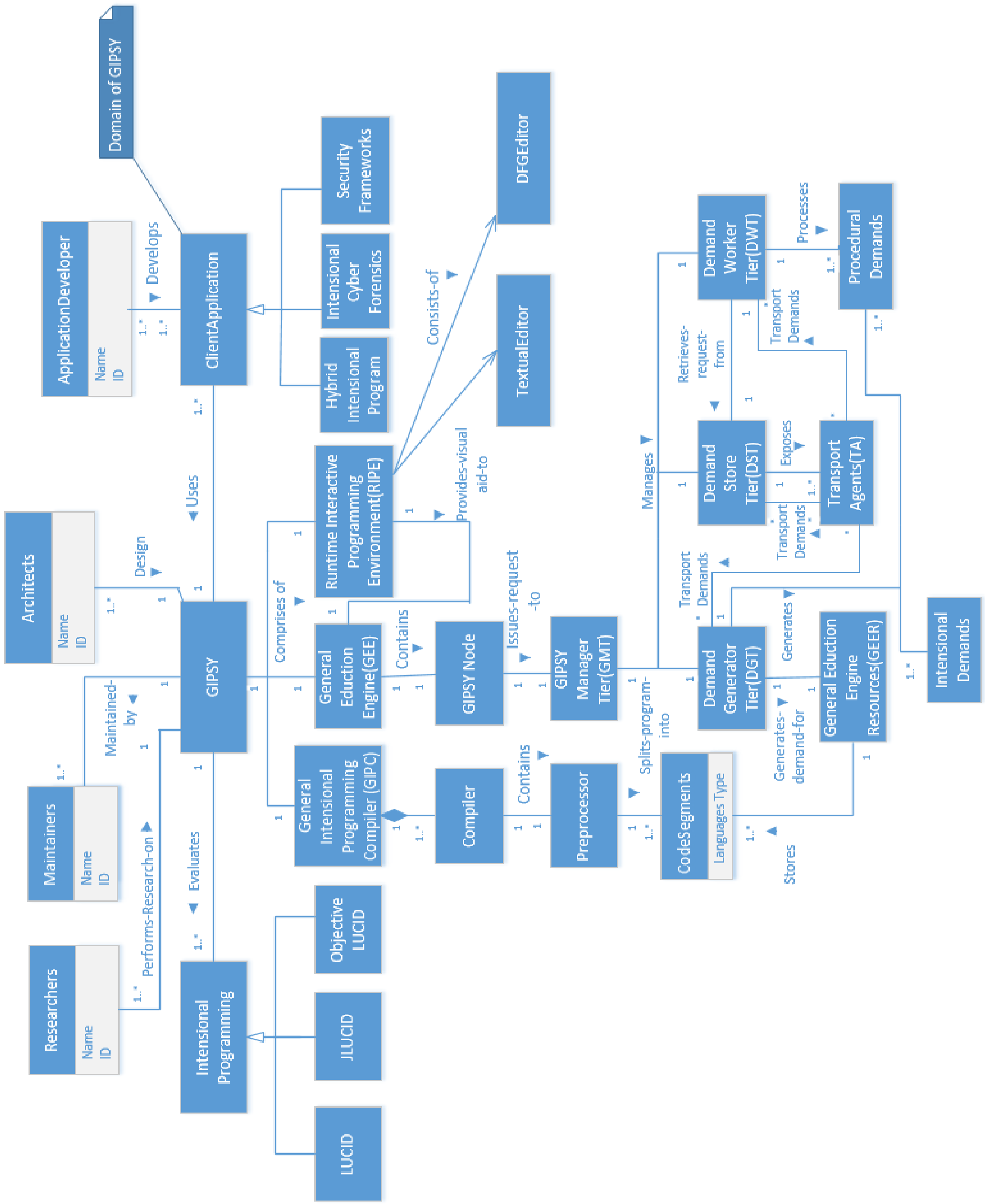

Figure 14: Domain Model Diagram for GIPSY

### 2.4 Fused DMARF-Over-GIPSY Run-time Architecture (DoGRTA)

The fused DMARF-Over-GIPSY Run-time Architecture (DoGRTA) can be seen in figure 15. In **DMARF,** the first pipeline stage is sample loading where different file types like wav, mp3, sine, text etc. can be loaded. To run **DMARF**'s sample loader in **GIPSY**'s multi-tier architecture, a **Sample Loader Demand Generator Tier** (SL-DGT) and a **Sample Loader Demand Worker Tier** (SL-DWT) are created. Here SL-DGT is an extension of **DGT** and SL-DWT is an extension of DWT. Here SL-DGT creates demands to examine the files which are basically source code files or compiled binary files. They are then stored as incomplete tasks in the **Demand Store Tier** (DST). The SL-DWT takes the unfinished works from the **DST** and then processes them using **Sample Loader** instance. When the output is generated, SL-DWT keeps them back into the **DST** as completed tasks. In the next preprocessing stage different kind of filters like low-pass, high-pass, band-pass, raw etc. can be chosen. This stage can run similarly in GIPSY's runtime environment. Similarly, **Preprocessor Demand Generator Tier** (P-DGT) and a **Preprocessor Demand Worker Tier** (P-DWT) are created to complete the preprocessing tasks. In the last two stages, feature extraction and training/classification where **Feature Extraction Demand Generator Tier** (FE-DGT) and a **Feature Extraction Demand Worker Tier** (FE-DWT) have been created to complete the feature extraction tasks. **Training/Classification Demand Generator Tier** and a **Training/Classification Demand Worker Tier** have been created to complete the training/classification tasks. The stakeholders in this system perform the same functions as discussed in the case studies.

### 2.5. Actual Architecture UML Diagrams

A UML class diagram describes the object and information structures used by your application, both internally and in communication with its users [39].

The conceptual classes give an abstraction whereas, the actual class architecture represent the structural aspect.

#### 2.5.1 DMARF

**Conceptual classes vs Actual Classes:**

The UML class diagram of DMARF can be seen in figures 16 and 17. Using the Conceptual model DMARF is denoted as an explicit component which is a subject of pattern recognition and a part of MARF. Using the actual system architecture DMARF is represented as a class with attributes and methods. DMARF is divided into two classes, front end and back end in both the architectures. BackEnd is dependent on FrontEnd. This relationship is not shown using conceptual classes as it deals only with the problem domain not the structure of the system. According to the conceptual classes Front end contains Sample Loader, Preprocessor and Feature Extraction. Input is provided to the Sample loader by a sample set denoted by a component SampleSet.In the actual class diagram frontend is associated with AudioSampleLoader, Preprocesser and Classification. Preprocessor is further classified into Endpoint, Dummy and Filter classes. This is represented using inheritance in class diagram. The classification class is also categorized as RandomClassification, Distance, Stochasctic and NeuralNetwork classes. Result set is obtained from the classification. The ResultSet is represented as an explicit class with methods. FeatureExtraction in domain model is followed by a classifier which has a Training and Classification components. Classification On the contrary FeatureExtraction in the actual system architecture is classified into sub classes namely, FFT, Segmentation, MinMaxAmplitudes, LPC, RandomFeatureExtraction, RawFeatureExtraction. These classes are linked using inheritance.

The BackEnd includes Monitoring, DisasterRecovery and Replication components. Backup is related to DisasterRecovery and Replication therefore an explicit component denoting the same is used. In the structural model Backend is generalized and methods are used to describe the functionalities for instance createBackup symbolizes that the class that creates a backup. Stakeholders are identified and a different class is used to define each stakeholder. ApplicationDeveloper as a component is directly associated with DMARF as he is responsible to make changes. Similarly, the Architects and Maintainers are responsible to design and maintain the system respectively. These components are denoted in the actual classes using classes that explain the same.

Therefore, the actual classes have a relative similarity with the conceptual classes however all the components from the conceptual classes are not present in the class diagram. For instance the stakeholders identified in the domain model are not represented in the class diagram as it deals with the system on the whole including the features of the software system. Actual classes include only the software components of the system while the conceptual classes help in visualizing the system on the whole including the stakeholders.

**Domain Model (DMARF-over-GIPSY):**

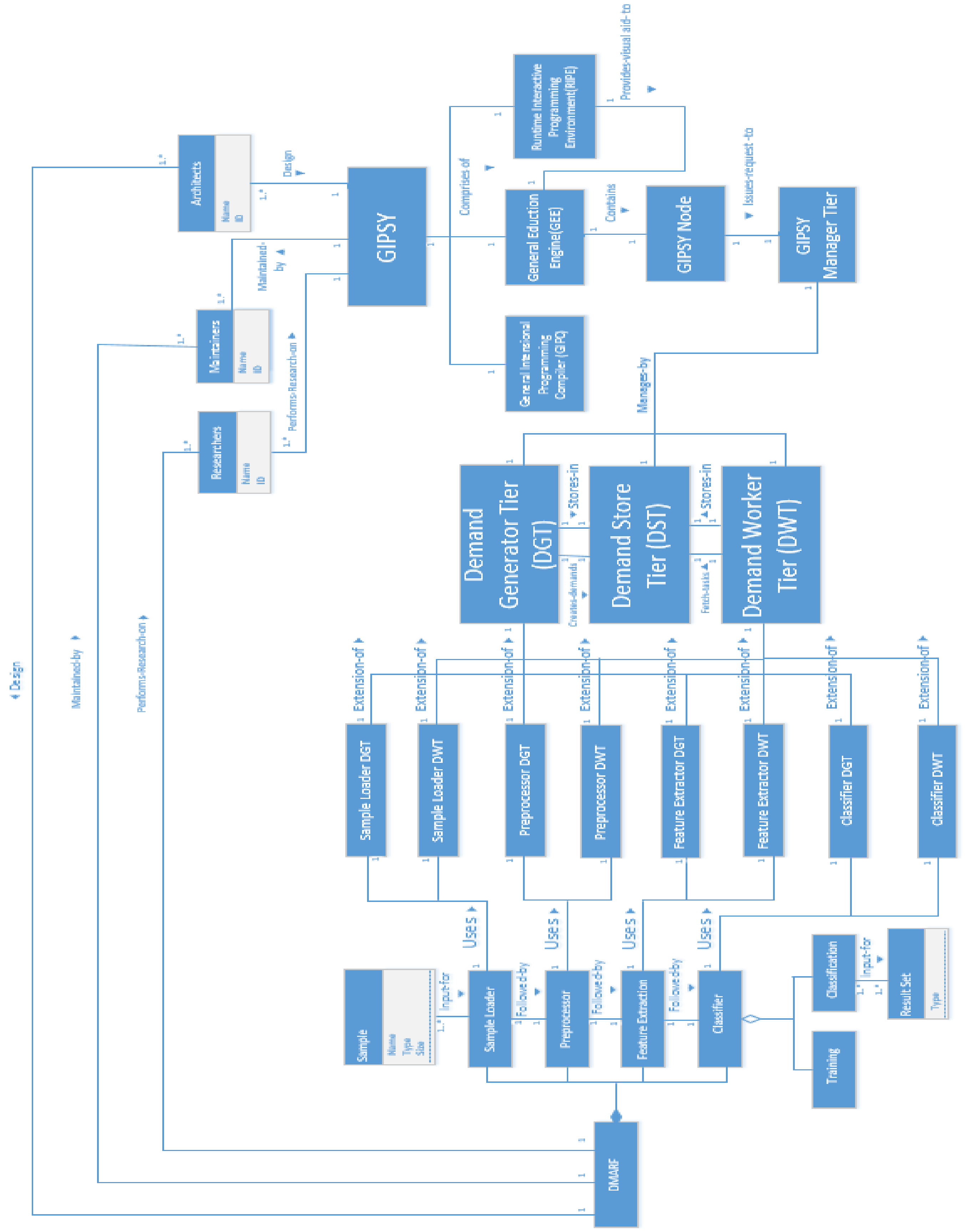

**Figure 15: Domain Model Diagram for Fused DMARF-Over-GIPSY**

## UML Class Diagram (DMARF)

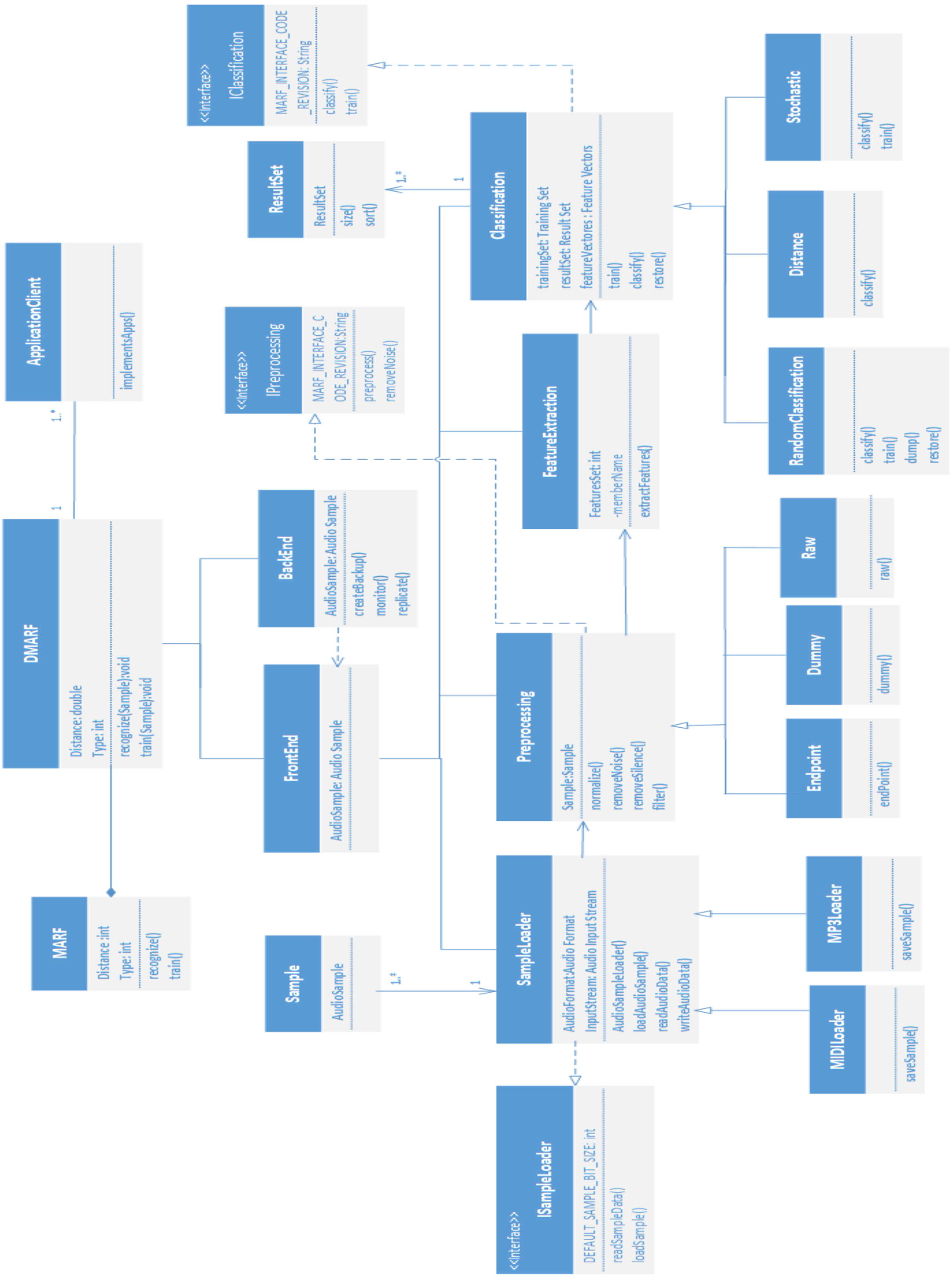

**Figure 16: Class Diagram for DMARF**

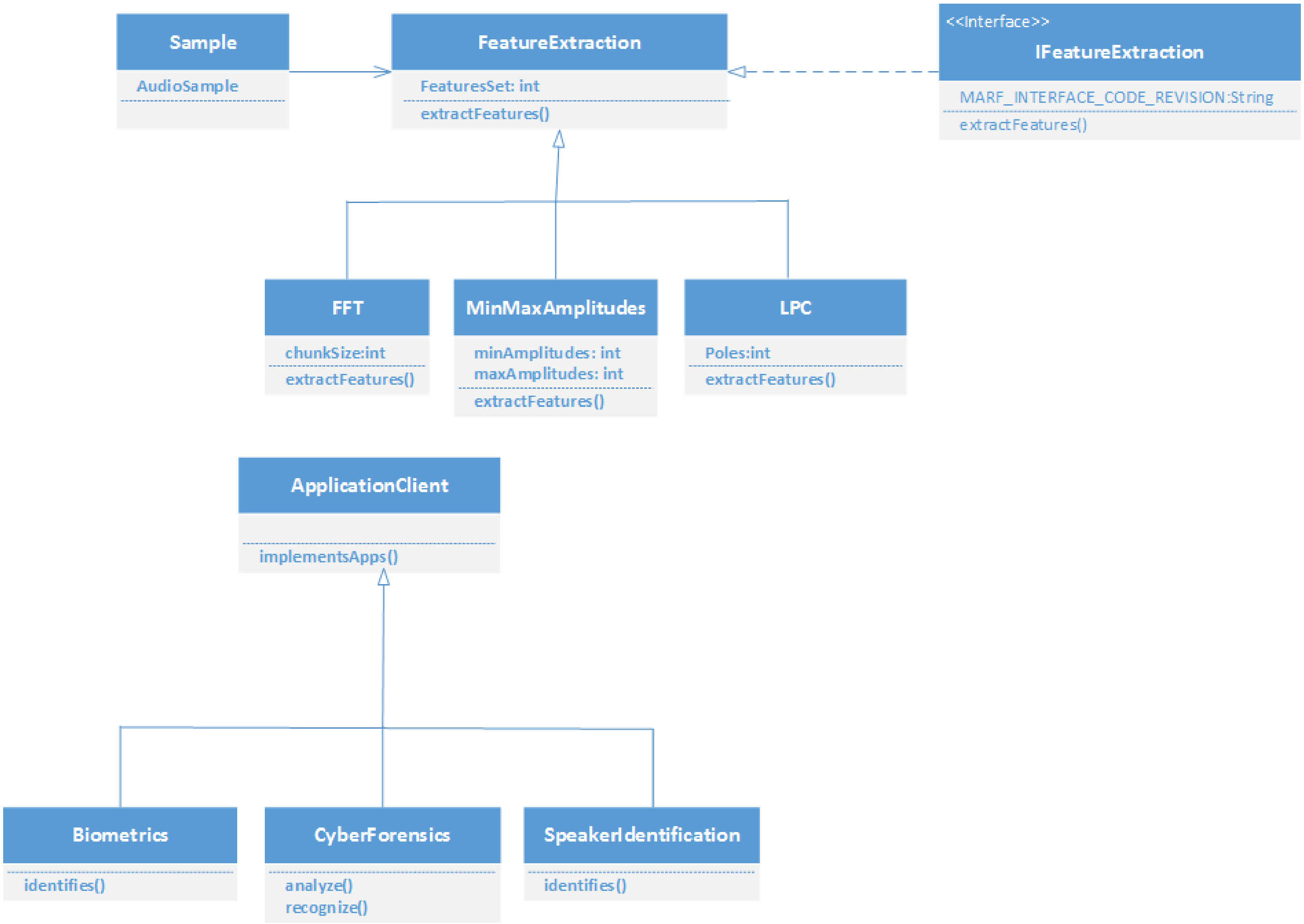

**Figure 17: Class Diagram for Feature Extraction and Client Applications in Detail**

### 2.5.2 GIPSY

**Conceptual classes vs Actual classes:**

The UML class diagram of DMARF can be seen in figures 18, 19 and 20. The "GIPSY" component from the conceptual classes is associated to the Intensional programming as GIPSY is a subject of Intensional Programming. In the actual architecture, GIPSY as a class is represented with attributes and methods that correspond to Intensional Programming. Instead of creating another class the actual class GIPSY is simplified to represent Intensional Programming that defines the structure of the system. Emphasis is laid on the structure of the system. GIPSY comprises of GIPC, GEE and RIPE. In the actual architecture (class diagram), GIPC is associated with an Intensional Compiler and a Preprocessor. The Intensional Compiler is extended to a translator. Additionally, the Intensional Compiler is further categorized into GIPLCompiler, LUCIDCompiler, JLUCIDCompiler, ObjectiveLucidComplier, LucxCompiler.

Preprocessor in the conceptual class model is represented as a generic component which is associated with a compiler. In the system architecture, the preprocessor contains methods to retrieve the code segments. The code segments are chunks of code represented as a different class in the actual system architecture. GEE comprises of a GIPSY node which sends a request to GIPSY manager Tier (GMT) that manages three tiers namely, Demand Generator Tier (DGT), Demand Worker Tier (DWT) and Demand Storage Tier (DST). In the actual system architecture, GEE is associated to a GIPSY node and later simplified into three tiers that are denoted by packages consisting of some classes. These are common to all the three tiers. To be specific, the Controller and Wrapper classes. The DGT Wrapper class is dependent on Procedural Demands and the DGT Controller class is dependent to generate Intensional Demands. RIPE consists of Textual Editor and DFG Editor. It is represented in a similar way in both the architectures.

Domain where GIPSY as a system is implemented is represented using different classes and components. Using conceptual classes the client applications are categorized as Hybrid Intensional Programs, Security Frameworks and Intensional Cyber Forensics. In the actual system architecture the same is followed including the working of the classes. Stakeholders are denoted using different classes and components in both the architectures. For instance the maintainers are responsible to maintain GIPSY as a system therefore a method is added to describe the same in the class diagram. On the other hand, here the domain model has only a generic component.

### UML Class Diagram for GIPSY

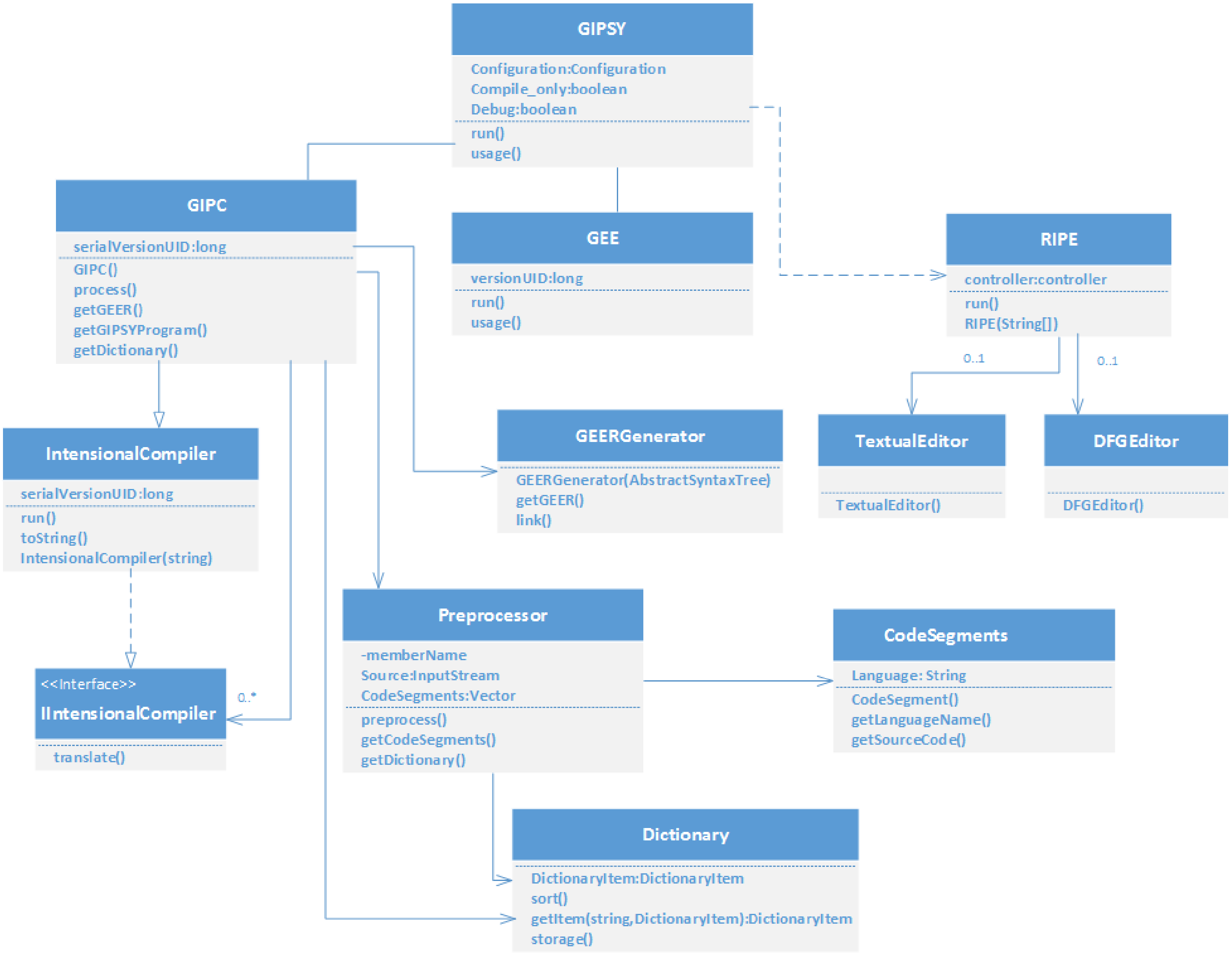

**Figure 18: Class Diagram for GIPSY**

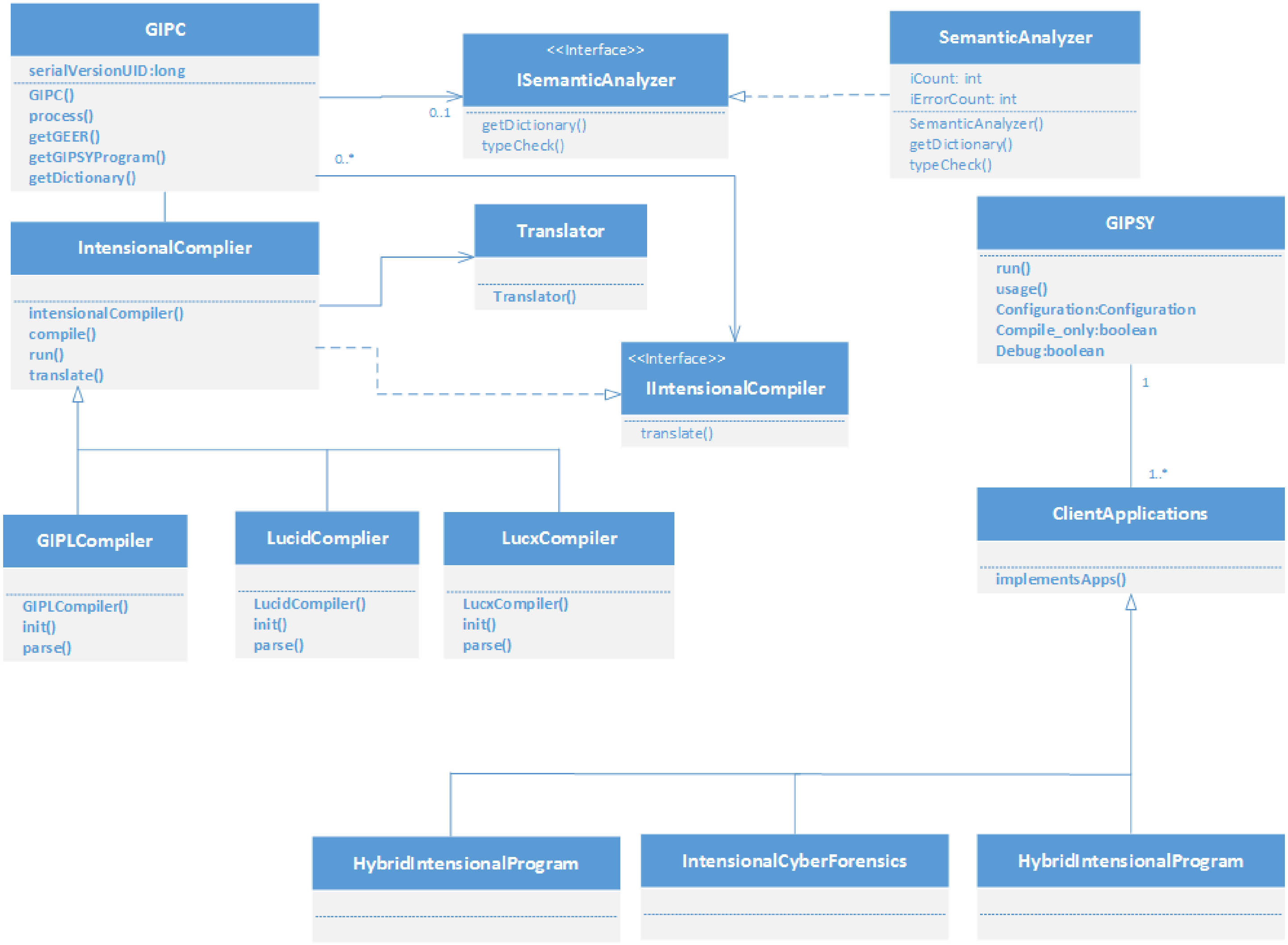

**Figure 19: Class Diagram for GIPC and Client Applications in Detail**

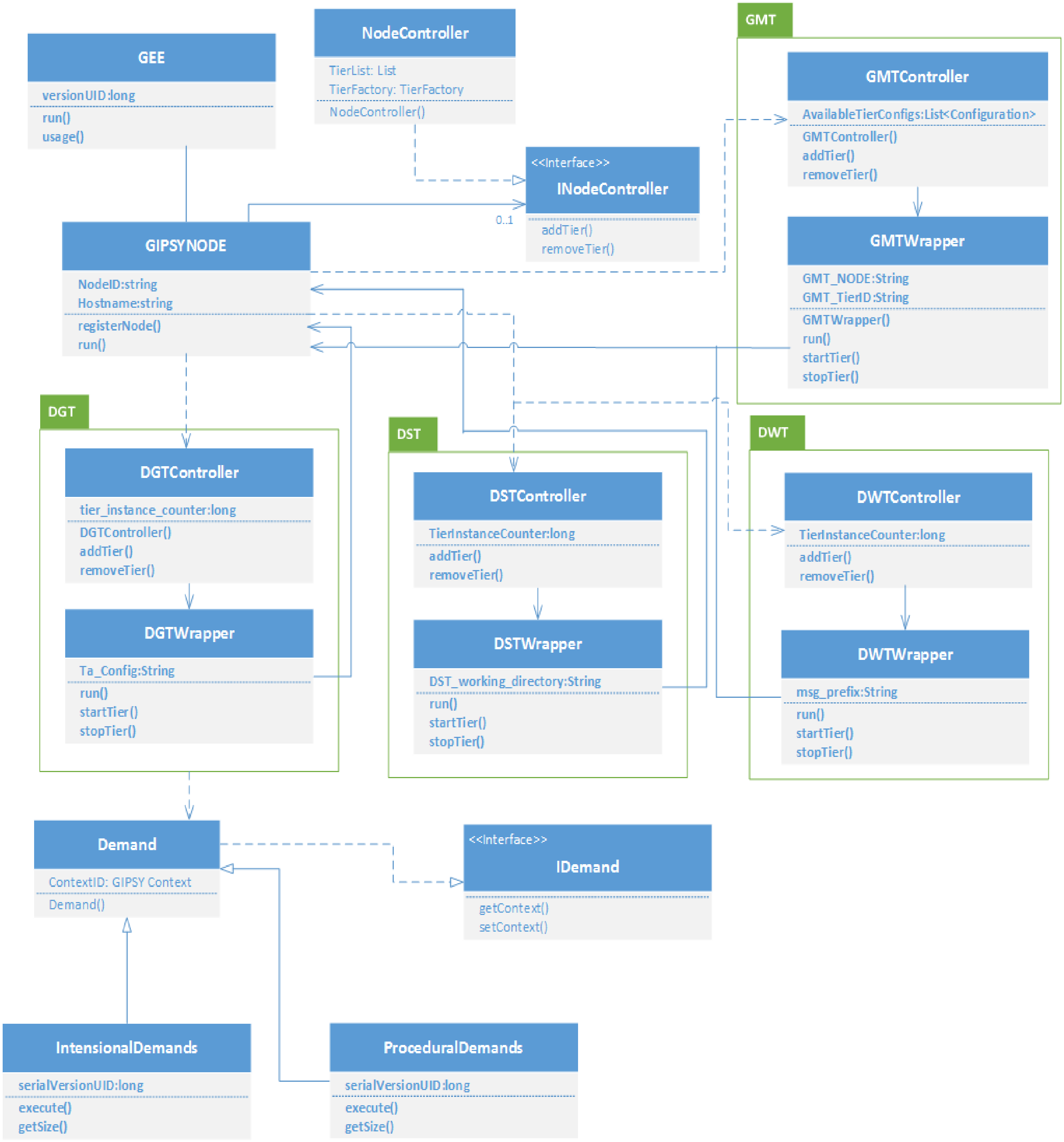

**Figure 20: Class Diagram for GEE in GIPSY**

### 2.5.3 ObjectAid UML

To re construct few classes for both the systems a reverse engineering tool is been used. This helped in constructing the UML diagrams for both the systems using the source code. The ObjectAid UML Explorer is an agile and lightweight code visualization tool for the Eclipse IDE [40]. ObjectAid UML can be used as a plugin in eclipse. Both the systems were first imported to Eclipse and later UML class diagrams were extracted.

The domain models were referred and simultaneously classes were chosen from the extracted project on Eclipse to obtain and simplified version of the class diagram. These were done by simply using "drag and drop" concept. The classes were dragged and dropped on to the workspace to see associations to the actual class. ObjectAid Class diagrams contain existing Java classes, interfaces, enumerations, annotations (collectively called *classifiers* henceforth in accordance with UML 2.0) as well as packages and package roots (i.e. JARs and source folders). Class diagrams only reflect the existing source code, which cannot be manipulated through the diagram [38].

### 2.5.4 Discrepancy between conceptual and actual classes

The conceptual classes when clubbed together makes a "domain model". Domain model deals with the entities within the problem domain. It does not deal with the software components. On the other hand, actual classes i.e. the UML Class Diagram mainly emphasis on the object schema in detail [41].

Domain models cannot be used to code whereas the actual classes can be used to develop a system. Actual classes of the both the systems were obtained using Reverse Engineering. ObjectAid UML was used as reference for both the systems.

The Domain Model covers all layers involved in modelling a business domain ensuring effective communication at all levels of engineering. Making it easy for Software Development [41].

The actual classes when clubbed together form a UML Class Diagram. The purpose of the class diagram is to model the static view of an application. It is also used for the following [41] –

- To describe responsibilities of a system
- As a base for component and deployment diagrams
- Forward and reverse engineering techniques

Therefore both the models are equally important during the software development cycle. Analyzing the problem domain and constructing the classes to create a static view of the system form a major part of the development cycle. Designing the architecture of the system becomes easy for the developer because of the simple conceptual view of the systems extracted using the domain model. The models together identify design patterns which help in improving the way the software is built where large scale problems of overall architecture come into picture. Most importantly it bridges the gap between how non-technical stakeholders view the architecture of the system when compared to the developers.

### 2.5.5 Relationship between two classes of DMARF and GIPSY

**All the code mentioned in this report has been taken from the case studies which can be found online at [43] and [44].**

**I. DMARF:**

In DMARF the `marf.Storage.Loaders` package has `AudioSampleLoader.java` class which takes audio sample input and to preprocess the sample it sends it to the class `Preprocessing.java` which is under package `marf.Preprocessing` and is shown in figure 21.

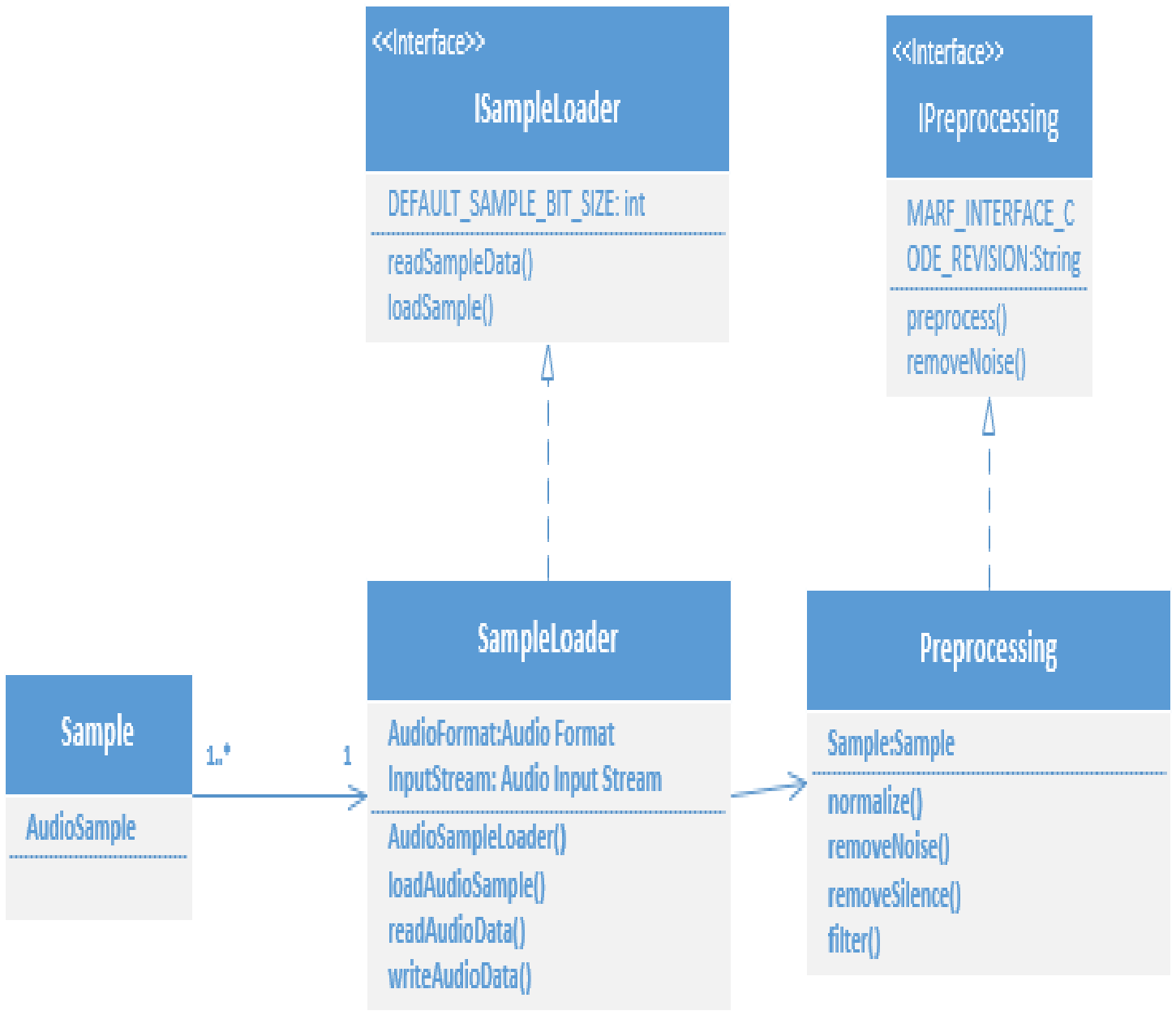

**Figure 21: Relationship between `AudioSampleLoader.java` and `Preprocessing.java`**

*Code Stub***:**

```
public abstract class AudioSampleLoader extends SampleLoader
{

	protected AudioFormat oAudioFormat = null;
	protected AudioInputStream oAudioInputStream = null;
	protected AudioFormat.Encoding oEncoding = AudioFormat.Encoding.PCM_SIGNED;

	public AudioSampleLoader()
	{
		super();
	}

	public Sample loadSample(byte[] patFileData)
	throws StorageException
	{
		return loadSample
		(
			new AudioInputStream
			(
				new BufferedInputStream
				(
					new ByteArrayInputStream(patFileData)
				),

				this.oAudioFormat,
				patFileData.length / this.oAudioFormat.getFrameSize()
			)
		);
	}

	public int readAudioData(double[] padAudioData)
	throws StorageException
	{
		return readSampleData(padAudioData);
	}
	public int writeAudioData(double[] padSampleData, int piWords)
	throws StorageException
	{
		return writeSampleData(padSampleData, piWords);
	}

}

public abstract class Preprocessing extends StorageManager
implements IPreprocessing
{
	public final static double DEFAULT_SILENCE_THRESHOLD = 0.001;

	protected Sample oSample = null;

	protected double dSilenceThreshold = DEFAULT_SILENCE_THRESHOLD;


	protected boolean bRemoveNoise = false;


	protected boolean bRemoveSilence = false;

public boolean removeNoise()
```

```
        throws PreprocessingException
        {
                LowPassFilter oNoiseRemover = new LowPassFilter(this.oSample);


                oNoiseRemover.bRemoveNoise = false;
                oNoiseRemover.bRemoveSilence = false;

                boolean bChanges = oNoiseRemover.preprocess();

                this.oSample.setSampleArray(oNoiseRemover.getSample().getSampleArray());
                oNoiseRemover = null;

                return bChanges;
        }

        public boolean removeSilence()
        throws PreprocessingException
        {
                this.oSample.setSampleArray(removeSilence(this.oSample.getSampleArray(),
                this.dSilenceThreshold));
                return true;
        }

        public final boolean normalize()
        throws PreprocessingException
        {
                return normalize(0);

        }
}
```

## II. GIPSY:

Here in `gipsy.GEE.multitier` package `NodeController.java` class implements `INodeController.java` class and under `gipsy.GEE.multitier.DGT` package `DGTController.java` class inherit `NodeController.java` class. `DGTController.java` class creates tiers of its type and then controls them.

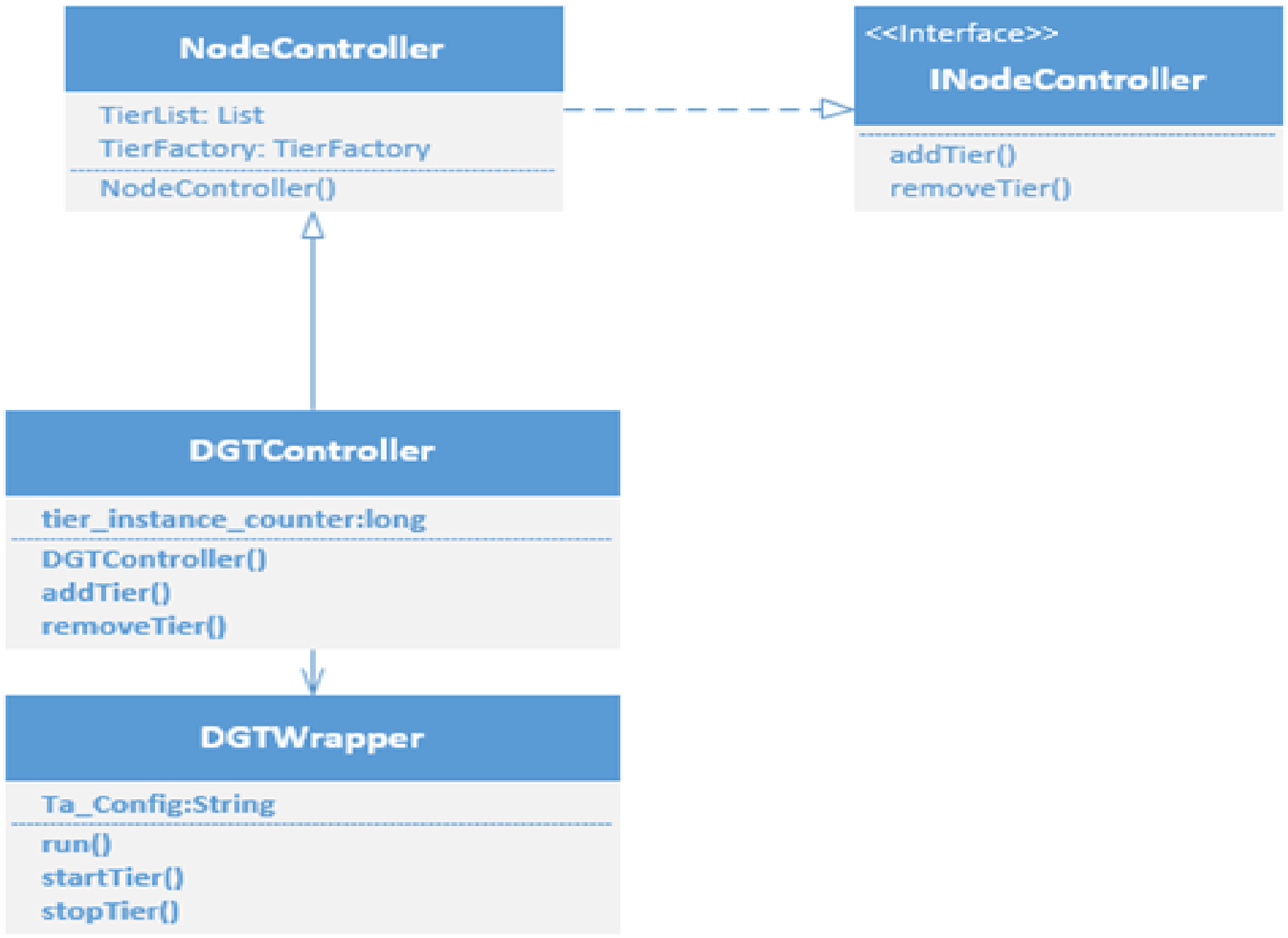

**Figure 22: Relationship between `NodeController.java` and `DGTController.java`**

*Code Stub:*

```
public abstract class NodeController
implements INodeController
{
	protected TierFactory oTierFactory;
	protected List<IMultiTierWrapper> oTierList = Collections.synchronizedList(new
ArrayList<IMultiTierWrapper>());
	protected List<Configuration> oAvailableTierConfigs = new ArrayList<Configuration>();
	protected TAExceptionHandler oTAExceptionHandler = null;

	public IMultiTierWrapper addTier(Configuration poTierConfig)
	throws MultiTierException
	{
		return null;
	}

	public void addTierConfig(Configuration poTierConfig)
	{
		this.oAvailableTierConfigs.add(poTierConfig);
	}

	public List<Configuration> getTierConfigs()
	{
		return this.oAvailableTierConfigs;
	}

	public void setTAExceptionHandler(TAExceptionHandler poTAExceptionHandler)
	{
		this.oTAExceptionHandler = poTAExceptionHandler;
	}

}
public class DGTController extends NodeController
{

	private Map<String, DGTWrapper> oActiveDGTs = new HashMap<String, DGTWrapper>();

	private long lTierInstanceCounter = 0;

public DGTController()

		this.oTierFactory = new DGTFactory();

	public void addTier(EDMFImplementation poDMFImp)
	{
		throw new NotImplementedException("addTier()");
	}

     public void addTier()
	{
		addTier(EDMFImplementation.JMS);
	}

public IMultiTierWrapper addTier(Configuration poTierConfig)
	throws MultiTierException
	{
		try
		{
			DGTWrapper oDGT = (DGTWrapper)
TierWrapperFactory.getInstance().createTier(poTierConfig);
```

```
                    this.oActiveDGTs.put(oDGT.getTierID(), oDGT);
                    oDGT.setTierID("" + this.lTierInstanceCounter);
                    this.lTierInstanceCounter++;
                    oDGT.setTAExceptionHandler(this.oTAExceptionHandler);
                    return oDGT;
            }
            catch(Exception oException)
            {
                    oException.printStackTrace(System.err);
                    throw new MultiTierException(oException);
            }
    }

    public void removeTier(EDMFImplementation poDMFImp)
    {
    }

public void removeTier()
    {
            throw new NotImplementedException("removeTier()");
    }

}
```

## 3. METHODOLOGY

Refactoring, identifying Design Patterns and Code Smells from system level architecture and making changes to the system is a very important duty of maintainer. Refactoring can include adding new features or it can also mean identifying anomalies and eliminating redundancies and unwanted smells from the code. There are many tools available in the online marketplace to perform refactoring. However, we use ObjectAid [38] and JDeodorant [42] to view the system level architecture and to identify code smells. Design patterns are identified by looking at the code manually. Identifying these design patterns help in gaining a broader understanding of the system architecture and aids in identifying bad smells. These bad smells can be removed and changed to improve the functionalities of the system by eliminating redundancies. The bad smells and changes made to specific classes are identified and discussed for both the DMARF and GIPSY case studies.

### 3.1 Refactoring

Refactoring is done to make changes to the system by removing bad smells and adding new functionalities to the case studies.

#### 3.1.1 Identification of Code Smells and System Level Refactorings

We use Logiscope [43] to analyze the code and identify bad smells with the help of the generated Kiviat graphs.

## I. MARF

**Case 1:**

The code smells have been identified in package `marf.Classification.NeuralNetwork`. The classes in which the bad smells are identified is `NeuralNetwork.java`. The red dots shown in the Kiviat graph as shown in figure 23 can be associated with the code smells for the class. Figure 24 shows the table generated by Logiscope which shows the values of the metrics along with their acceptable values.

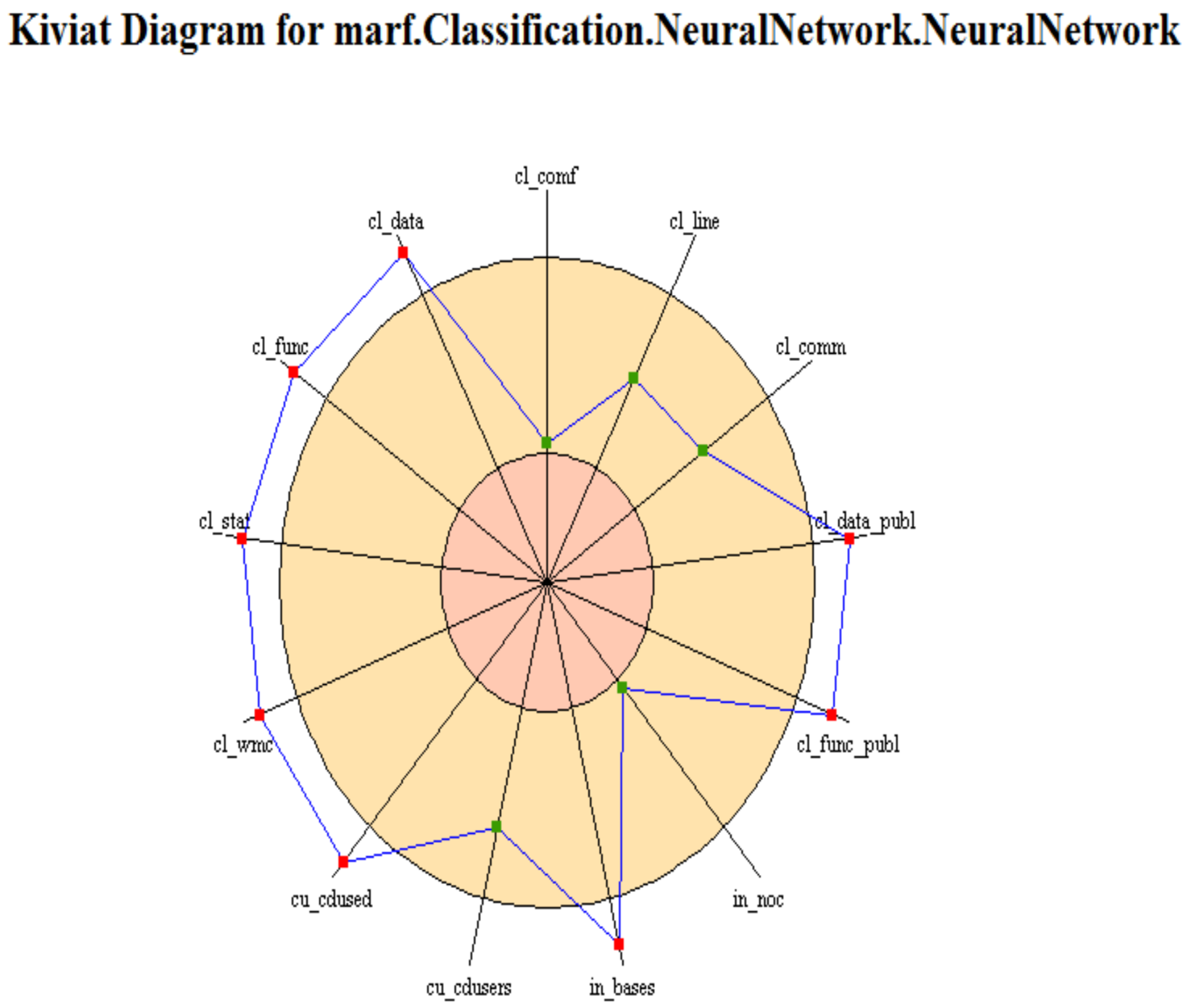

**Figure 23: Kiviat Diagram for `NeuralNetwork.java`**

| Metric : marf.Classification.NeuralNetwork.NeuralNetwork | Value | Min | Max | Status |
|---|---|---|---|---|
| cl_comf: Class comment rate | 0.26 | 0.20 | +oo | 0 |
| cl_comm: Number of lines of comment | 356 | -oo | +oo | 0 |
| cl_data: Total number of attributes | 17 | 0 | 7 | -1 |
| cl_data_publ: Number of public attributes | 8 | 0 | 0 | -1 |
| cl_func: Total number of methods | 27 | 0 | 25 | -1 |
| cl_func_publ: Number of public methods | 21 | 0 | 15 | -1 |
| cl_line: Number of lines | 1347 | -oo | +oo | 0 |
| cl_stat: Number of statements | 372 | 0 | 100 | -1 |
| cl_wmc: Weighted Methods per Class | 115 | 0 | 60 | -1 |
| cu_cdused: Number of direct used classes | 34 | 0 | 10 | -1 |
| cu_cdusers: Number of direct users classes | 3 | 0 | 5 | 0 |
| in_bases: Number of base classes | 6 | 0 | 3 | -1 |
| in_noc: Number of children | 0 | 0 | 3 | 0 |

**Figure 24: Table depicting metric values for `NeuralNetwork.java`**

The bad smells identified with respect to the architecture of the system are listed as follows –

1. Large Number of Methods
2. Too many constants
3. Most of the child classes have the same methods
4. Too many functions
5. The number of directly used classes are very high which increases the coupling of these classes with the other classes leading to the increase of the complexity of the system
6. To many functions and data which are public
7. Too many base classes

Figure 25 shows the class diagram for the class.

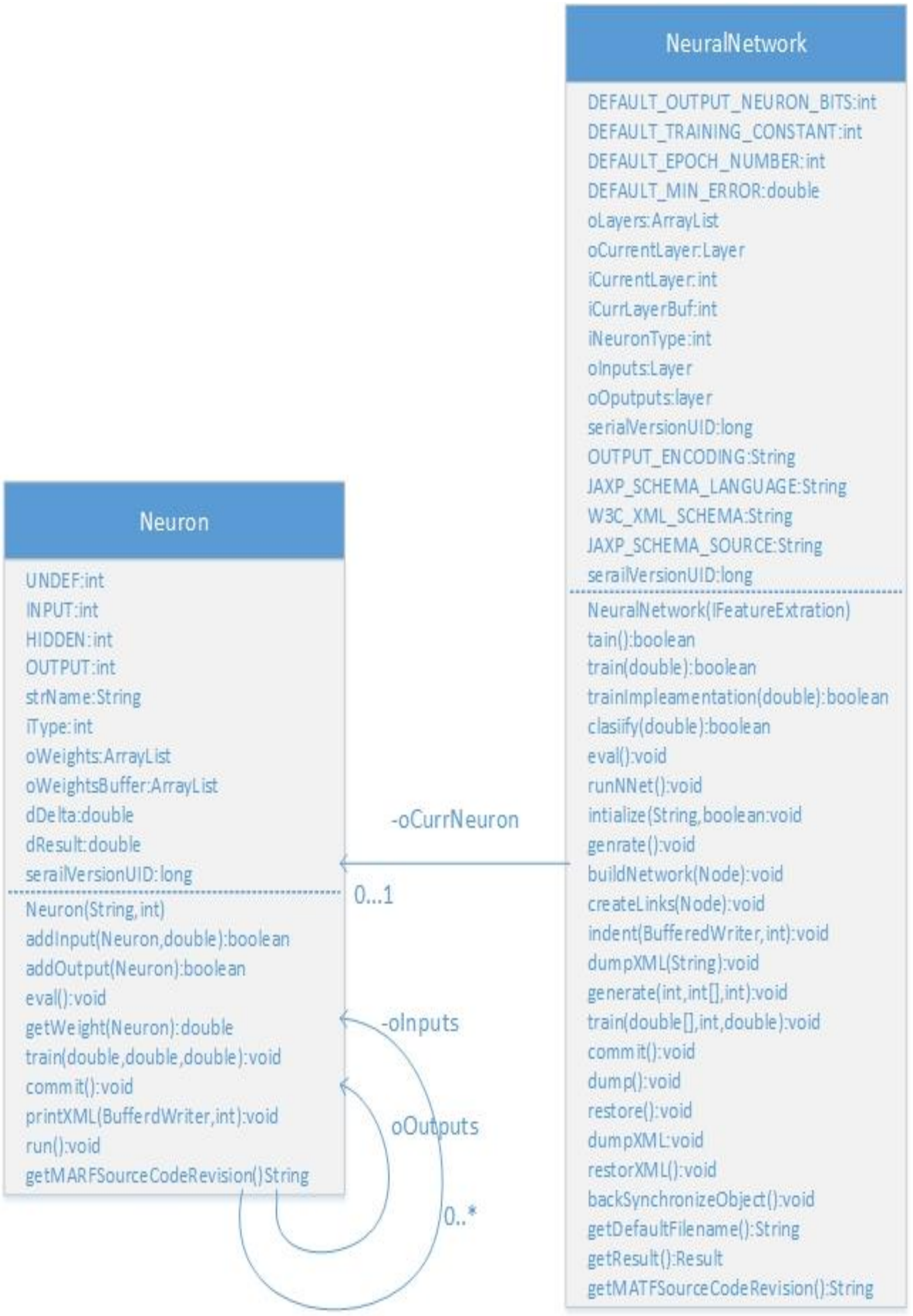

**Figure 25: `NeuralNetwork.java` class before Refactoring**

Université Concordia University

*Code Stub:*

**Class `NeuralNetwork.java`**

```
public class NeuralNetwork {
extends Classification
{
    public static final int DEFAULT_OUTPUT_NEURON_BITS = 32;
    public static final double DEFAULT_TRAINING_CONSTANT = 1;
    public static final int DEFAULT_EPOCH_NUMBER = 64;
    public static final double DEFAULT_MIN_ERROR = 0.1;
    private ArrayList<Layer> oLayers = new ArrayList<Layer>();
    private transient Layer oCurrentLayer;
    private transient int iCurrenLayer = 0;
    private transient int iCurrLayerBuf = 0;
    private transient Neuron oCurrNeuron;
    private transient int iNeuronType = Neuron.UNDEF;
    private Layer oInputs = new Layer();
    private Layer oOutputs = new Layer();
    public static final String OUTPUT_ENCODING = "UTF-8";
    private Layer oOutputs = new Layer();
    public static final String JAXP_SCHEMA_LANGUAGE =
"http://java.sun.com/xml/jaxp/properties/schemaLanguage";
    public static final String W3C_XML_SCHEMA =
            "http://www.w3.org/2001/XMLSchema";
    public static final String JAXP_SCHEMA_SOURCE =
            "http://java.sun.com/xml/jaxp/properties/schemaSource";
    private static final long serialVersionUID = 6116721242820120028L;

    public NeuralNetwork(IFeatureExtraction poFeatureExtraction)
      {
            super(poFeatureExtraction);
            this.iCurrentDumpMode = DUMP_GZIP_BINARY;
            this.strFilename = getDefaultFilename();
      }
    public final boolean train() throws ClassificationException
    public final boolean train(double[] padFeatureVector) throws ClassificationException
    private final boolean trainImplementation(double[] padFeatureVector)
      throws ClassificationException
    public final boolean classify(double[] padFeatureVector)
      throws ClassificationException
    public void generate()throws ClassificationException
    private final void buildNetwork(Node poNode)
    private final void createLinks(Node poNode)throws ClassificationException
     public final void setInputs(final double[] padInputs)throws ClassificationException
    public double[] getOutputResults()

    public static final void indent(BufferedWriter poWriter, final int piTabsNum)
      throws IOException

    public final void dumpXML(final String pstrFilename)
      throws StorageException

    public final void generate(int piNumOfInputs, int[] paiHiddenLayers,
    int piNumOfOutputs) throws ClassificationException
   public final void train(final double[] padInput, int piExpectedLength,
   final double pdTrainConst) throws ClassificationException
   public final void commit()

   private final int interpretAsBinary()

   public void dump()throws StorageException
   public void restore()throws StorageException
   public void dumpXML()throws StorageException
```

```
public void restoreXML()throws StorageException
protected String getDefaultFilename()
public Result getResult()
private static class NeuralNetworkErrorHandler implements ErrorHandler }
```

**Case 2:**

The code smells have been identified in package `marf.` The classes in which the bad smells are identified is `MARF.java`. Figures 26 and 27 show the Kiviat graph and the table of the values for the parameters of the Kiviat Graph.

**Kiviat Diagram for marf.MARF**

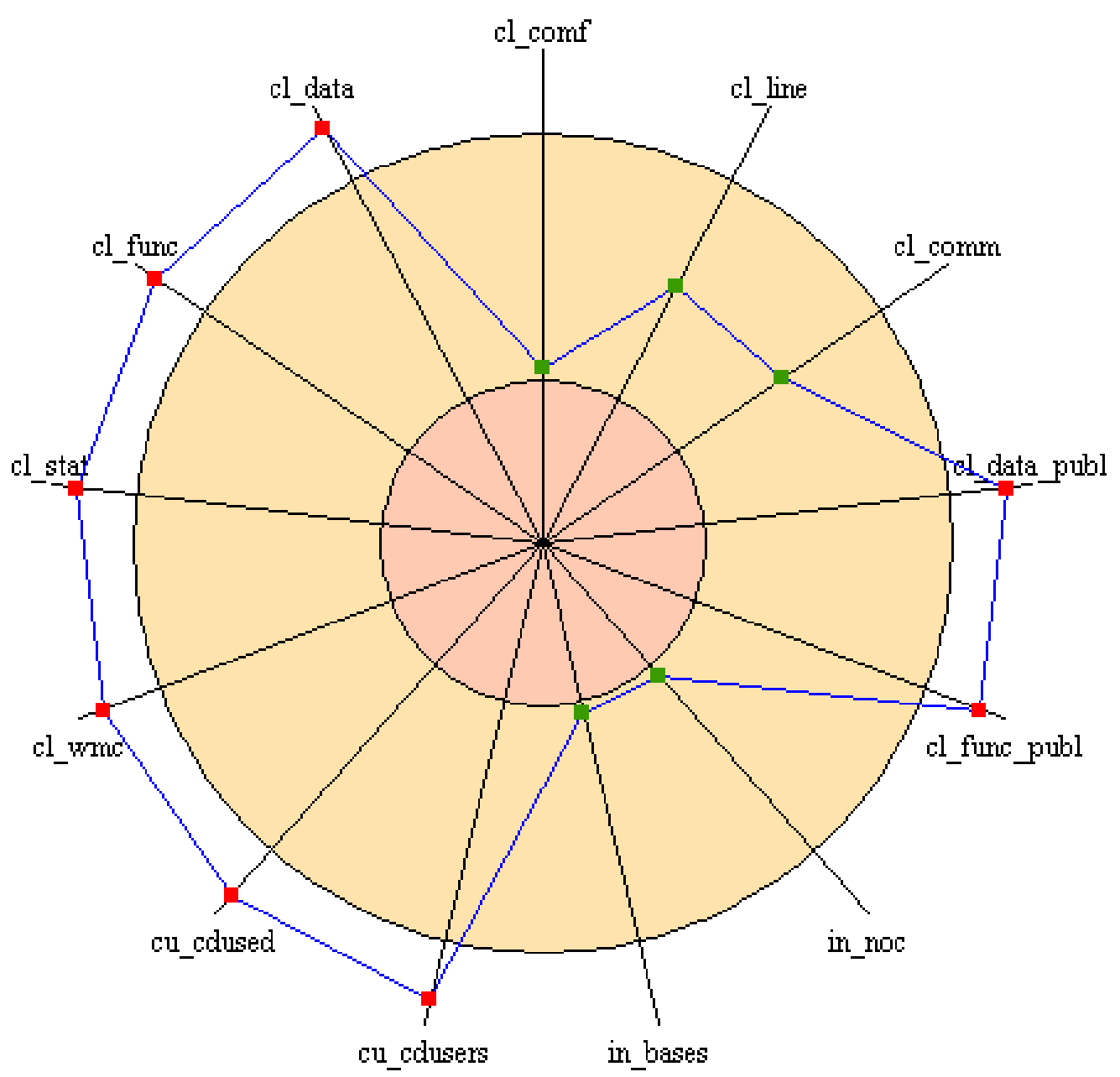

**Figure 26: Kiviat Diagram for `MARF.java`**

| Metric : marf.MARF | Value | Min | Max | Status |
|---|---|---|---|---|
| cl_comf: Class comment rate | 0.49 | 0.20 | +oo | 0 |
| cl_comm: Number of lines of comment | 897 | -oo | +oo | 0 |
| cl_data: Total number of attributes | 80 | 0 | 7 | -1 |
| cl_data_publ: Number of public attributes | 61 | 0 | 0 | -1 |
| cl_func: Total number of methods | 55 | 0 | 25 | -1 |
| cl_func_publ: Number of public methods | 52 | 0 | 15 | -1 |
| cl_line: Number of lines | 1831 | -oo | +oo | 0 |
| cl_stat: Number of statements | 213 | 0 | 100 | -1 |
| cl_wmc: Weighted Methods per Class | 79 | 0 | 60 | -1 |
| cu_cdused: Number of direct used classes | 26 | 0 | 10 | -1 |
| cu_cdusers: Number of direct users classes | 38 | 0 | 5 | -1 |
| in_bases: Number of base classes | 0 | 0 | 3 | 0 |
| in_noc: Number of children | 0 | 0 | 3 | 0 |

**Figure 27: Table depicting metric values for `MARF.java`**

The bad smells identified with respect to the architecture of the system are listed as follows –

1. Large Number of Methods
2. Too many constants
3. Most of the child classes have the same methods
4. Too many functions
5. The number of directly used classes are very high which increases the coupling of these classes with the other classes leading to the increase of the complexity of the system
6. The number of user classes are more which increases coupling
7. To many functions and data which are public
8. Too many base classes

Figure 28 shows the class diagram for the class.

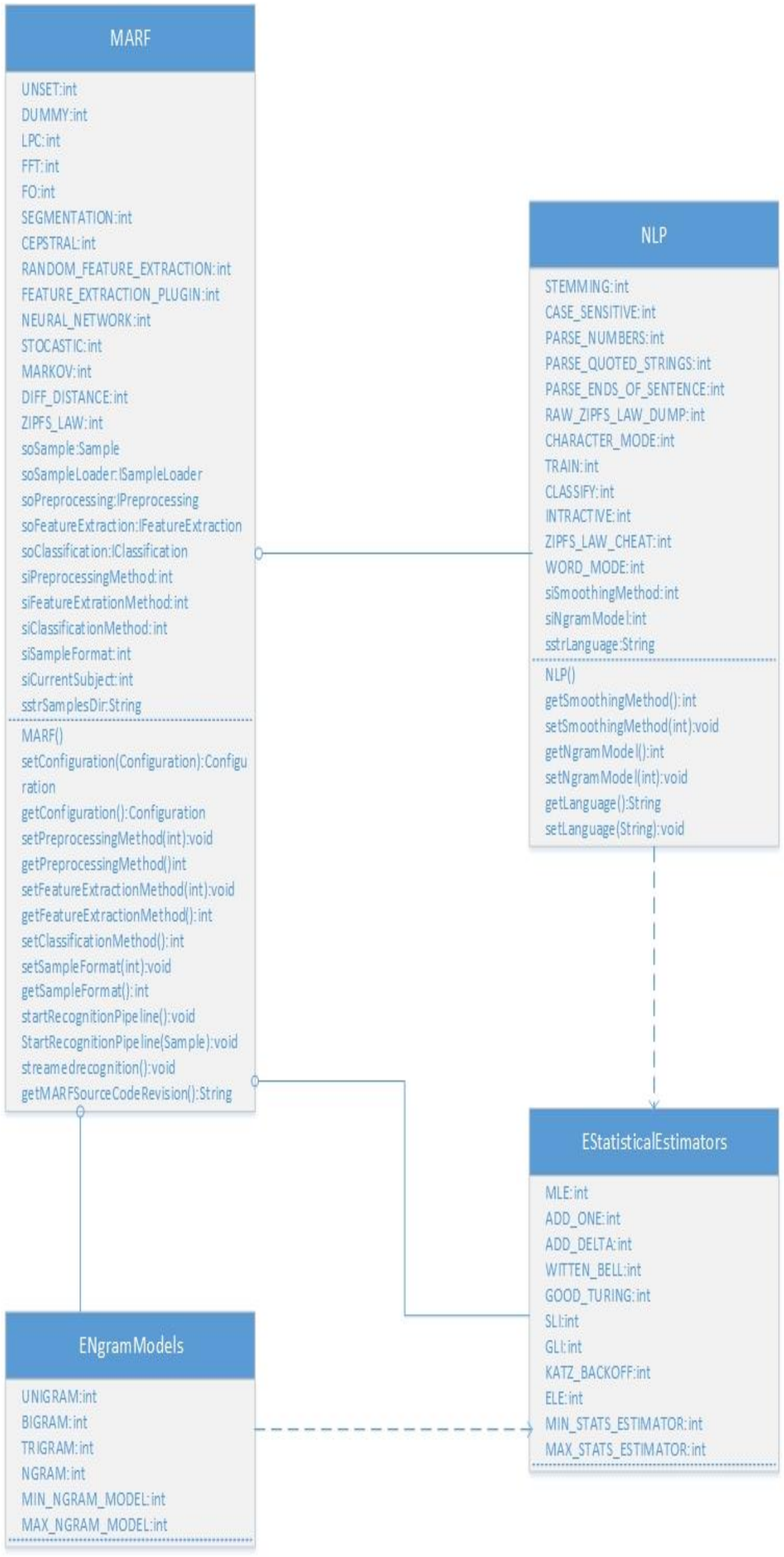

**Figure 28: `MARF.java class without Refactoring`**

*Code Stub:*

```
public class MARF {
public static final int DUMMY                           = 100;
public static final int HIGH_FREQUENCY_BOOST_FFT_FILTER = 101;
public static final int BANDPASS_FFT_FILTER             = 102;
public static final int ENDPOINT                        = 103;
public static final int LOW_PASS_FFT_FILTER             = 104;
public static final int HIGH_PASS_FFT_FILTER            = 105;
public static final int HIGH_PASS_BOOST_FILTER          = 106;
public static final int RAW                             = 107;
public static final int PREPROCESSING_PLUGIN            = 108;
public static final int LOW_PASS_CFE_FILTER             = 109;
public static final int HIGH_PASS_CFE_FILTER            = 110;
public static final int BAND_PASS_CFE_FILTER            = 111
public static final int BAND_STOP_CFE_FILTER            = 112;
public static final int BAND_STOP_FFT_FILTER            = 113;
public static final int MAX_PREPROCESSING_METHOD = BAND_STOP_FFT_FILTER;
public static final int MIN_PREPROCESSING_METHOD = DUMMY;
public static final int LPC                           = 300;
public static final int FFT                           = 301;
public static final int F0                            = 302;
public static final int SEGMENTATION                  = 303;
public static final int CEPSTRAL                      = 304;
public static final int RANDOM_FEATURE_EXTRACTION     = 305;
public static final int MIN_MAX_AMPLITUDES            = 306;
public static final int FEATURE_EXTRACTION_PLUGIN     = 307;
public static final int FEATURE_EXTRACTION_AGGREGATOR = 308;
public static final int MAX_FEATUREEXTRACTION_METHOD = FEATURE_EXTRACTION_AGGREGATOR;
public static final int MIN_FEATUREEXTRACTION_METHOD = LPC;
public static final int NEURAL_NETWORK            = 500;
public static final int STOCHASTIC                = 501;
public static final int MARKOV                    = 502;
public static final int EUCLIDEAN_DISTANCE        = 503;
public static final int CHEBYSHEV_DISTANCE        = 504;
public static final int MANHATTAN_DISTANCE        = 504;
public static final int CITYBLOCK_DISTANCE        = 504;
public static final int MINKOWSKI_DISTANCE        = 505;
public static final int MAHALANOBIS_DISTANCE      = 506;
public static final int RANDOM_CLASSIFICATION     = 507;
public static final int DIFF_DISTANCE             = 508;
public static final int CLASSIFICATION_PLUGIN     = 509;
public static final int ZIPFS_LAW                 = 510;
public static final int HAMMING_DISTANCE          = 511;
public static final int COSINE_SIMILARITY_MEASURE = 512;
public static final int MAX_CLASSIFICATION_METHOD = COSINE_SIMILARITY_MEASURE;
public static final int MIN_CLASSIFICATION_METHOD = NEURAL_NETWORK;
public static final int WAV    = MARFAudioFileFormat.WAV;
public static final int ULAW   = MARFAudioFileFormat.ULAW;
public static final int MP3    = MARFAudioFileFormat.MP3;
public static final int SINE   = MARFAudioFileFormat.SINE;
public static final int AIFF   = MARFAudioFileFormat.AIFF;
public static final int AIFFC  = MARFAudioFileFormat.AIFFC;
public static final int AU     = MARFAudioFileFormat.AU;
public static final int SND    = MARFAudioFileFormat.SND;
public static final int MIDI   = MARFAudioFileFormat.MIDI;
public static final int CUSTOM = MARFAudioFileFormat.CUSTOM;
public static final int TEXT   = MARFAudioFileFormat.TEXT;
public static final Map<Integer, String> MODULE_NAMES_MAPPING = new HashMap<Integer,
String>();
      static
      {
}
```

```
public static synchronized final Configuration setConfiguration(Configuration
poConfiguration)throws MARFException
public static synchronized final Configuration getConfiguration()
public static synchronized final void setPreprocessingMethod(final int
piPreprocessingMethod)throws MARFException
public static synchronized final ISampleLoader getSampleLoader()
public static synchronized final IPreprocessing getPreprocessing()
public static synchronized final IFeatureExtraction getFeatureExtraction()
public static synchronized final IClassification getClassification()
public static final void recognize()
public static final void recognize(Sample poSample)
public static final void train()
BDApublic static final void train(Sample poSample)
BDAprivate static final void startRecognitionPipeline()
BDAprivate static final void startRecognitionPipeline(Sample poSample)throws
MARFException }
```

**Case 3:**

The code smells have been identified in package `marf.storage`. The classes in which the bad smells are identified is `ResultSet.java`. Figures 29 and 30 show the Kiviat graph and the table of the values for the parameters of the Kiviat Graph.

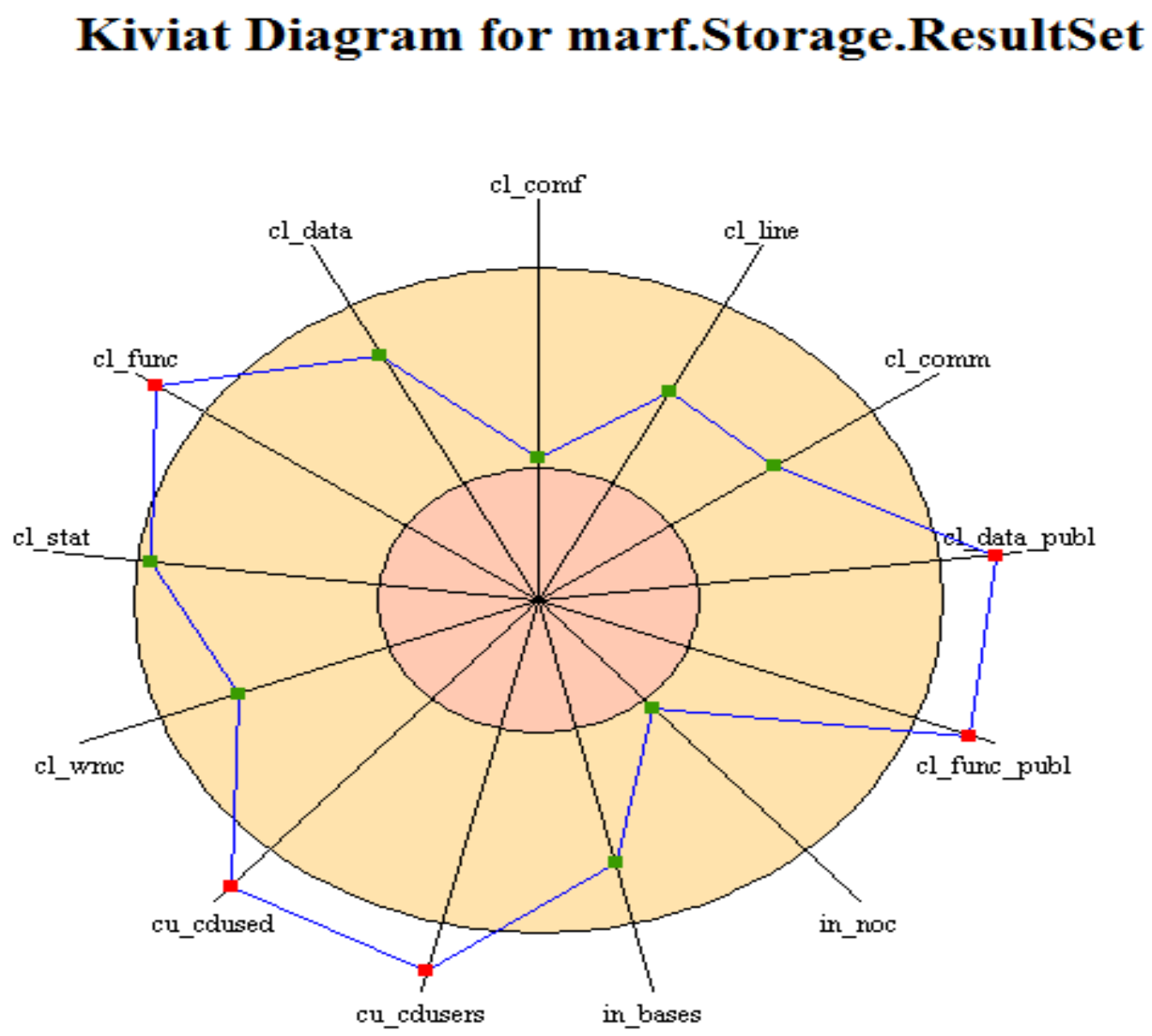

**Figure 29: `ResultSet.java` class without Refactoring**

| Metric : marf.Storage.ResultSet | Value | Min | Max | Status |
|---|---|---|---|---|
| cl_comf: Class comment rate | 0.48 | 0.20 | +oo | 0 |
| cl_comm: Number of lines of comment | 220 | -oo | +oo | 0 |
| cl_data: Total number of attributes | 5 | 0 | 7 | 0 |
| cl_data_publ: Number of public attributes | 1 | 0 | 0 | -1 |
| cl_func: Total number of methods | 28 | 0 | 25 | -1 |
| cl_func_publ: Number of public methods | 28 | 0 | 15 | -1 |
| cl_line: Number of lines | 454 | -oo | +oo | 0 |
| cl_stat: Number of statements | 94 | 0 | 100 | 0 |
| cl_wmc: Weighted Methods per Class | 38 | 0 | 60 | 0 |
| cu_cdused: Number of direct used classes | 13 | 0 | 10 | -1 |
| cu_cdusers: Number of direct users classes | 30 | 0 | 5 | -1 |
| in_bases: Number of base classes | 2 | 0 | 3 | 0 |
| in_noc: Number of children | 0 | 0 | 3 | 0 |

**Figure 30: Table depicting metric values for `ResultSet.java`**

The bad smells identified with respect to the architecture of the system are listed as follows –

1. Too many functions
2. The number of directly used classes are very high which increases the coupling of these classes with the other classes leading to the increase of the complexity of the system
3. The number of user classes are more which increases coupling
4. Too many functions and data which are public

Figure 31 shows the class diagram for the class.

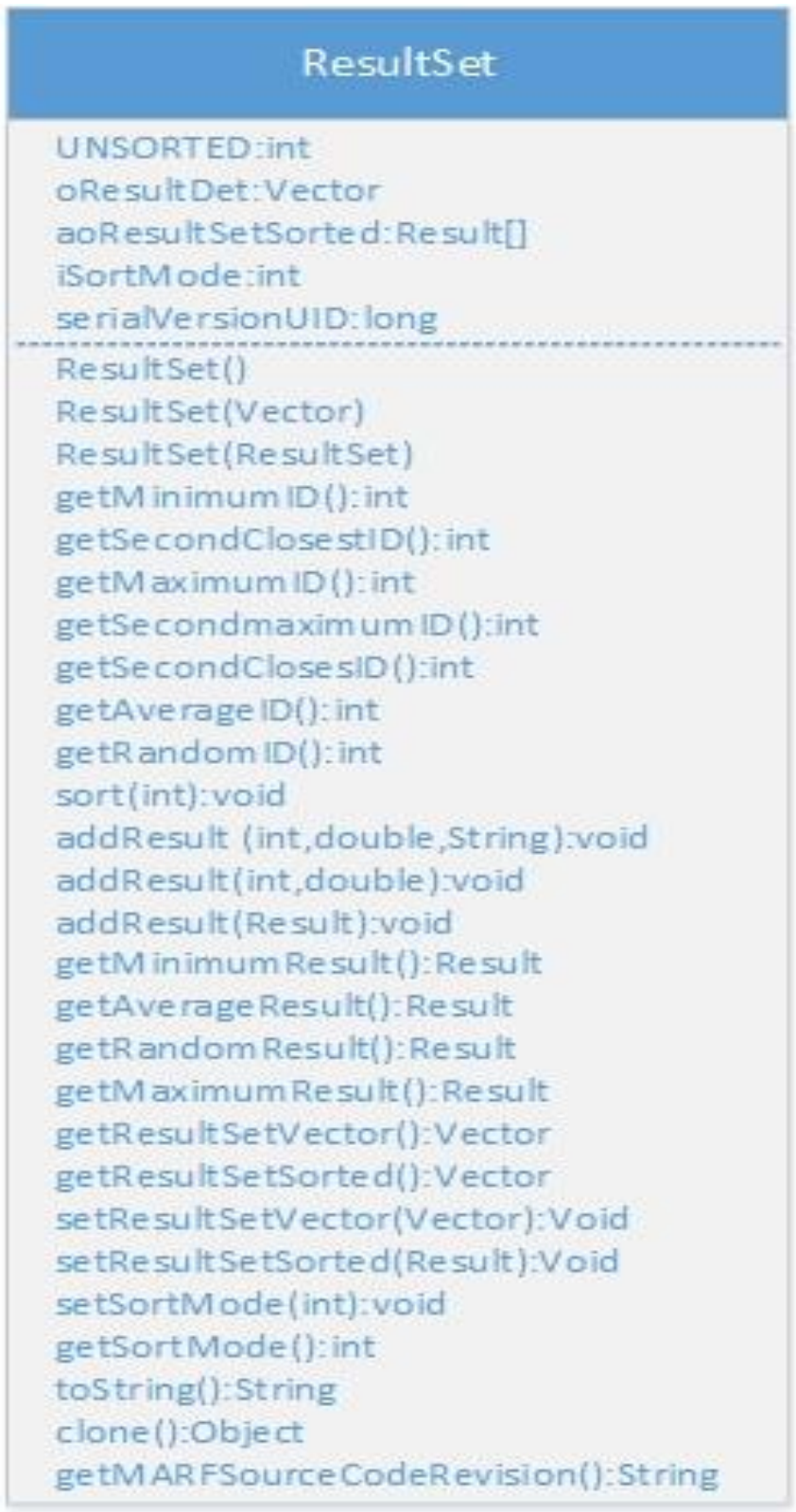

**Figure 31: `ResultSet.java` class without Refactoring**

***Code Stub:***

```
public class ResultSet implements Serializable, Cloneable
{

	public static final int UNSORTED = -1;
	protected Vector oResultSet = null;
	protected Result[] aoResultSetSorted = null;
	protected int iSortMode = UNSORTED;
	private static final long serialVersionUID = -3133714664001380852L;
	public ResultSet()
	public ResultSet(Vector poResultSet)
	public ResultSet(final ResultSet poResultSet)
	public final int getMininumID()
	public final int getSecondMininumID()
	public final int getMaximumID()
	public final int getSecondMaximumID()
	public final int getSecondClosestID()
	public final int getAverageID()
	public final int getRandomID()
	public final void sort(final int piMode)
	public final void addResult(int piID, double pdOutcome, String pstrDescription)
	public final void addResult(int piID, double pdOutcome)
	public final void addResult(Result poResult)
	public Result getMinimumResult()
	public Result getAverageResult()
	public Result getRandomResult()
	public Result getMaximumResult()
	public Vector getResultSetVector()
	public Result[] getResultSetSorted()
	public void setResultSetVector(Vector poResultSet)
	public void setResultSetSorted(Result[] paoResultSetSorted)
	public void setSortMode(int piSortMode)
	public final int getSortMode()
	public String toString()
	public int size()
	public Object clone()
	public static String getMARFSourceCodeRevision()
	}
```

**Case 4:**

The code smells have been identified in package `marf.nlp.Parsing.GrammarCompiler`. The classes in which the bad smells are identified is `GrammarCompiler.java`. Figures 32 and 33 show the Kiviat graph and the table of the values for the parameters of the Kiviat Graph.

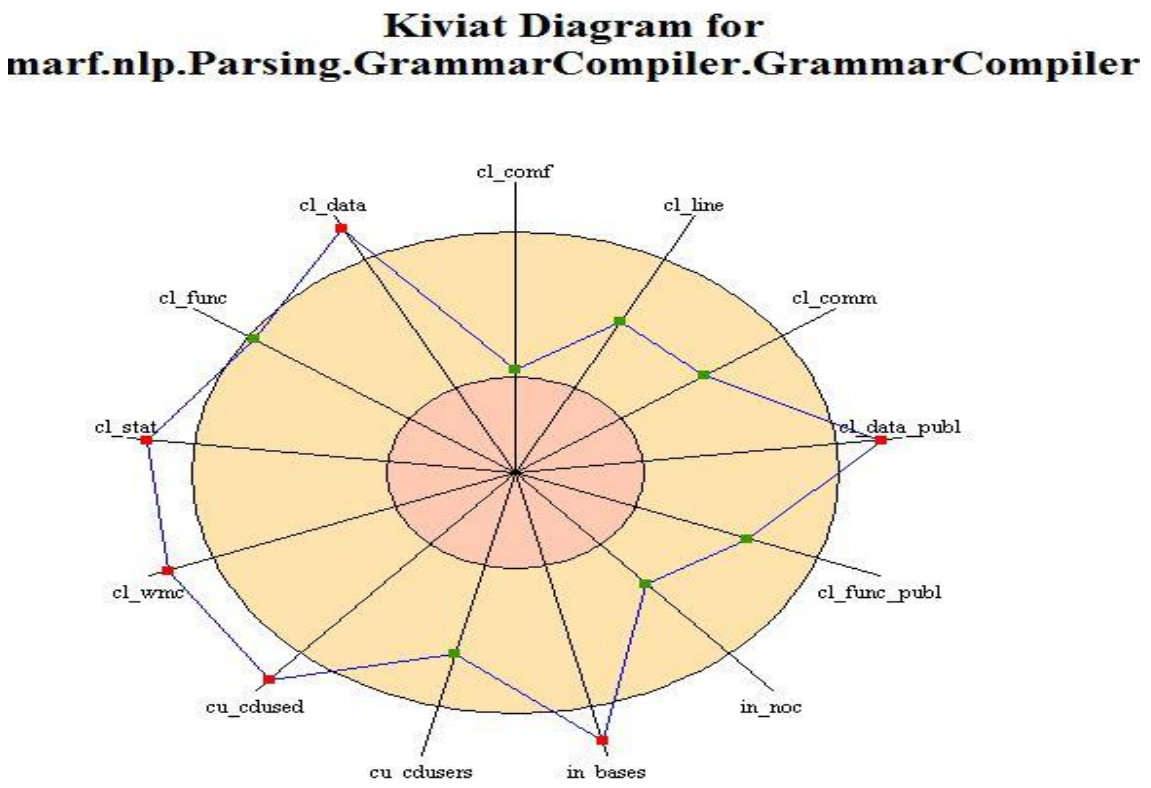

**Figure 32: `GrammarCompiler.java` class without Refactoring**

| Metric : marf.nlp.Parsing.GrammarCompiler.GrammarCompiler | Value | Min | Max | Status |
|---|---|---|---|---|
| cl_comm: Number of lines of comment | 427 | -oo | +oo | 0 |
| cl_data: Total number of attributes | 11 | 0 | 7 | -1 |
| cl_data_publ: Number of public attributes | 3 | 0 | 0 | -1 |
| cl_func: Total number of methods | 24 | 0 | 25 | 0 |
| cl_func_publ: Number of public methods | 9 | 0 | 15 | 0 |
| cl_line: Number of lines | 1181 | -oo | +oo | 0 |
| cl_stat: Number of statements | 245 | 0 | 100 | -1 |
| cl_wmc: Weighted Methods per Class | 78 | 0 | 60 | -1 |
| cu_cdused: Number of direct used classes | 29 | 0 | 10 | -1 |
| cu_cdusers: Number of direct users classes | 3 | 0 | 5 | 0 |
| in_bases: Number of base classes | 4 | 0 | 3 | -1 |
| in_noc: Number of children | 1 | 0 | 3 | 0 |

**Figure 33: Table depicting metric values for `GrammarCompiler.java`**

The bad smells identified with respect to the architecture of the system are listed as follows –

1. Too many functions
2. Used classes are very high, more complexity
3. More used classes, increased coupling
4. Too much public data
5. Too many base classes

Figure 34 shows the class diagram for the class.

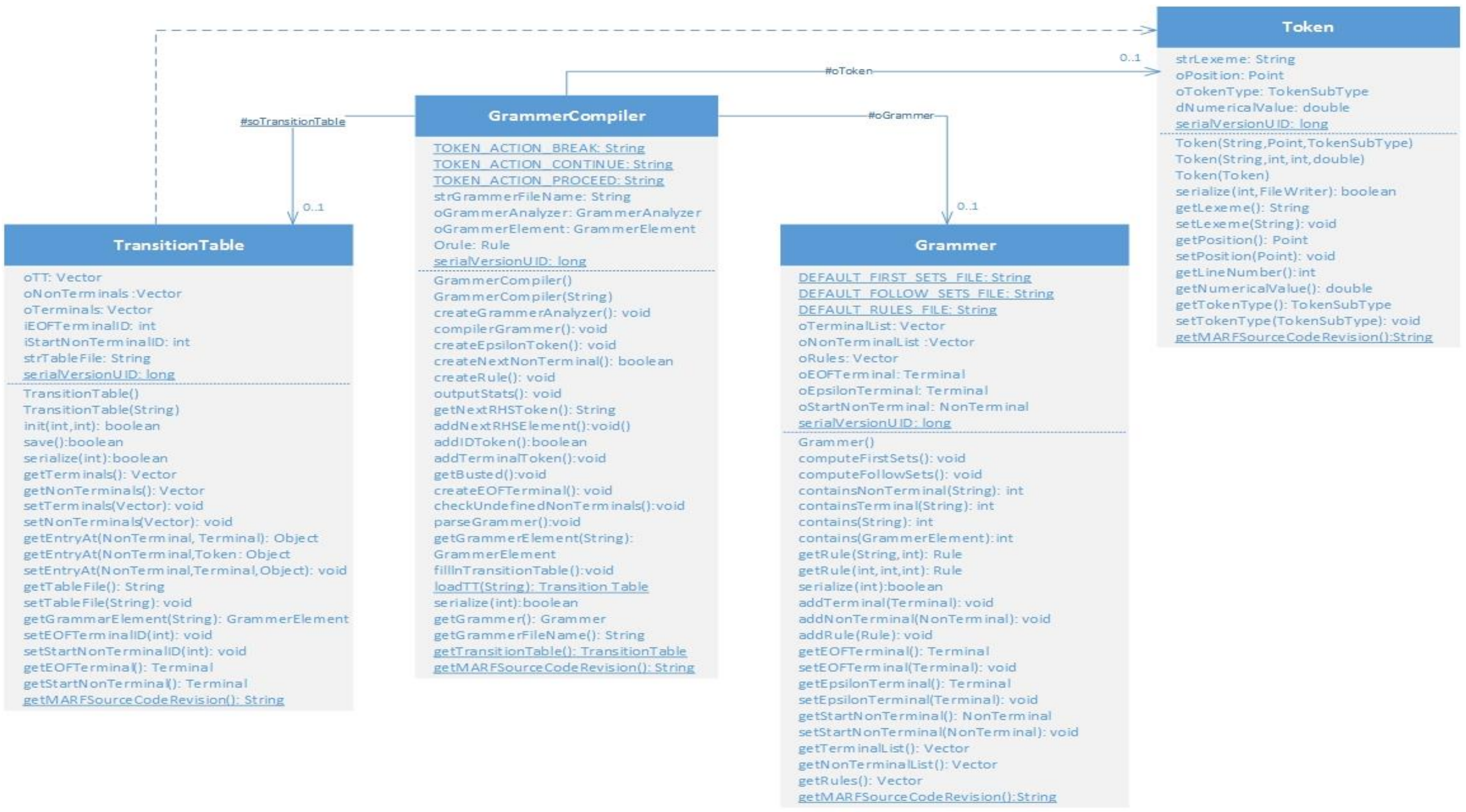

**Figure 34: `GrammarCompiler.java` class without Refactoring**

*Code Stub:*

```
public class GrammarCompiler extends StorageManager {
public static final String TOKEN_ACTION_BREAK = "break";
public static final String TOKEN_ACTION_CONTINUE = "continue";
public static final String TOKEN_ACTION_PROCEED = "proceed";
protected Grammar oGrammar = null;
protected String strGrammarFileName = "";
protected GrammarAnalyzer oGrammarAnalyzer = null;
protected static TransitionTable soTransitionTable = null;
protected GrammarElement oGrammarElement;
protected Token oToken;
protected Rule oRule;
public GrammarCompiler() throws CompilerError
public GrammarCompiler(String pstrGrammarFileName)
protected void createGrammarAnalyzer()
public void compileGrammar() throws CompilerError
protected void createEpsilonToken()
protected boolean createNextNonTerminal()
protected void createRule()
protected void outputStats()
protected String getNextRHSToken()
protected void addNextRHSElement()
protected boolean addIDToken()
protected void addTerminalToken()
protected void getBusted()
protected void createEOFTerminal()
protected void parseGrammar()
protected GrammarElement getGrammarElement(String pstrName)
private void fillInTransitionTable()
public static TransitionTable loadTT(String pstrTTFileName)
public boolean serialize(int piOperation)
public final Grammar getGrammar()
public static final TransitionTable getTransitionTable()
public static String getMARFSourceCodeRevision() }
```

## II. GIPSY

**Case 1:** The code smells have been identified in the package `gipsy.GIPC.DFG.DFGGenerator`. The classes in which the bad smells are identified are –

1. `DFGCodeGenerator.java`
2. `DFGTranCodeGenerator.java`

We use Logiscope [43] to analyze the code and identify bad smells with the help of the generated Kiviat graphs as shown in figure 37, 38, 39 and 40

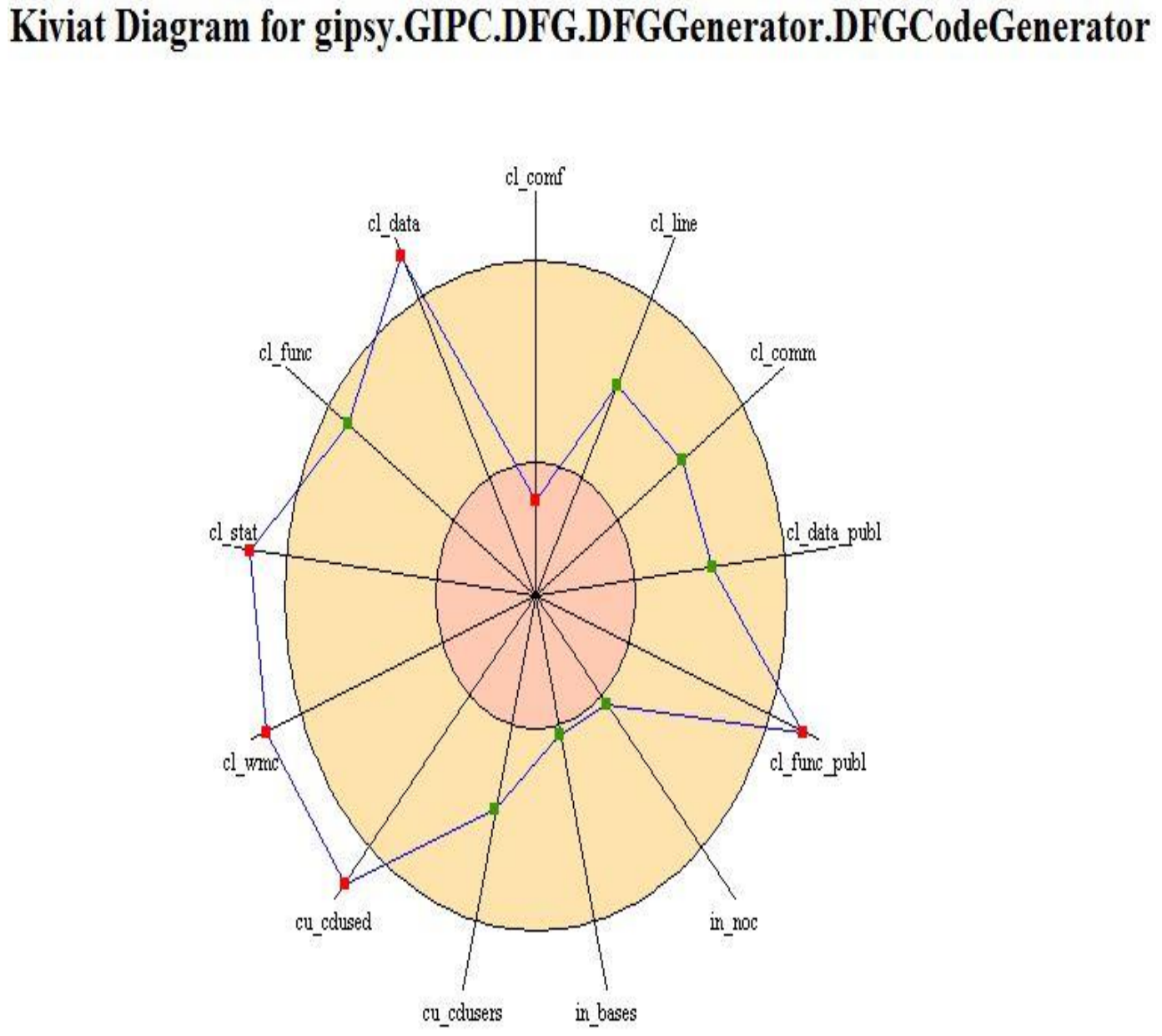

**Figure 37: Kiviat Diagram for `DFGCodeGenerator.java`**

| Metric : gipsy.GIPC.DFG.DFGGenerator.DFGCodeGenerator | Value | Min | Max | Status |
|---|---|---|---|---|
| cl_comm: Number of lines of comment | 85 | -oo | +oo | 0 |
| cl_data: Total number of attributes | 26 | 0 | 7 | -1 |
| cl_data_publ: Number of public attributes | 0 | 0 | 0 | 0 |
| cl_func: Total number of methods | 21 | 0 | 25 | 0 |
| cl_func_publ: Number of public methods | 20 | 0 | 15 | -1 |
| cl_line: Number of lines | 665 | -oo | +oo | 0 |
| cl_stat: Number of statements | 242 | 0 | 100 | -1 |
| cl_wmc: Weighted Methods per Class | 90 | 0 | 60 | -1 |
| cu_cdused: Number of direct used classes | 12 | 0 | 10 | -1 |
| cu_cdusers: Number of direct users classes | 2 | 0 | 5 | 0 |
| in_bases: Number of base classes | 0 | 0 | 3 | 0 |
| in_noc: Number of children | 0 | 0 | 3 | 0 |

**Figure 38: Table depicting metric values for `DFGCodeGenerator.java`**

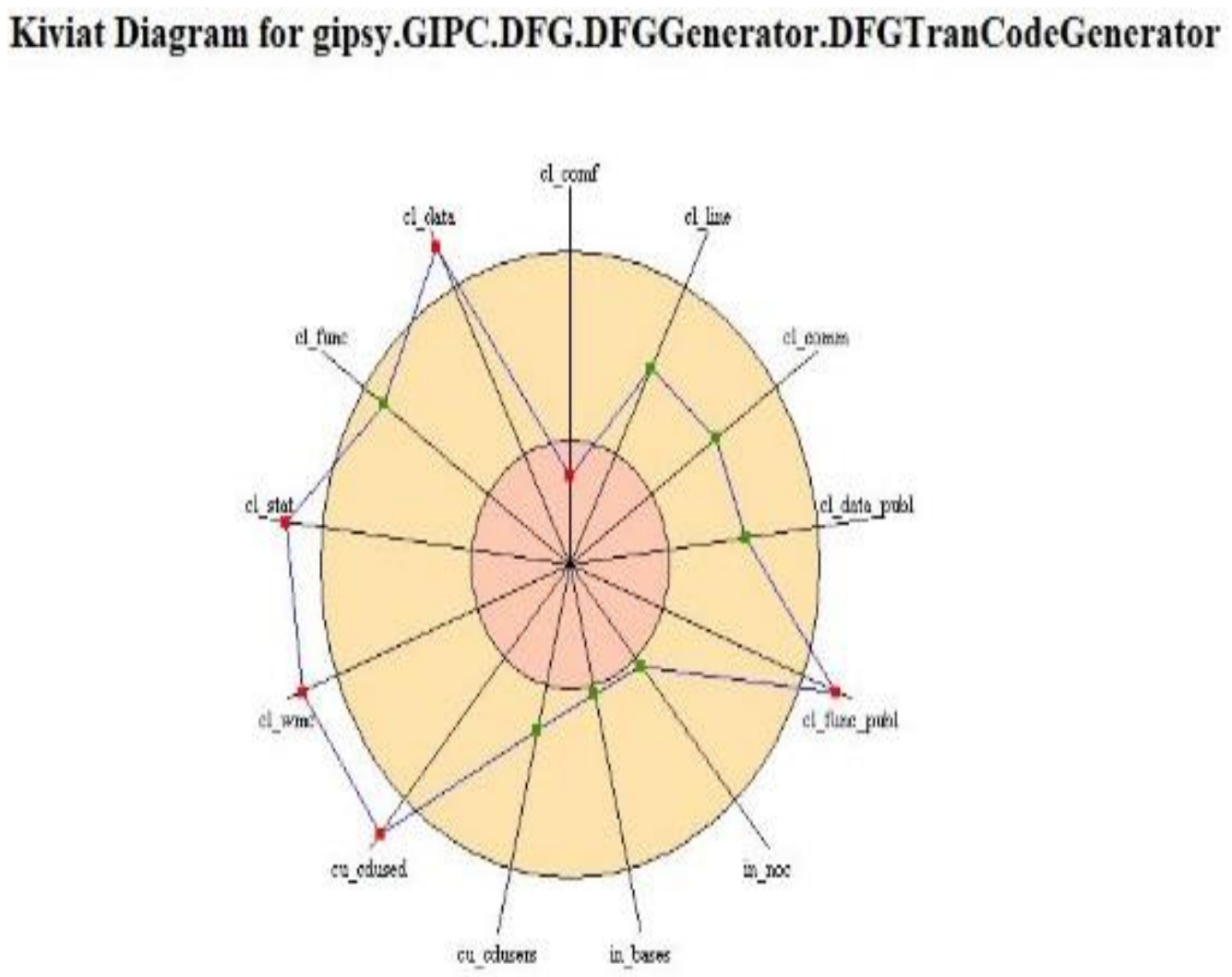

**Figure 39: Kiviat Diagram for `DFGTranCodeGenerator.java`**

| Metric : gipsy.GIPC.DFG.DFGGenerator.DFGTranCodeGenerator | Value | Min | Max | Status |
|---|---|---|---|---|
| cl_comm: Number of lines of comment | 91 | -oo | +oo | 0 |
| cl_data: Total number of attributes | 28 | 0 | 7 | -1 |
| cl_data_publ: Number of public attributes | 0 | 0 | 0 | 0 |
| cl_func: Total number of methods | 21 | 0 | 25 | 0 |
| cl_func_publ: Number of public methods | 20 | 0 | 15 | -1 |
| cl_line: Number of lines | 527 | -oo | +oo | 0 |
| cl_stat: Number of statements | 233 | 0 | 100 | -1 |
| cl_wmc: Weighted Methods per Class | 91 | 0 | 60 | -1 |
| cu_cdused: Number of direct used classes | 14 | 0 | 10 | -1 |
| cu_cdusers: Number of direct users classes | 1 | 0 | 5 | 0 |
| in_bases: Number of base classes | 0 | 0 | 3 | 0 |
| in_noc: Number of children | 0 | 0 | 3 | 0 |

**Figure 40: Table depicting metric values for `DFGTranCodeGenerator.java`**

Figure 41 shows the class diagram of the class before refactoring. The bad smells identified in these classes are listed and described as follows–

1. They have a large number of attributes
2. Both classes have the same attributes which was causing redundancy in the system
3. Both classes have many methods which have the same functionalities again increasing redundancy
4. The number of directly used classes are very high which increases the coupling of these classes with the other classes leading to the increase of the complexity of the system
5. The number of public methods in the classes are more causing increase in dependency with these classes and other classes in the system

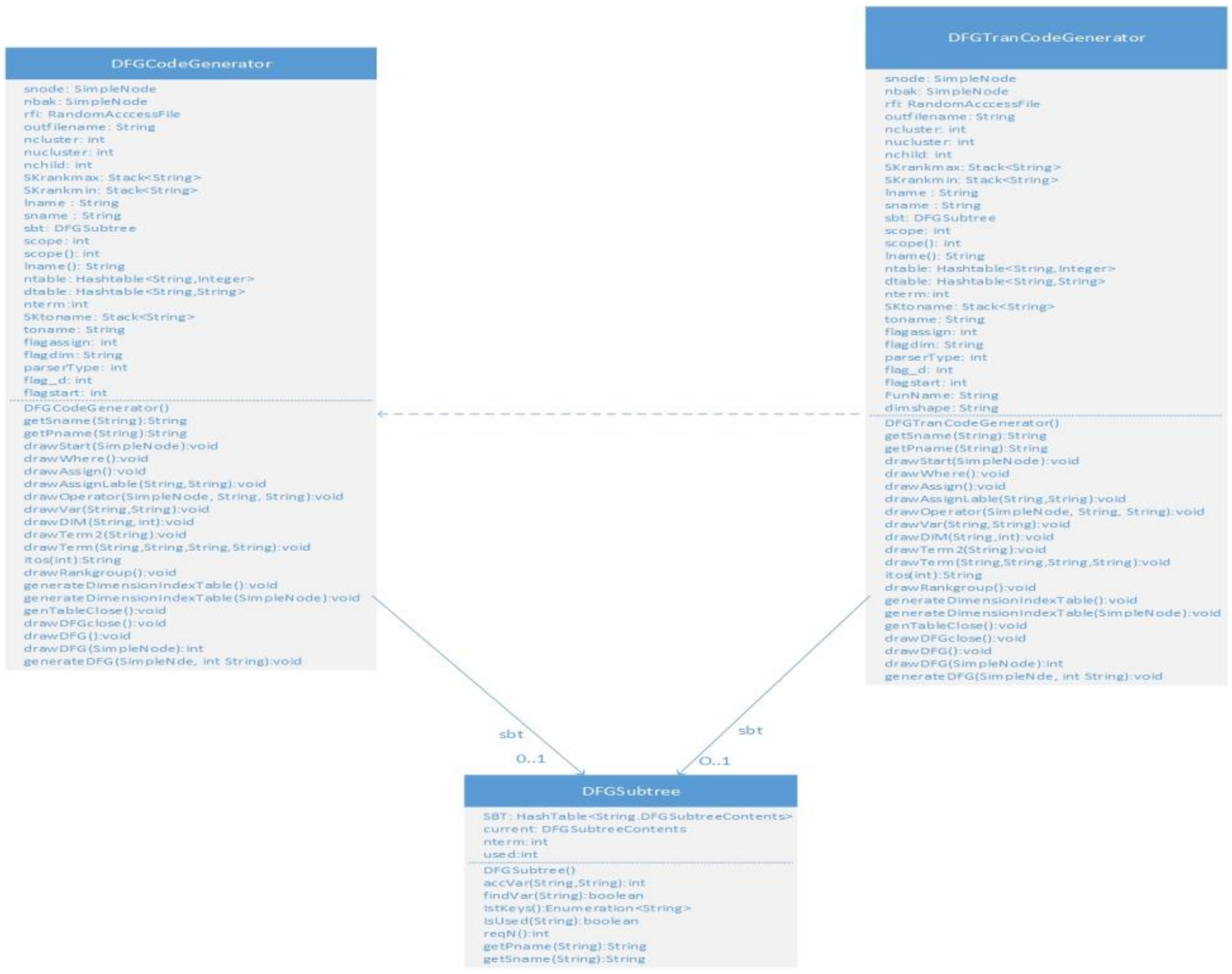

**Figure 41: `DFGCodeGenerator.java` and `DFGTranCodeGenerator.java` class without Refactoring**

*Code Stub:*

**Class `DFGCodeGenerator.java`**

```
public class DFGCodeGenerator{
private SimpleNode snode;
private SimpleNode nbak;
private RandomAccessFile rf1;
private String outfilename;
private int ncluster = 0;
private int nchild = 0;
Stack<String> SKrankmax = new Stack<String>();
Stack<String> SKrankmin = new Stack<String>();
private String lname = "0";
private String sname = "";
DFGSubtree sbt = new DFGSubtree();
private int scope = 0;
private int scope0 = 0;
private String lname0 = "0";
private Hashtable<String, Integer> ntable = new Hashtable<String, Integer>();
private Hashtable<String, String> dtable = new Hashtable<String, String>();
private int nterm = 0;
Stack<String> SKtoname = new Stack<String>();
private String toname = "";
private int flagassign = 0;
private String flagdim = "";
private int funnum = 1;
private int parserType = 0;
private int flag_d = 0;
private int flagstart = 0;
```

```
public String getSname(String pstrLname)
public String getPname(String pstrLname)
public void drawStart(SimpleNode n)
public void drawWhere()
public void drawAssign()
public void drawAssignLable(String image, String outpoint)
public void drawOperator(SimpleNode n, String oper, String filename1)
public void drawVar(String iname, String outpoint)
public void drawDIM(String fl, int ndim2)
public void drawTerm2(String nimage)
public void drawTerm(String fn, String fl, String tn, String hl)
private String itos(int num)
public void drawRankgroup()
public void drawDFG() throws DFGException
public int drawDFG(SimpleNode n)
throws DFGException
public void drawDFGclose()
throws DFGException
public void generateDimensionIndexTable()
public void generateDimensionIndexTable(SimpleNode n)
public void genTableclose()
public void generateDFG(SimpleNode simplenode, int parserType1, String filename)
throws DFGException}
```

### Class **DFGTranCodeGenerator.java**

```
public class DFGCodeGenerator{
private SimpleNode snode;
private SimpleNode nbak;
private RandomAccessFile rf1;
private String outfilename;
private int ncluster = 0;
private int nchild = 0;
Stack<String> SKrankmax = new Stack<String>();
Stack<String> SKrankmin = new Stack<String>();
private String lname = "0";
private String sname = "";
DFGSubtree sbt = new DFGSubtree();
private int scope = 0;
private int scope0 = 0;
private String lname0 = "0";
private String FunName = "";
private String dimshape = "";
private Hashtable<String, Integer> ntable = new Hashtable<String, Integer>();
private Hashtable<String, String> dtable = new Hashtable<String, String>();
private int nterm = 0;
Stack<String> SKtoname = new Stack<String>();
private String toname = "";
private int flagassign = 0;
private String flagdim = "";
private int funnum = 1;
private int parserType = 0;
private int flag_d = 0;
private int flagstart = 0;
public String getSname(String pstrLname)
public String getPname(String pstrLname)
```

```
public void drawStart(SimpleNode n)
public void drawWhere()
public void drawAssign()
public void drawAssignLable(String image, String outpoint)
public void drawOperator(SimpleNode n, String oper, String filename1)
public void drawVar(String iname, String outpoint)
public void drawDIM(String fl, int ndim2)
public void drawTerm2(String nimage)
public void drawTerm(String fn, String fl, String tn, String hl)
private String itos(int num)
public void drawRankgroup()
public void drawDFG() throws DFGException
public int drawDFG(SimpleNode n)
throws DFGException
public void drawDFGclose()
throws DFGException
public void genTable()
public void genTable(SimpleNode n)
public void genTableclose()
public void generateDFG(SimpleNode simplenode, int parserType1, String filename)
throws DFGException
}
```

**Case 2:**

The code smells have been identified in package `gipsy.GEE.multitier.GMT`. The classes in which the bad smells are identified is `GMTWrapper.java`. Figures 42 and Figure 43 show the Kiviat graph and the table of the values for the parameters of the Kiviat Graph.

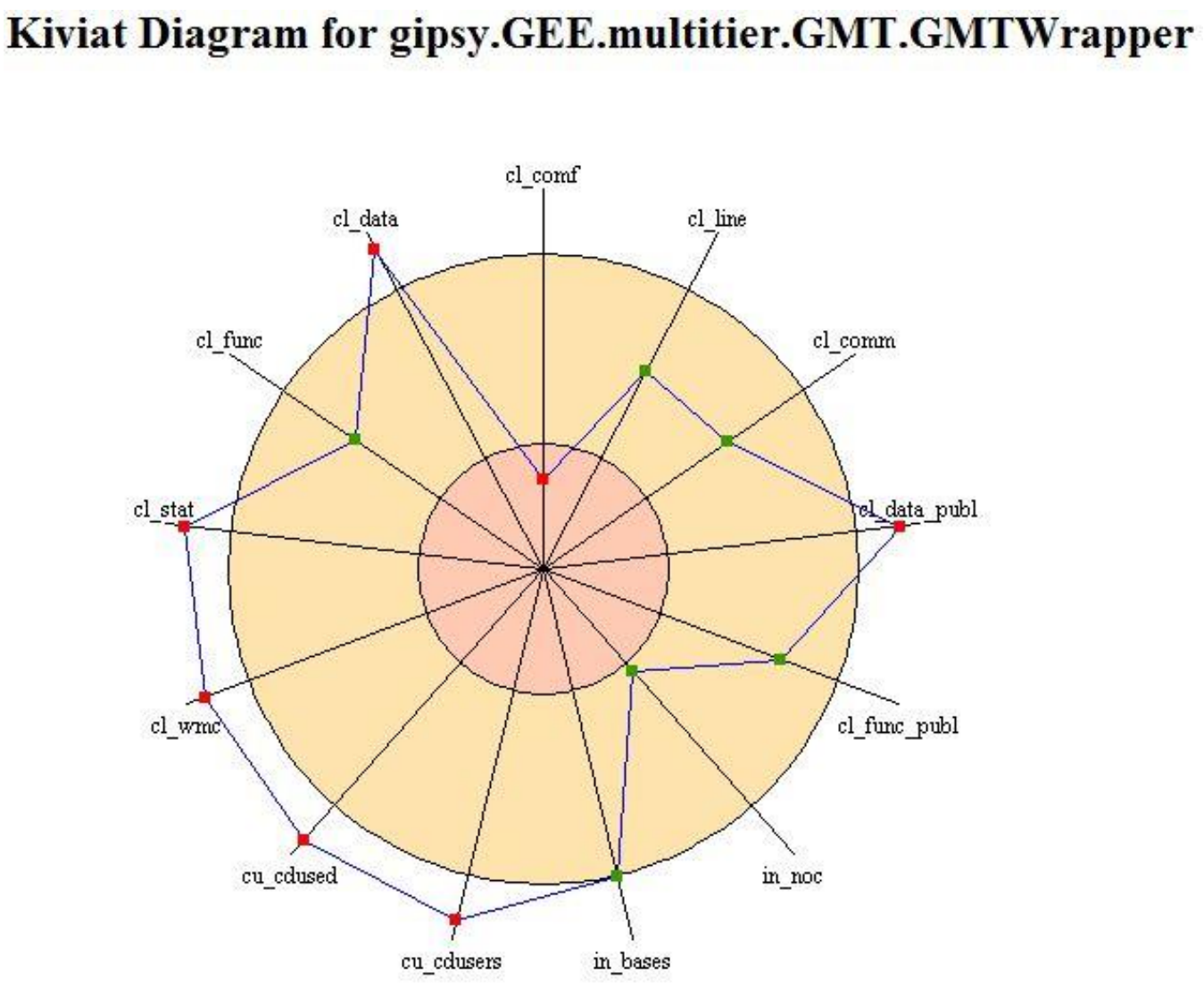

**Figure 42: Kiviat Diagram for `GMTWrapper.java`**

| Metric : gipsy.GEE.multitier.GMT.GMTWrapper | Value | Min | Max | Status |
|---|---|---|---|---|
| cl_comm: Number of lines of comment | 148 | -oo | +oo | 0 |
| cl_data: Total number of attributes | 18 | 0 | 7 | -1 |
| cl_data_publ: Number of public attributes | 9 | 0 | 0 | -1 |
| cl_func: Total number of methods | 13 | 0 | 25 | 0 |
| cl_func_publ: Number of public methods | 10 | 0 | 15 | 0 |
| cl_line: Number of lines | 893 | -oo | +oo | 0 |
| cl_stat: Number of statements | 284 | 0 | 100 | -1 |
| cl_wmc: Weighted Methods per Class | 87 | 0 | 60 | -1 |
| cu_cdused: Number of direct used classes | 32 | 0 | 10 | -1 |
| cu_cdusers: Number of direct users classes | 8 | 0 | 5 | -1 |
| in_bases: Number of base classes | 3 | 0 | 3 | 0 |
| in_noc: Number of children | 0 | 0 | 3 | 0 |

**Figure 43: Table depicting metric values for `GMTWrapper.java`**

Figure 44 shows the class diagram of the class before refactoring. The bad smells identified in these classes are listed and described as follows–

1. They have a large number of attributes
2. It has few comments which leads to decrease in understandability
3. Both classes have the same attributes which was causing redundancy in the system
4. Both classes have many methods which have the same functionalities again increasing redundancy
5. The number of directly used classes are very high which increases the coupling of these classes with the other classes leading to the increase of the complexity of the system
6. The number of public methods in the classes are more causing increase in dependency with these classes and other classes in the system
7. It has many methods which have the same functionalities again increasing redundancy

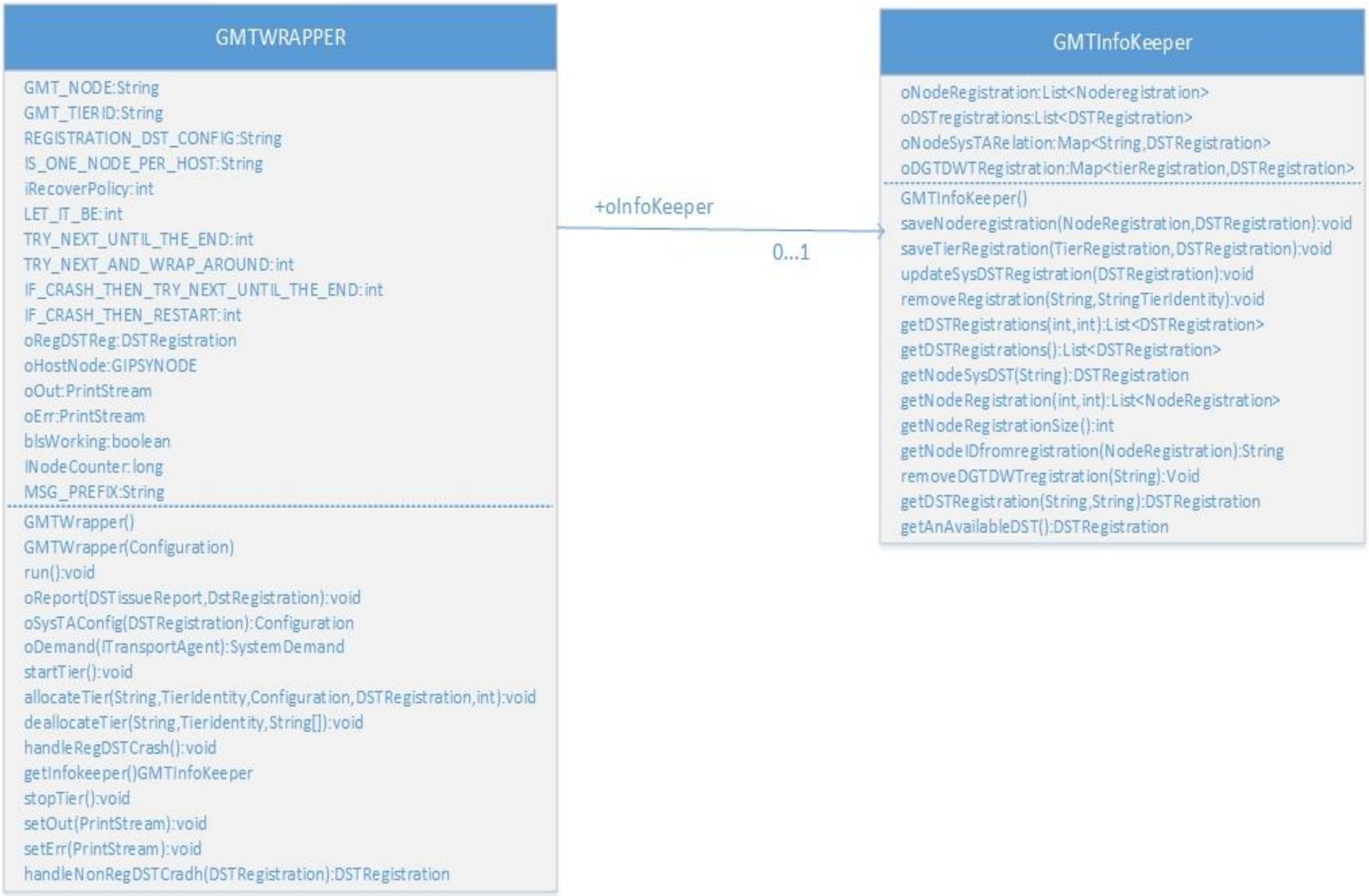

**Figure 44: `GMTWrapper.java` class without Refactoring**

*Code Stub:*

```
package gipsy.GEE.multitier.GMT;
import gipsy.Configuration;

public class GMTWrapperextends GenericTierWrapper

{

public static final String GMT_NODE = "gipsy.GEE.multitier.GMT.Node";

protected static final String GMT_TIERID = "gipsy.GEE.multitier.GMT.tierID";

public static final String REGISTRATION_DST_CONFIG = "gipsy.GEE.multitier.registrationDST.config";

private static final String IS_ONE_NODE_PER_HOST = "gipsy.GEE.multitier.GMT.isOneNodePerHost";

public GMTInfoKeeper oInfoKeeper = new GMTInfoKeeper();

public volatile int iRecoverPolicy = 0;

public static final int LET_IT_BE = 0;

public static final int TRY_NEXT_UNTIL_THE_END = 1;

public static final int TRY_NEXT_AND_WRAP_AROUND = 2;

public static final int IF_CRASH_THEN_TRY_NEXT_UNTIL_THE_END = 3;

public static final int IF_CRASH_THEN_RESTART = 4;

private DSTRegistration oRegDSTReg = null;

private GIPSYNode oHostNode = null;

private PrintStream oOut = null;
```

```
private PrintStream oErr = null;
private volatile boolean bIsWorking = true;
private long lNodeCounter = 0;
private static final String MSG_PREFIX = "[" + Trace.getEnclosingClassName() + "] ";
public GMTWrapper()
public GMTWrapper(Configuration poGMTConfig)
public void run()
public void startTier()
public void allocateTier(String pstrNodeID,
public void deallocateTier(String pstrNodeID, TierIdentity poTierIdentity,
private void handleRegDSTCrash()
public GMTInfoKeeper getInfoKeeper()
public void stopTier()
public void setOut(PrintStream poOut)
public void setErr(PrintStream poErr)
private DSTRegistration getAnAvailableDST()
private DSTRegistration handleNonRegDSTCrash(DSTRegistration poCrashedDSTReg)
}
```

**Case 3:**

The code smells have been identified in package `gipsy.GIPC`. The classes in which the bad smells are identified is `GIPC.java`. Figures 45 and Figure 46 show the Kiviat graph and the table of the values for the parameters of the Kiviat Graph.

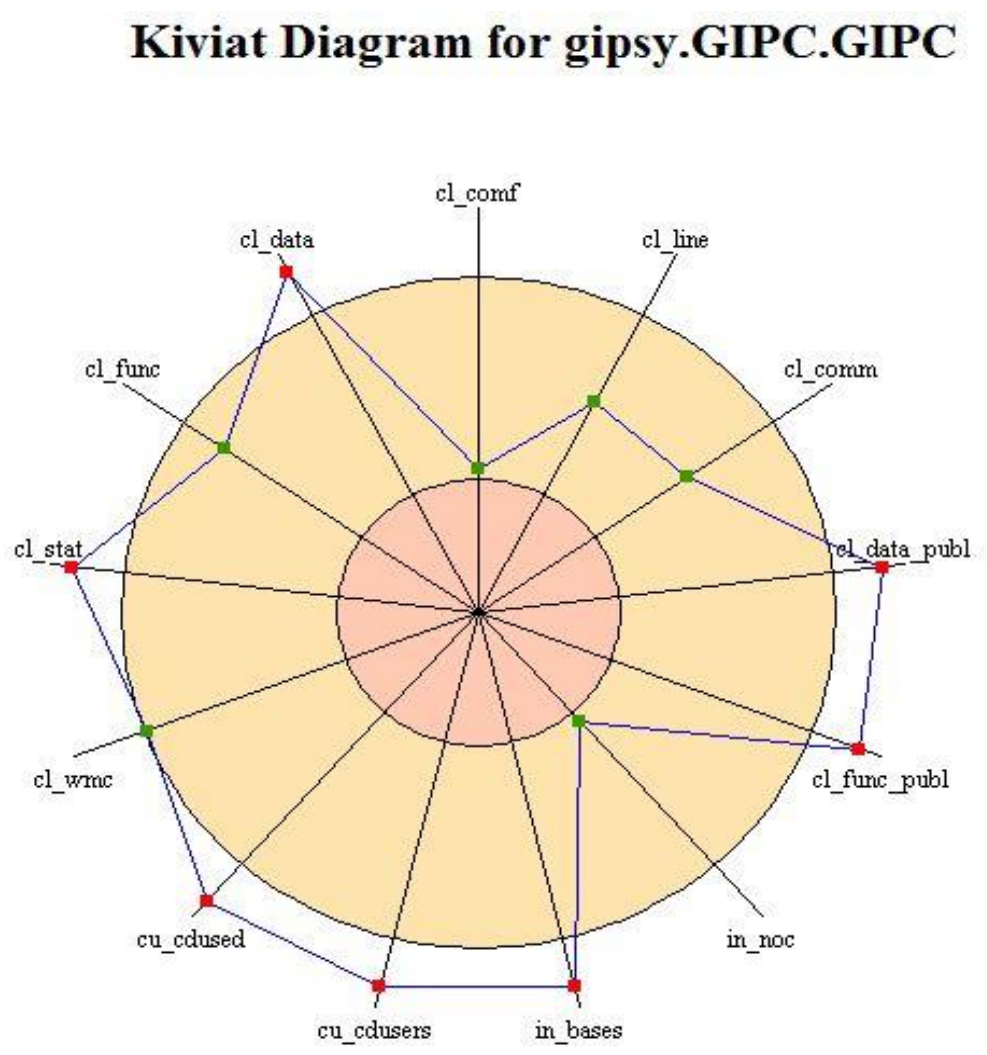

**Figure 45: Kiviat Diagram for GIPC.java**

| Metric : gipsy.GIPC.GIPC | Value | Min | Max | Status |
|---|---|---|---|---|
| cl_comm: Number of lines of comment | 280 | -oo | +oo | 0 |
| cl_data: Total number of attributes | 50 | 0 | 7 | -1 |
| cl_data_publ: Number of public attributes | 36 | 0 | 0 | -1 |
| cl_func: Total number of methods | 19 | 0 | 25 | 0 |
| cl_func_publ: Number of public methods | 16 | 0 | 15 | -1 |
| cl_line: Number of lines | 907 | -oo | +oo | 0 |
| cl_stat: Number of statements | 233 | 0 | 100 | -1 |
| cl_wmc: Weighted Methods per Class | 59 | 0 | 60 | 0 |
| cu_cdused: Number of direct used classes | 45 | 0 | 10 | -1 |
| cu_cdusers: Number of direct users classes | 12 | 0 | 5 | -1 |
| in_bases: Number of base classes | 5 | 0 | 3 | -1 |
| in_noc: Number of children | 0 | 0 | 3 | 0 |

**Figure 46: Table depicting metric values for GIPC`.java`**

The bed smells identified in class are given as follow.

1. Class has many number of attributes
2. Used classes are very high, more complexity
3. More used classes, increased coupling
4. Too many base classes
5. Too many public functions and data which make it insecure

Figure 47 shows the class diagram of class before refactoring.

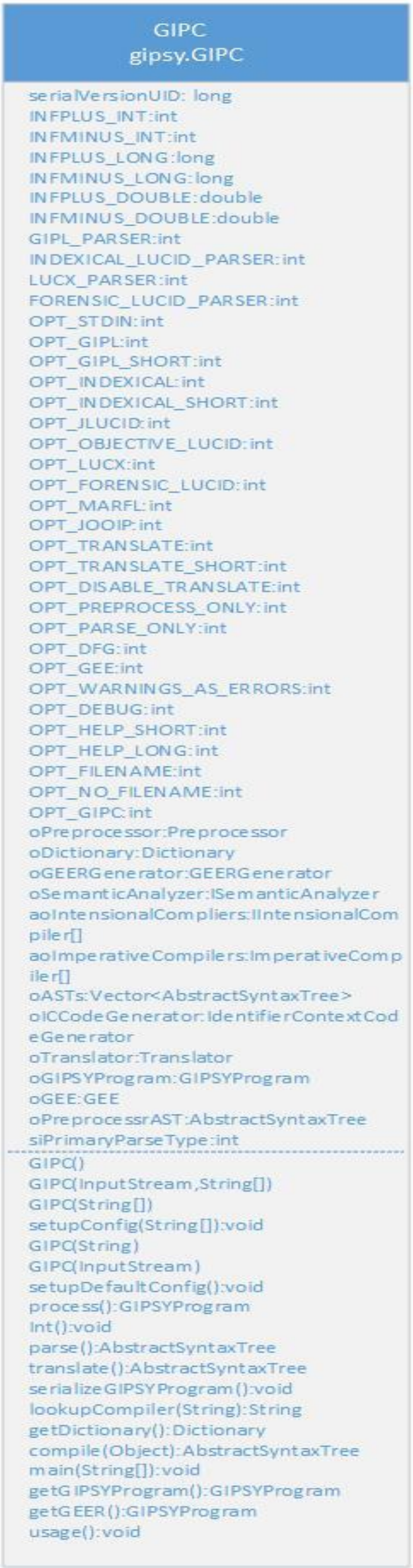

**Figure 47: `GIPC.java` class without Refactoring**

*Code Stub*

```
package gipsy.GIPC;
import gipsy.GEE.GEE;

public class GIPC
extends IntensionalCompiler
{

private static final long serialVersionUID = 5412492426375587515L;

public static final int INFPLUS_INT = Integer.MAX_VALUE;
public static final int INFMINUS_INT = Integer.MIN_VALUE;

public static final long INFPLUS_LONG = Long.MAX_VALUE;
public static final long INFMINUS_LONG = Long.MIN_VALUE;

public static final double INFPLUS_DOUBLE = Double.MAX_VALUE;
public static final double INFMINUS_DOUBLE = Double.MIN_VALUE;

public static final int GIPL_PARSER = 0;

public static final int INDEXICAL_LUCID_PARSER = 1;
public static final int LUCX_PARSER = 2;
public static final int FORENSIC_LUCID_PARSER = 3;
public static final int OPT_STDIN = 1;

public static final int OPT_GIPL = 2;


public static final int OPT_GIPL_SHORT = 3;

public static final int OPT_INDEXICAL = 4;

public static final int OPT_INDEXICAL_SHORT = 5;
public static final int OPT_JLUCID = 6;

public static final int OPT_OBJECTIVE_LUCID = 7;
public static final int OPT_LUCX = 8;

public static final int OPT_FORENSIC_LUCID = 9;

public static final int OPT_MARFL = 10;

public static final int OPT_JOOIP = 11;

public static final int OPT_TRANSLATE = 12;
public static final int OPT_TRANSLATE_SHORT = 13;

public static final int OPT_DISABLE_TRANSLATE = 14;

public static final int OPT_PREPROCESS_ONLY = 15;

public static final int OPT_PARSE_ONLY = 16;

public static final int OPT_DFG = 17;
public static final int OPT_GEE = 18;

public static final int OPT_WARNINGS_AS_ERRORS = 19;

public static final int OPT_DEBUG = 20;

public static final int OPT_HELP_SHORT = 21;

public static final int OPT_HELP_LONG = 22;
```

```
public static final int OPT_FILENAME = 23;
public static final int OPT_NO_FILENAME = 24;
public static final int OPT_GIPC = 25;

private Preprocessor oPreprocessor = null;

private Dictionary oDictionary = null;

private OptionProcessor oOptionProcessor = new OptionProcessor();

private GEERGenerator oGEERGenerator = null;

private ISemanticAnalyzer oSemanticAnalyzer = null;

private IIntensionalCompiler[] aoIntensionalCompilers = null;

private IImperativeCompiler[] aoImperativeCompilers = null;

private Vector<AbstractSyntaxTree> oASTs = new Vector<AbstractSyntaxTree>();

private IdentifierContextCodeGenerator oICCodeGenerator = null;

private Translator oTranslator = null;

private GIPSYProgram oGIPSYProgram = null;

private GEE oGEE = null;

private AbstractSyntaxTree oPreprocessortAST = null;

public static int siPrimaryParserType;

public GIPC()

public GIPC(InputStream poSourceCodeStream, String[] argv)

public GIPC(String[] argv)

protected void setupConfig(String[] argv)

public GIPC(String pstrFilename)

public GIPC(InputStream poInputStream)

protected void setupDefaultConfig()

public GIPSYProgram process()

public void init()

public AbstractSyntaxTree parse()

public AbstractSyntaxTree translate()

public void serializeGIPSYProgram()

public String lookupCompiler(String pstrLanguageName)

public Dictionary getDictionary()

public AbstractSyntaxTree compile(Object poExtraArgs)

public static final void main(String[] argv)

public GIPSYProgram getGIPSYProgram()

public GIPSYProgram getGEER()

private static final void usage()
```

## Case 4:

The code smells have been identified in package `gipsy.RIPE.editors.RuneTimeGraphEditor.ui.` The classes in which the bad smells are identified is `GIPSYGMTOperator.java`. Figures 47 and Figure 48 show the Kiviat graph and the table of the values for the parameters of the Kiviat Graph.

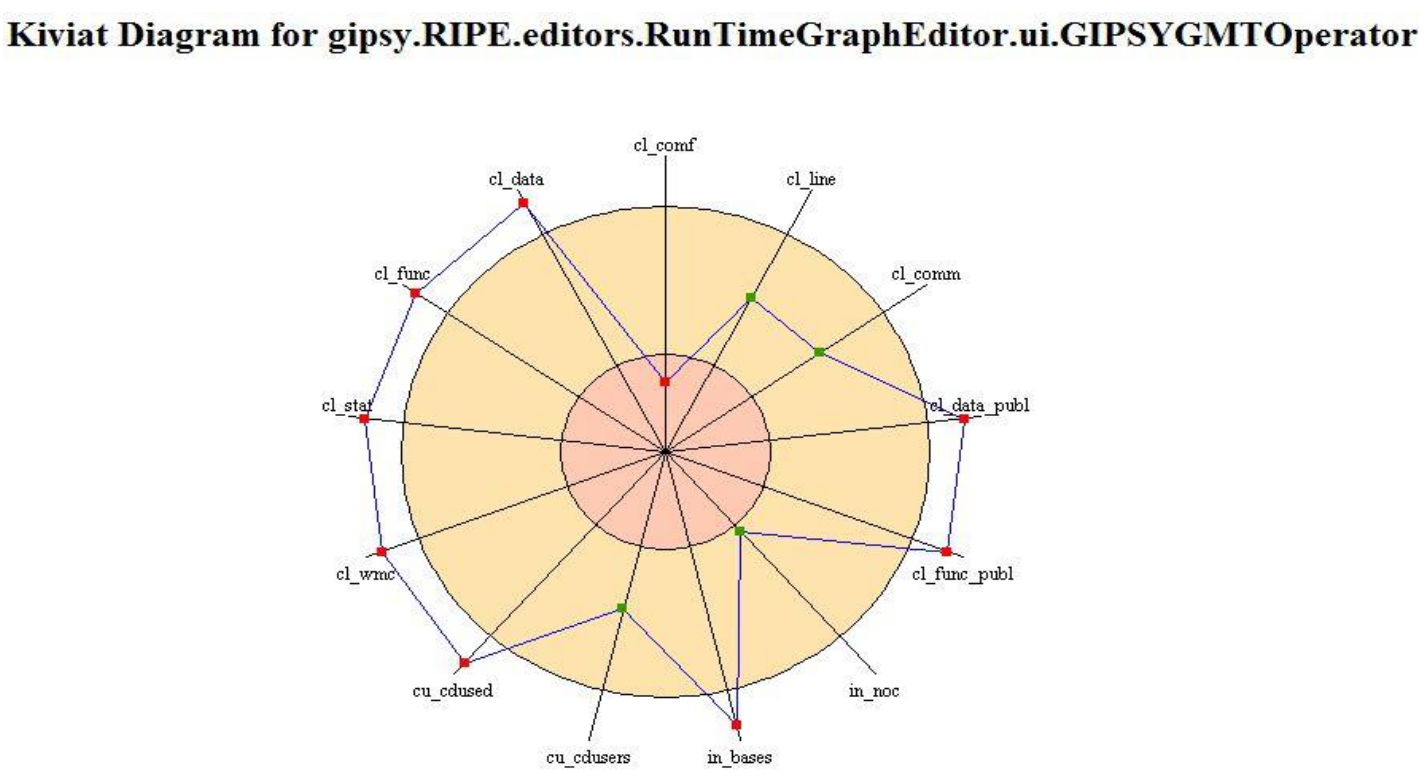

**Figure 47: Kiviat Diagram for `GIPSYGMTOperator.java`**

| Metric : gipsy.RIPE.editors.RunTimeGraphEditor.ui.GIPSYGMTOper... | Value | Min | Max | Status |
|---|---|---|---|---|
| cl_comm: Number of lines of comment | 160 | -oo | +oo | 0 |
| cl_data: Total number of attributes | 44 | 0 | 7 | -1 |
| cl_data_publ: Number of public attributes | 8 | 0 | 0 | -1 |
| cl_func: Total number of methods | 37 | 0 | 25 | -1 |
| cl_func_publ: Number of public methods | 19 | 0 | 15 | -1 |
| cl_line: Number of lines | 859 | -oo | +oo | 0 |
| cl_stat: Number of statements | 311 | 0 | 100 | -1 |
| cl_wmc: Weighted Methods per Class | 69 | 0 | 60 | -1 |
| cu_cdused: Number of direct used classes | 51 | 0 | 10 | -1 |
| cu_cdusers: Number of direct users classes | 2 | 0 | 5 | 0 |
| in_bases: Number of base classes | 6 | 0 | 3 | -1 |
| in_noc: Number of children | 0 | 0 | 3 | 0 |

**Figure 48**: Table depicting metric values for. `GIPSYGMTOperator.java`

The bed smells identified in class are given as follow.

1. They have a large number of attributes
2. It has few comments which leads to decrease in understandability
3. Both classes have the same attributes which was causing redundancy in the system
4. Both classes have many methods which have the same functionalities again increasing redundancy
5. The number of directly used classes are very high which increases the coupling of these classes with the other classes leading to the increase of the complexity of the system
6. The number of public methods in the classes are more causing increase in dependency with these classes and other classes in the system
7. It has many methods which have the same functionalities again increasing redundancy

Figure 49 shows the class diagram of class before refactoring.

.

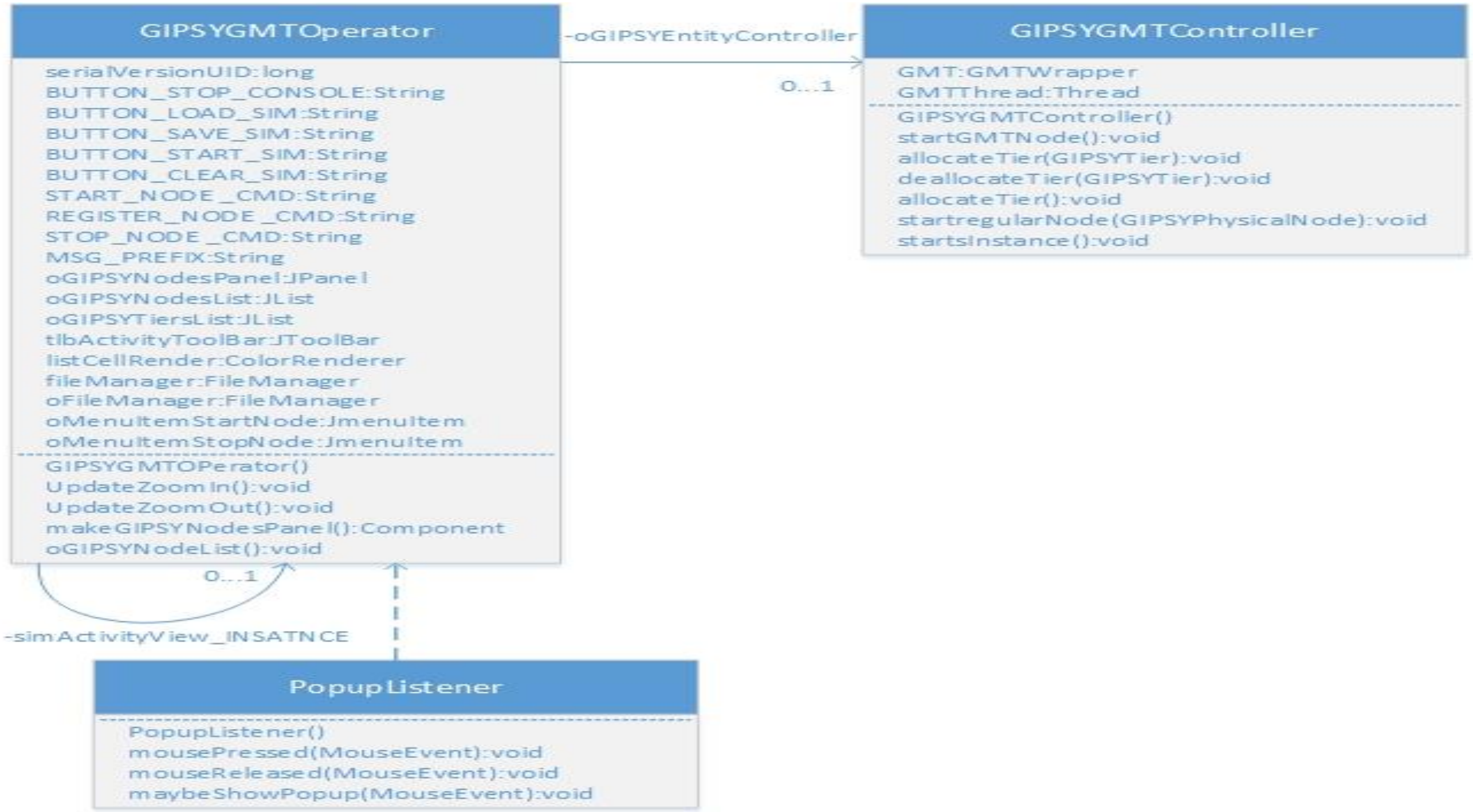

**Figure 49: `GIPSYGMTOperator.java` class without Refactoring**

*Code Stub:*

```
public class GIPSYGMTOperator extends JPanel implements ToolBarSwitchView,
        ActionListener, MouseListener, Serializable, ItemListener
{
private static final long serialVersionUID = -4698624349613823294L;
public static final String BUTTON_STOP_CONSOLE = "Stop Instances";
public static final String BUTTON_LOAD_SIM = "Load Graph";
public static final String BUTTON_SAVE_SIM = "Save Graph";
public static final String BUTTON_START_SIM = "Start Instances";
public static final String BUTTON_CLEAR_SIM = "Clear View";
public static final String START_NODE_CMD = "Start Node";
public static final String REGISTER_NODE_CMD = "Register Node";
public static final String STOP_NODE_CMD = "STOP Node";

private static final String MSG_PREFIX = "[ + Trace.getEnclosingClassName() + "]";

private static GIPSYGMTOperator simActivityView_INSTANCE;
private Image map;
private JButton btnStopSim;
private JButton btnLoadSim;
private JButton btnSaveSim;
private JButton btnStartSim;
private JButton btnClearSim;
private JButton minus;
private JButton plus;
private JPanel leftPanel;
private JPanel oGIPSYNodesPanel;
private JPanel statesPanel;
private JPanel oSimPanel;
private JPanel imgPanel;
private JList oGIPSYNodeList;
private DefaultListModel oGIPSYNodeListModel;
private JList oGIPSYTiersList;
private DefaultListModel oStatesListModel;
private JToolBar tlbActivityToolBar;
private JPanel oMainPanel;
private GraphPanel bOperatorView;
private ColorRenderer listCellRender;
private FileManager fileManager;
private ActionsLog oActionsLog;
private boolean bIsRunning;
private float zoomPoint;

private boolean bIsCountinue;
private FileManager oFileManager;
private GIPSYGMTController oGIPSYEntityController;
private JPopupMenu oNodeListPopUpMenu;

private JMenuItem oMenuItemStartNode;
private JMenuItem oMenuItemRegisterNode;
private JMenuItem oMenuItemStopNode;
private GIPSYPhysicalNode oCurrentSelectedNode;

private GIPSYGMTOperator()

private void initializeComponents()

private void startInstance()

private Component makeGIPSYNodesPanel()

public static GIPSYGMTOperator getInstance() }
```

#### 3.1.2 Specific Refactoring

In this section, specific refactoring is suggested on the classes in which bad smells were identified.

**I. MARF**

**Case 1: `NeuralNetwork.java` in `marf.Classification.NeuralNetwork` package**

In order to make the class less complex the declared constants need to be separate in a new class which will be named as `NeuralNetworkConstants.java`, where the literal constants in the code will also need change to name constants and these constant will also be defined in new class. So all the constants are isolated from the code in `NeuralNetwork.java`. The outcome of this refactoring is the reduction in complexity of the class. This will make class more understandable, easy to debug and easy to maintain. Furthermore if the values of the constants need to be modified, it needs to be changed in only one place.

**Case 2: `MARF.java` in `marf` package**

`MARF.java` class is very large and complex because of too many declarations of attributes and methods. So to make it more understandable and maintainable extract class refactoring technique should be applied and remove some methods and attributes to other new classes. This class also has a large method named startRecognitionPipeline(). Now to make this method less complex extract method refactoring technique should be applied.

**Case 3: `ResultSet.java` in `marf.Storage` package**

`ResultSet.java` class is too large because of many methods. So to make it less complex extract class refactoring technique should be applied and remove some methods to other new classes. It also has a large method toString(). Extract method refactoring technique should be applied here.

**Case 4: `GrammarCompiler.java` in marf.nlp.Parsing package**

`GrammarCompiler.java` class has many methods which actually are more interested in other classes. Some of them are `fillInTransitionTable()`, `getBusted()`,`getGrammarElement()`,`addTerminalToken()`,`addIDToken()`, `createEOFTerminal()`. Those methods using other classes objects to perform their own operations. So here move method refactoring technique should be applied.

**II. GIPSY**

**Case 1:** `CodeGenerator.java` in `gipsy.GIPC.DFG.DFGGenerator` package

First, we will take the common attributes from both the classes and put those attributes in a new class `CodeGenerator.java` and make this class as the parent class of both `DFGCodeGenerator.java` and `DFGTranCodeGenerator.java`. Second, we will take the common methods with same functionalities in both classes and put them in the parent class.

The outcome of this refactoring is the reduction of redundancy in the system, increase in reusability as both classes now have a common parent and inherit their properties from it and each class has now their own specific functionality which reduces coupling. Although, there will be a requirement of changing some lines of code in other classes. For example, in class `GIPC.java,` the creation of objects of the classes in question should be changed according to the hierarchy that is a result of the refactoring applied. The changes made to the original system can be viewed by checking figure 31.

**Case 2: `GMTWrapper.java` in `gipsy.GEE.multitier.GMT` package**

`GMTWrapper.java` class is very long and complex. To make it more understandable extract class refactoring technique should be applied and remove some of the attribute and method declarations to other classes. Moreover, the run() method in this class is very large which is hard to understand. So extract method refactoring technique should be applied here. Furthermore, the method getAnAvailableDST() is more interested in class GMTInfoKeeper.java. So here move method refactoring technique should be applied.

**Case 3: `GIPC.java` in `gipsy.GIPC` package**

`GIPC.java` class is very large because of many attribute and method declarations. So to make this class more maintainable extract class refactoring should be applied. The process() method of this class is also long. So here extract method refactoring technique should be applied.

**Case 4: `GIPSYGMTOperator.java` in `gipsy.RIPE.editors.RunTimeGraphEditor.ui` package**

`GIPSYGMTOperator.java` class is very complicated because of its many methods and attribute declarations.

So here extract class refactoring technique should be applied. This class also has large methods like makeGIPSYNodesPanel() which need to be shorter. So extract method refactoring technique should be applied here. Moreover, this class has a method named startInstance() which is more interested in `GIPSYGMTController.java` class. So here move method refactoring should be applied.

### 4.1.3 Identification of Design Patterns

**All the code mentioned in this report has been taken from the case studies which can be found online at [43] and [44].**

**A. DMARF:**

**I. Strategy Pattern**: Strategy pattern is a behavioral design pattern which is used when we have multiple algorithms for a specified task and client decides the actual implementation during the run-time. Strategy pattern is also known as Policy pattern. Multiple algorithms are defined and based upon the input parameter, the algorithm will be executed [45].

In the marf.Classification.Distance package there exists a strategy pattern. Here, Distance.java class is an abstract class which is inherited by six different classes. It has a method named abstract distance() which the other six classes implement to calculate distance using their own algorithmic logic. The abstract class Distance.java is a common interface for the other classes to calculate distance. The UML diagram for the pattern is seen in figure 50. The relevant code of the class is given as follows

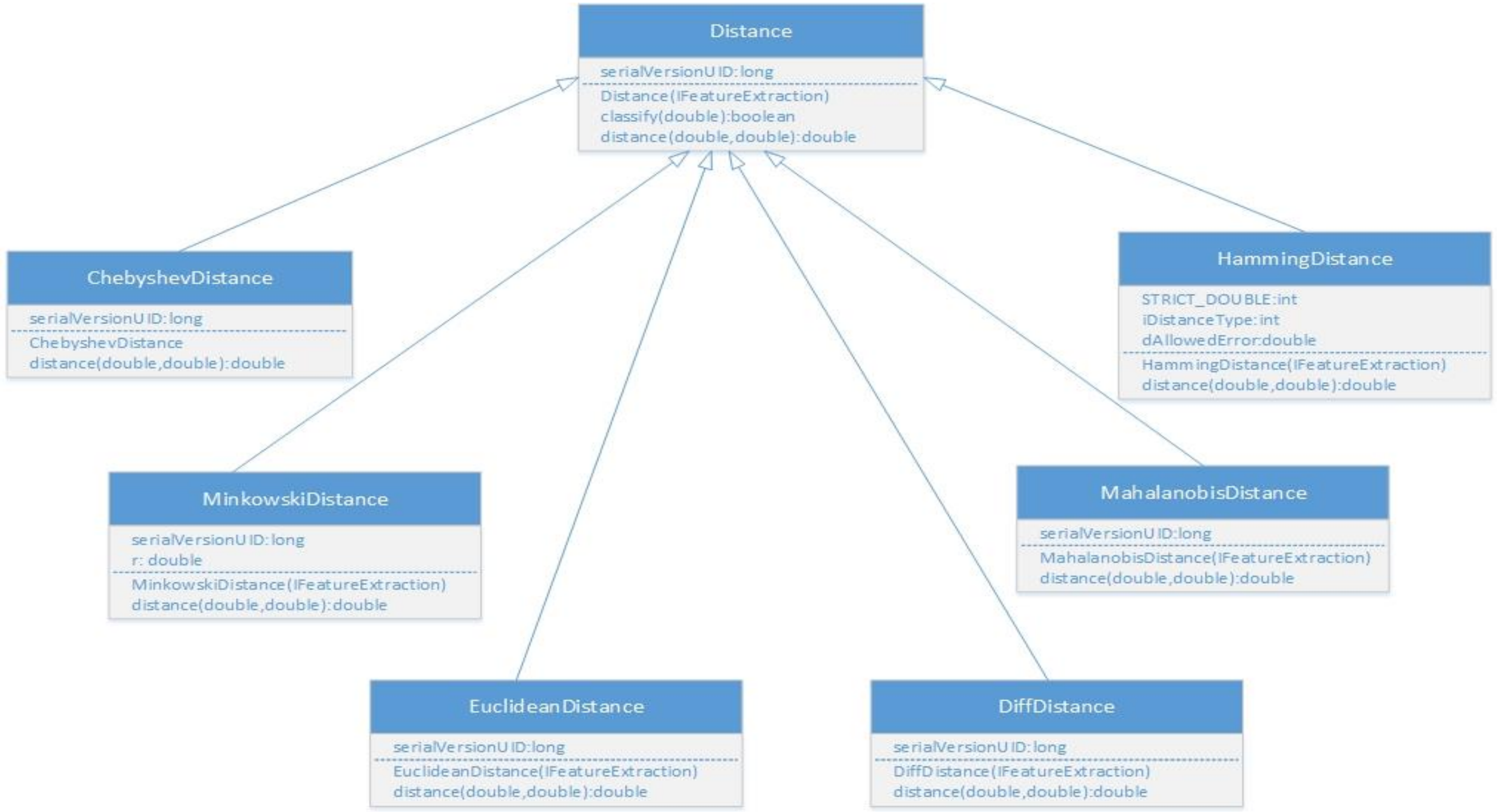

**Figure 50: Strategy pattern**

**Pattern as seen in the code:**

```
public abstract class Distance
extends Classification
{
public Distance(IFeatureExtraction poFeatureExtraction)
	{
		super(poFeatureExtraction);
	}
```

```
public abstract double distance(final double[] padVector1, final double[] padVector2);

}

public class EuclideanDistance
extends Distance
{

public EuclideanDistance(IFeatureExtraction poFeatureExtraction)
	{
		super(poFeatureExtraction);
	}

public final double distance(final double[] paVector1, final double[] paVector2)
	{
		double dDistance = 0;

		for(int f = 0; f < paVector1.length; f++)
		{
			dDistance += (paVector1[f] - paVector2[f]) * (paVector1[f] - paVector2[f]);
		}

		return dDistance;
	}

}

public class ChebyshevDistance extends Distance
{

public final double distance(final double[] padVector1, final double[] padVector2)

}

public class DiffDistance extends Distance
{

public final double distance(final double[] padVector1, final double[] padVector2)

}

public class HammingDistance extends Distance
{

public final double distance(final double[] padVector1, final double[] padVector2)

}

public class MahalanobisDistance extends Distance
{

public final double distance(final double[] paVector1, final double[] paVector2)
	}

public class MinkowskiDistance extends Distance
{

public final double distance(final double[] paVector1, final double[] paVector2)
	{

	}

}
```

**II. Abstract Factory Pattern:** Abstract factory pattern is a creational design pattern and is similar to factory pattern. In factory pattern, we have a single factory class that returns the different sub classes based on the inputs and it uses nested if-else or switch case statements. But in Abstract factory pattern, it gets rid of using if-else and switch statements and defines each factory method in the sub classes and the abstract factory class will return the sub class based on the input factory class [45].

The Abstract Factory pattern can be found in the `marf.Preprocessing` package. The package contains a class named `PreprocessingFactory.java` which contain method `IPreprocessing create()` which helps different kinds of objects to be instantiated at run time. In this scenario `marf.Preprocessing.CFEFilters` and `marf.Preprocessing.FFTfilter` packages have a family of classes which are related. There `CFEFilter` and `FFTFilter` classes are abstract and other related classes of their own package extend them. The UML diagram for the pattern is seen in figure 51. The relevant code of the class is given as follows.

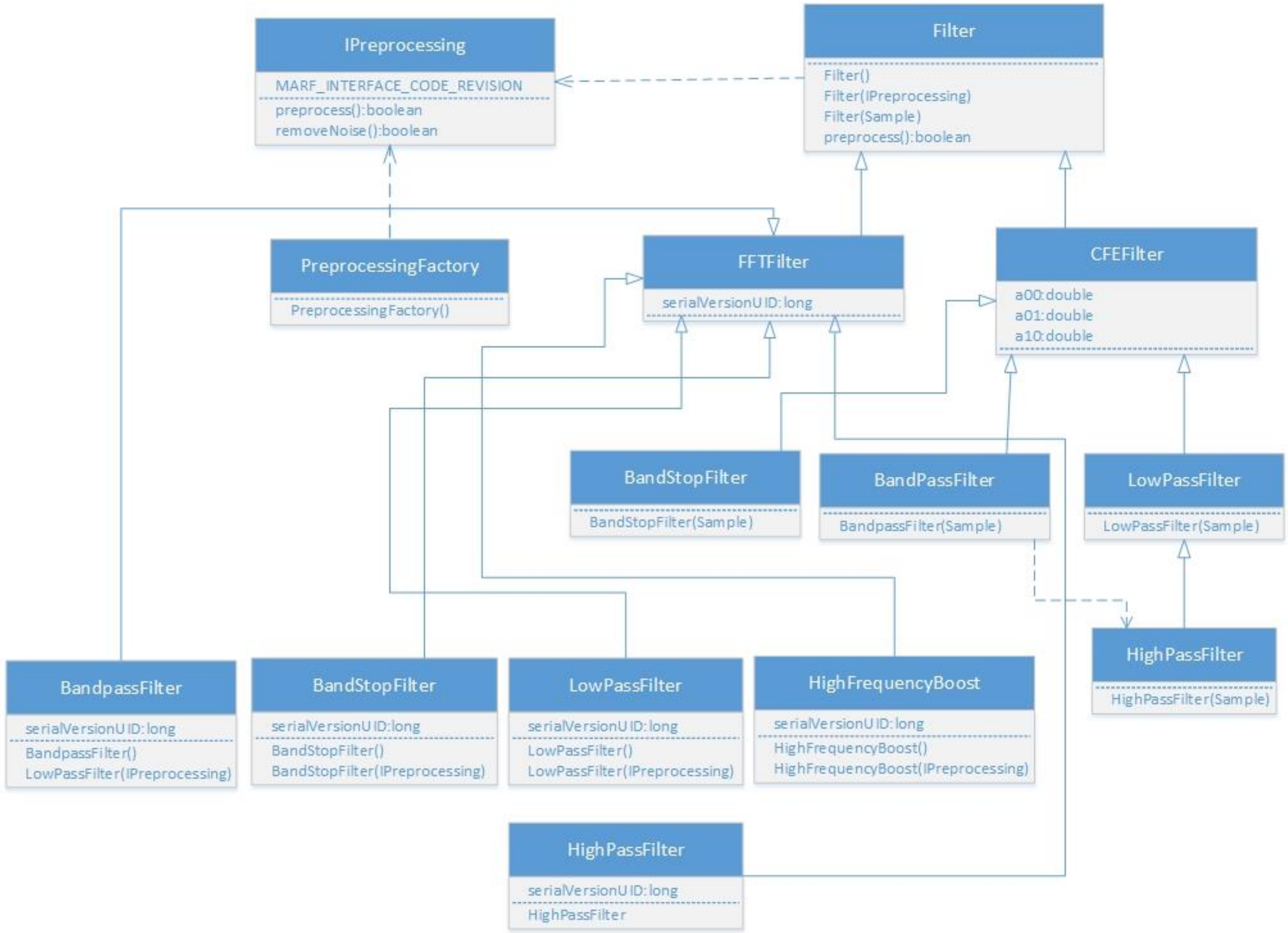

**Figure 51: Abstract factory Pattern**

**Pattern as seen in the code:**

```
public final class PreprocessingFactory
{

public static final IPreprocessing create(final int piPreprocessingMethod, Sample poSample)
       throws PreprocessingException
       {
              IPreprocessing oPreprocessing = null;

              switch(piPreprocessingMethod)
```

```
{

        case MARF.BANDPASS_FFT_FILTER:
        {
                oPreprocessing = new BandpassFilter(poSample);
                break;
        }

        case MARF.HIGH_FREQUENCY_BOOST_FFT_FILTER:
        {       oPreprocessing = new HighFrequencyBoost(poSample);

                break;
        }

        case MARF.LOW_PASS_FFT_FILTER:
        {
                oPreprocessing = new LowPassFilter(poSample);
                break;
        }

        case MARF.HIGH_PASS_FFT_FILTER:
        {
                oPreprocessing = new HighPassFilter(poSample);
                break;
        }

        case MARF.HIGH_PASS_BOOST_FILTER:
        {
                oPreprocessing = new HighFrequencyBoost(new HighPassFilter(poSample));
                break;
        }

        case MARF.BAND_STOP_FFT_FILTER:
        {
                oPreprocessing=newmarf.Preprocessing.FFTFilter.BandStopFilter(poSample);
                break;
        }

        case MARF.RAW:
        {
                oPreprocessing = new Raw(poSample);
                break;
        }

                MARF.LOW_PASS_CFE_FILTER:
        {
                OPreprocessing=new
                marf.Preprocessing.CFEFilters.LowPassFilter(poSample);
                break;
        }

        case MARF.HIGH_PASS_CFE_FILTER:
        {
                oPreprocessing=new
                marf.Preprocessing.CFEFilters.HighPassFilter(poSample);
                break;
        }

        case MARF.BAND_PASS_CFE_FILTER:
        {
```

```
                oPreprocessing=new
                marf.Preprocessing.CFEFilters.BandPassFilter(poSample);
                break;
            }

            case MARF.BAND_STOP_CFE_FILTER:
            {
                oPreprocessing=new
                marf.Preprocessing.CFEFilters.BandStopFilter(poSample);
                break;
            }

        return oPreprocessing;
    }
```

**III. Decorator:** A Decorator is a structural pattern that is also called as a wrapper class which adds additional responsibilities to an object at run-time. They help in extending the functionality of a single class instead of implementing a number of sub-classes [45].

The marf.net.server.corba.Preprocessing package has IPreprocessingCORBAPOATIE class which implements the decorator pattern by using IPreprocessingCORBAOperations interface class's operations. The UML diagram for the pattern is seen in figure 52. The relevant code of the class is given as follows.

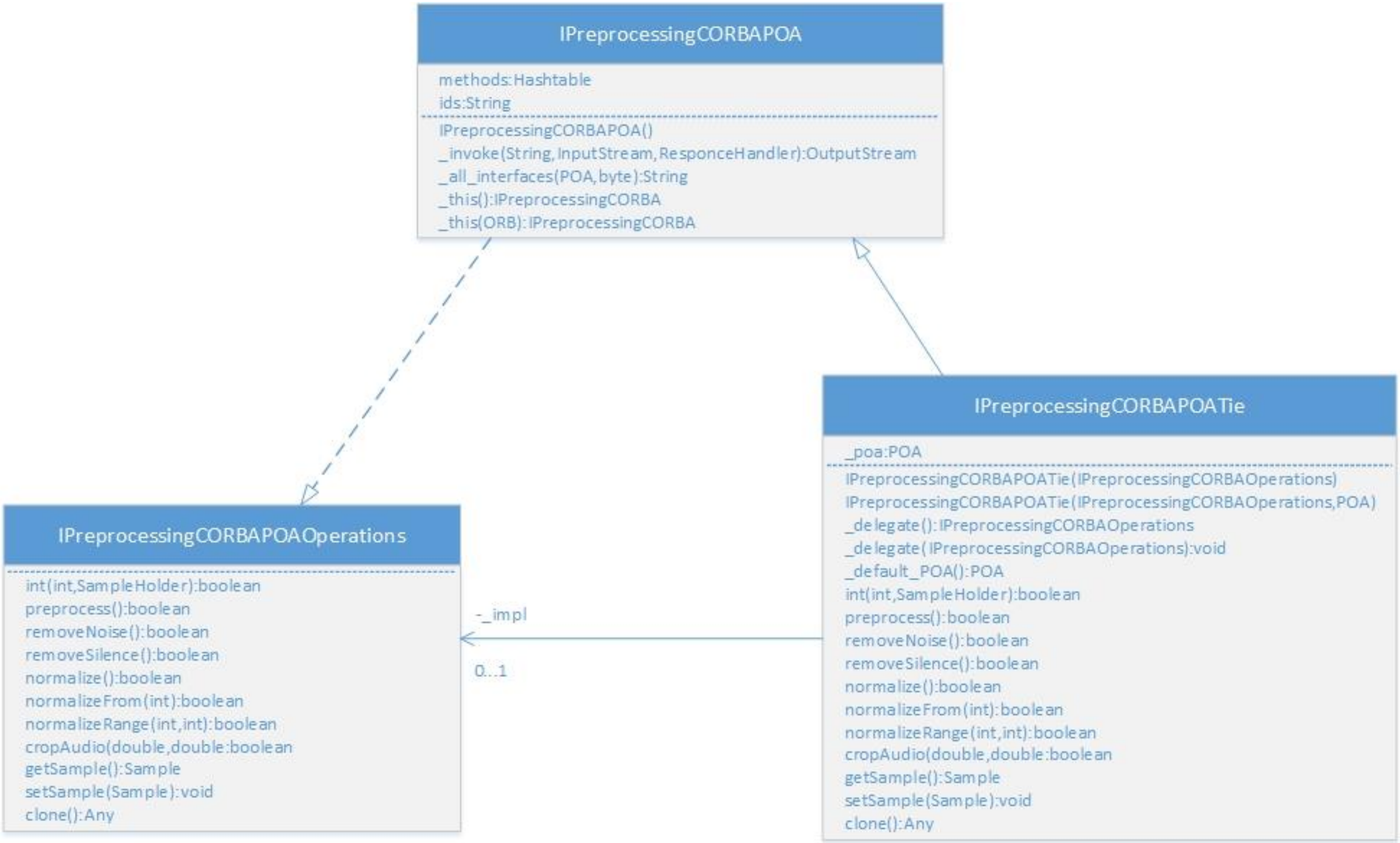

**Figure 52: Decorator Pattern**

**Pattern as seen in the code:**

```
public interface IPreprocessingCORBAOperations
{
  boolean init (int piPreprocessingMethod, marf.net.server.corba.Storage.SampleHolder poSample)
throws marf.net.server.corba.CORBACommunicationException;
  boolean preprocess () throws marf.net.server.corba.CORBACommunicationException;
  boolean removeNoise () throws marf.net.server.corba.CORBACommunicationException;
  boolean removeSilence () throws marf.net.server.corba.CORBACommunicationException;
  boolean normalize () throws marf.net.server.corba.CORBACommunicationException;
  boolean normalizeFrom (int piIndexFrom) throws
marf.net.server.corba.CORBACommunicationException;
  boolean normalizeRange (int piIndexFrom, int piIndexTo) throws
marf.net.server.corba.CORBACommunicationException;
  boolean cropAudio (double pdStartingFrequency, double pdEndFrequency) throws
marf.net.server.corba.CORBACommunicationException;
  marf.net.server.corba.Storage.Sample getSample ();
  void setSample (marf.net.server.corba.Storage.Sample poSample);
  org.omg.CORBA.Any _clone () throws marf.net.server.corba.CORBACommunicationException;
}

public class IPreprocessingCORBAPOATie extends IPreprocessingCORBAPOA
{

  public IPreprocessingCORBAPOATie (
marf.net.server.corba.Preprocessing.IPreprocessingCORBAOperations delegate ) {
      this._impl = delegate;
  }
  public IPreprocessingCORBAPOATie (
marf.net.server.corba.Preprocessing.IPreprocessingCORBAOperations delegate ,
org.omg.PortableServer.POA poa ) {
      this._impl = delegate;
      this._poa      = poa;
  }
  public marf.net.server.corba.Preprocessing.IPreprocessingCORBAOperations _delegate() {
      return this._impl;
  }
  public void _delegate (marf.net.server.corba.Preprocessing.IPreprocessingCORBAOperations
delegate ) {
      this._impl = delegate;
  }
  public org.omg.PortableServer.POA _default_POA() {
      if(_poa != null) {
          return _poa;
      }
      else {
          return super._default_POA();
      }
  }
  public boolean init (int piPreprocessingMethod, marf.net.server.corba.Storage.SampleHolder
poSample) throws marf.net.server.corba.CORBACommunicationException
  {
    return _impl.init(piPreprocessingMethod, poSample);
  }

  public boolean preprocess () throws marf.net.server.corba.CORBACommunicationException
  {
    return _impl.preprocess();
  }
  public boolean removeNoise () throws marf.net.server.corba.CORBACommunicationException
  {
```

```
    return _impl.removeNoise();
  }

  public boolean removeSilence () throws marf.net.server.corba.CORBACommunicationException
  {
    return _impl.removeSilence();
  }

  public boolean normalize () throws marf.net.server.corba.CORBACommunicationException
  {
    return _impl.normalize();
  }

  public boolean normalizeFrom (int piIndexFrom) throws
marf.net.server.corba.CORBACommunicationException
  {
    return _impl.normalizeFrom(piIndexFrom);
  }

  public boolean normalizeRange (int piIndexFrom, int piIndexTo) throws
marf.net.server.corba.CORBACommunicationException
  {
    return _impl.normalizeRange(piIndexFrom, piIndexTo);
  }

  public boolean cropAudio (double pdStartingFrequency, double pdEndFrequency) throws
marf.net.server.corba.CORBACommunicationException
  {
    return _impl.cropAudio(pdStartingFrequency, pdEndFrequency);
  }

  public marf.net.server.corba.Storage.Sample getSample ()
  {
    return _impl.getSample();
  }

  public void setSample (marf.net.server.corba.Storage.Sample poSample)
  {
    _impl.setSample(poSample);
  }

  public org.omg.CORBA.Any _clone () throws marf.net.server.corba.CORBACommunicationException
  {
    return _impl._clone();
  }

  private marf.net.server.corba.Preprocessing.IPreprocessingCORBAOperations _impl;
  private org.omg.PortableServer.POA _poa;

}
```

**IV.Factory Pattern:** Factory design pattern is used when there is a super class with multiple sub classes and based on the inputs, it returns the output from one of the sub classes. This pattern usually takes the responsibility of the instantiation of the class from the client to the factory class. In order to do so, a singleton pattern can be applied on a factory class or make the factory method as static [45].

The `marf.FeatureExtraction` package has a factory pattern in `FeatureExtractionFactory.java`. In this class new objects can be created at run time for feature extraction. Here `FeatureExtraction.java` class is abstract which other classes inherit. The UML diagram for the pattern is seen in figure 53. The relevant code of the class is given as follows.

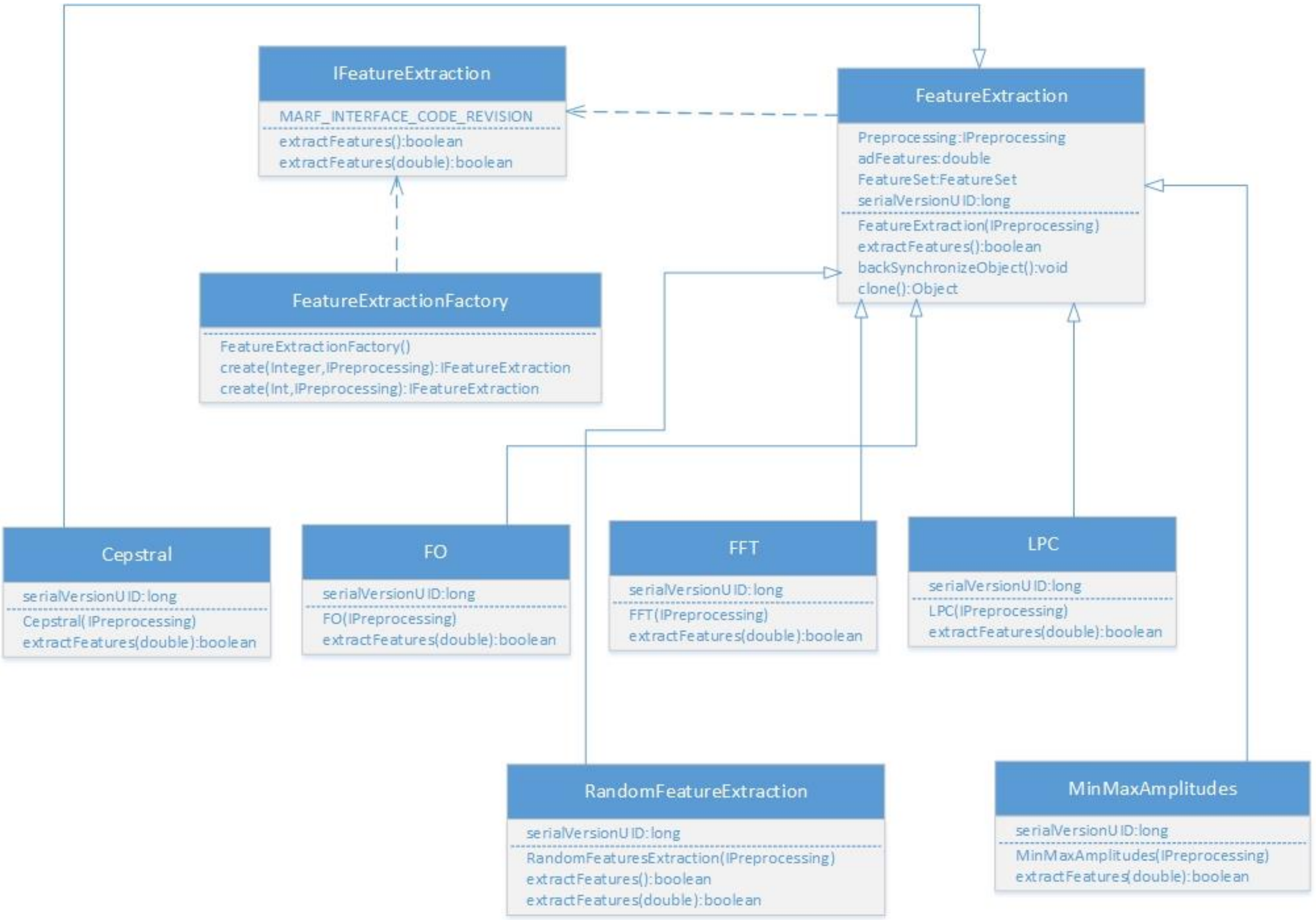

**Figure 53: Factory Pattern**

**Pattern as seen in Code:**

```
public final class FeatureExtractionFactory
{
	private FeatureExtractionFactory()
	{
	}
public static final IFeatureExtraction create(final int piFeatureExtractionMethod, IPreprocessing
poPreprocessing)
	throws FeatureExtractionException
	{
		IFeatureExtraction oFeatureExtraction = null;

		switch(piFeatureExtractionMethod)
		{
			case MARF.LPC:
				oFeatureExtraction = new LPC(poPreprocessing);
				break;

			case MARF.FFT:
```

```
                        oFeatureExtraction = new FFT(poPreprocessing);
                        break;

                case MARF.F0:
                        oFeatureExtraction = new F0(poPreprocessing);
                        break;


                case MARF.CEPSTRAL:
                        oFeatureExtraction = new Cepstral(poPreprocessing);
                        break;

                case MARF.RANDOM_FEATURE_EXTRACTION:
                        oFeatureExtraction = new RandomFeatureExtraction(poPreprocessing);
                        break;

                case MARF.MIN_MAX_AMPLITUDES:
                        oFeatureExtraction = new MinMaxAmplitudes(poPreprocessing);
                        break;

            return oFeatureExtraction;
        }

}
```

## B. GIPSY

**I. Observer Pattern:** Observer design pattern is a behavioral pattern which is useful when you are in the state of an object and want to get notified whenever there is any change. In this pattern, the object that watches the state of another object is called the Observer and the object that is being watched is called Subject. Java Message Service uses observer pattern to allow the applications to subscribe and publish the data to the other applications [45].

In GIPSY there is observer pattern because `gipsy.tests.GEE.simulator.ResultPool.java` class works as a subscriber and `gipsy.GEE.IDP.demands.IDemand.java` works as a publisher here. Here ResultPool class has updateGUI() method which changes if anything happens to IDemand class.

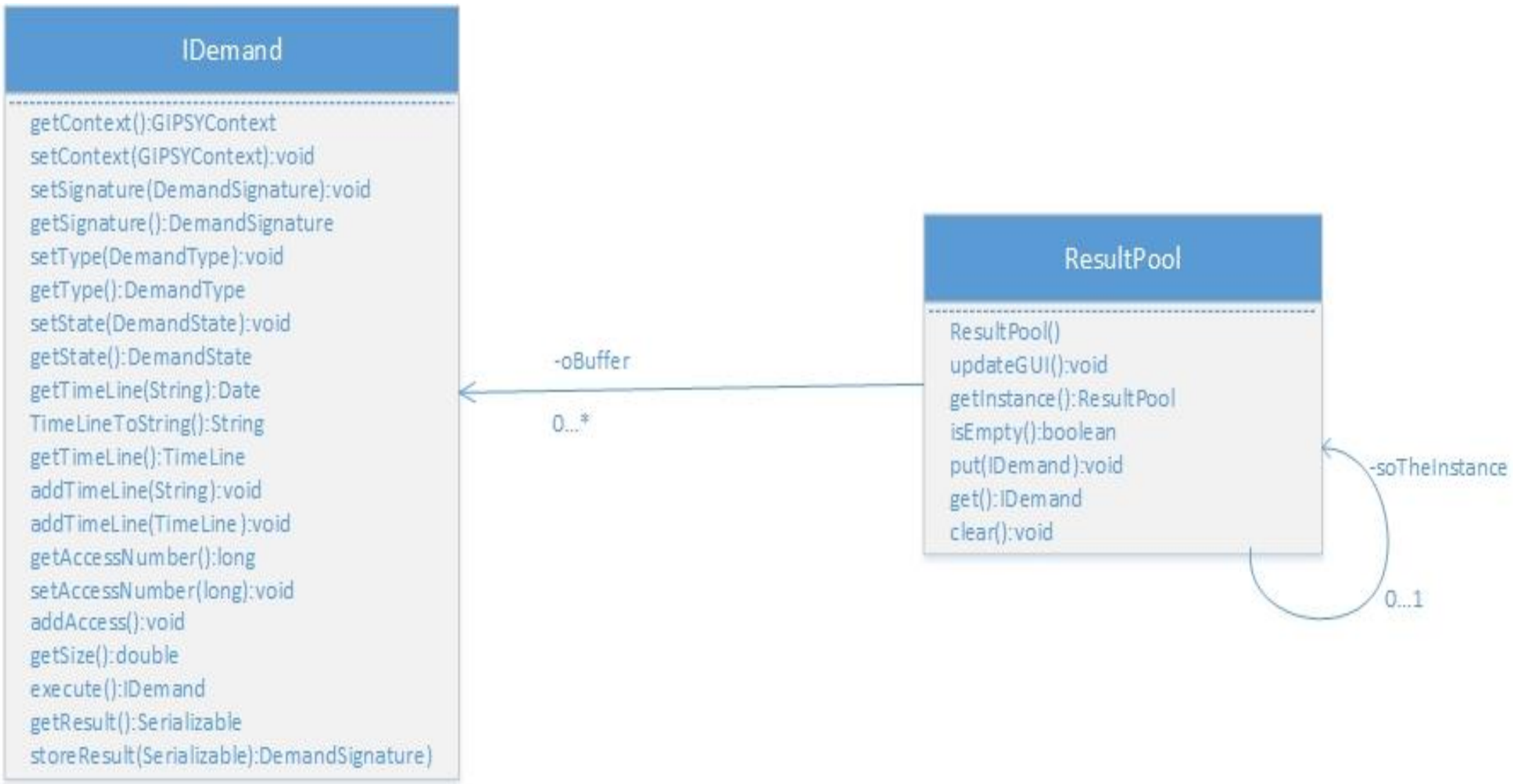

**Figure 54: Observer Pattern**

}

**Pattern as seen in Code:**

```
public interface IDemand extends ISequentialThread, Cloneable
{

public GIPSYContext getContext();



	public void setContext(GIPSYContext poContext);
	void setSignature(DemandSignature poSignatureID);
	DemandSignature getSignature();
	void setType(DemandType poType);
	DemandType getType();
	void setState(DemandState poState);
	DemandState getState();
	Date[] getTimeLine(String pstrTierID);
	String timeLineToString();
	TimeLine getTimeLine();
	void addTimeLine(String pstrTierID);
	void addTimeLine(TimeLine poTimeLine);
	long getAccessNumber();
	void setAccessNumber(long plAccessNumber);
	void addAccess();
	double getSize();
	IDemand execute();
	Serializable getResult();
	DemandSignature storeResult(Serializable poResult);
}


public class ResultPool
{

	private List<IDemand> oBuffer = null;
	private static ResultPool soTheInstance = null;

	private ResultPool()
	{
		this.oBuffer = new LinkedList<IDemand>();
	}

	public synchronized void updateGUI()
	{
		StringBuilder oStrBuilder = new StringBuilder();

		for(int i = 0; i < this.oBuffer.size(); ++i)
		{
			oStrBuilder.append(this.oBuffer.get(i).getSignature().toString());
			oStrBuilder.append(GlobalDef.CR);
			oStrBuilder.append(GlobalDef.LF);
		}
	GlobalDef.soDGTDialog.getTAResults().setText(oStrBuilder.toString());
	}

}
```

**II. Adapter Pattern:** Adapter design pattern is a structural design pattern and it's used so that two unrelated interfaces can work together. The object that joins these unrelated interfaces is called adaptor. It is adapting between the classes and objects and also bridges between several objects [45].

In GIPSY there is adapter pattern in package `gipsy.apps.marfcat`. There `gipsy.apps.marfcat.MARFCATDGT` works as an adaptee and `gipsy.apps.marfcat.MARFCATDGTApp` works as an adapter. Basically here adapter pattern is needed because another app MARFCAT is going to run on GIPSY's multitier architecture.

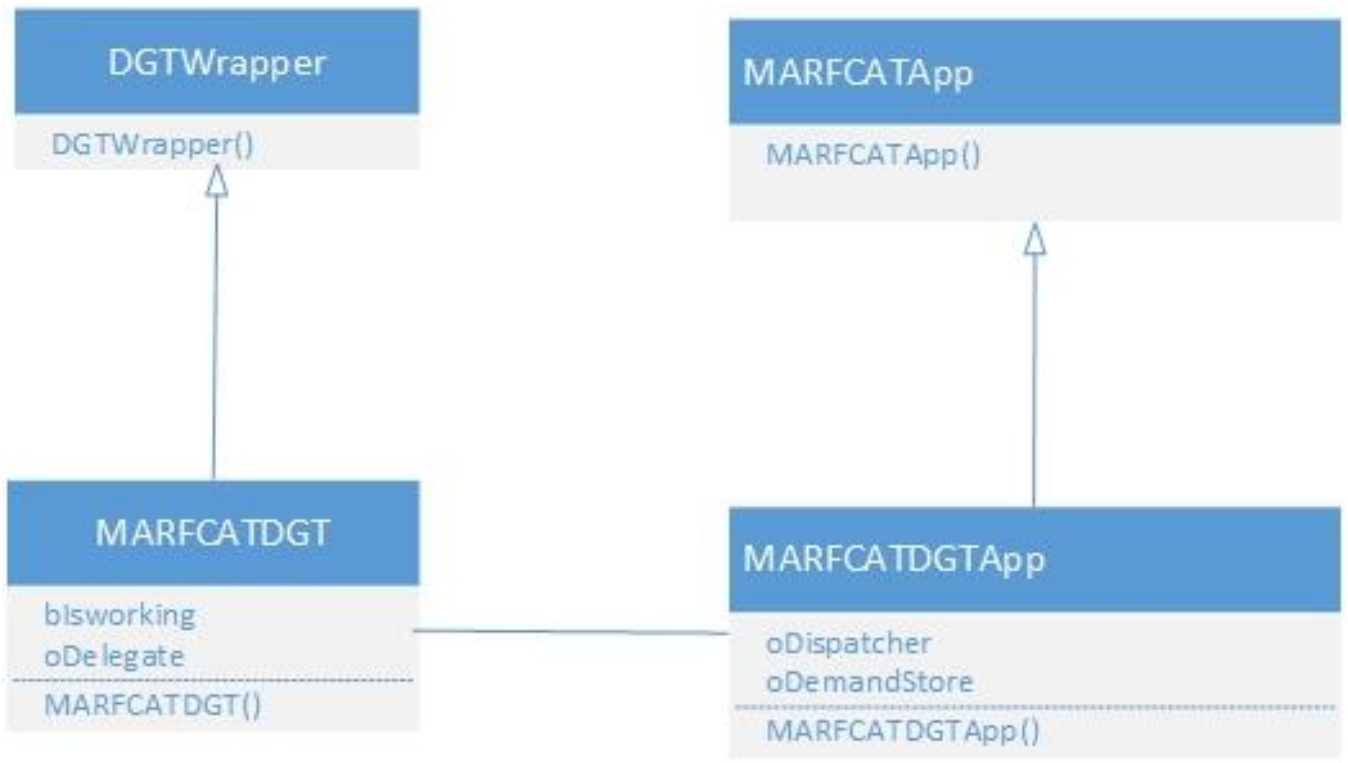

**Figure 55: Adapter Pattern**

**Pattern as seen in Code:**

```
public class MARFCATDGT extends DGTWrapper
{

	protected boolean bIsWorking = false;

	protected IMARFCATDelegateApp oDelegate = null;

	public MARFCATDGT();

}

public static void main(String[] argv)
	throws Exception
	{
		Debug.enableDebug();

		MARFCATDGT oDGT = new MARFCATDGT();
		oDGT.startTier();
		oDGT.wait();
	}

public void run()
	{
		this.bIsWorking = true;

		while(this.bIsWorking)
		{
			this.oDelegate.execute();

			System.out.println("MARFCAT Delegate Completed.");

			this.bIsWorking = false;
		}
	}
```

```
public class MARFCATDGTApp extends MARFCATApp implements IMARFCATDelegateApp
       {

              private IDemandDispatcher oDispatcher = null;

              private LocalDemandStore oDemandStore = null;


              public MARFCATDGTApp(IDemandDispatcher poDispatcher)
              {
                     this.oDispatcher = poDispatcher;
              }

public void doTestingWork(String[] pastrArgv, Integer piExpectedID)
{}
public void execute()
              {

}
```

**III. Prototype Pattern:** The prototype pattern is a creational pattern which is used to create customized objects without knowing any details about the class [45].

In GIPSY `gipsy.Configuration.java` class is used as a prototype by client class `gipsy.GEE.multitier.GIPSYNode.java`. Here `gipsy.GEE.multitier.GIPSYNode.java` class declares `oRegDSTTAConfig` and `oSysDSTTAConfig` as instances of `gipsy.Configiguaration.java` class.

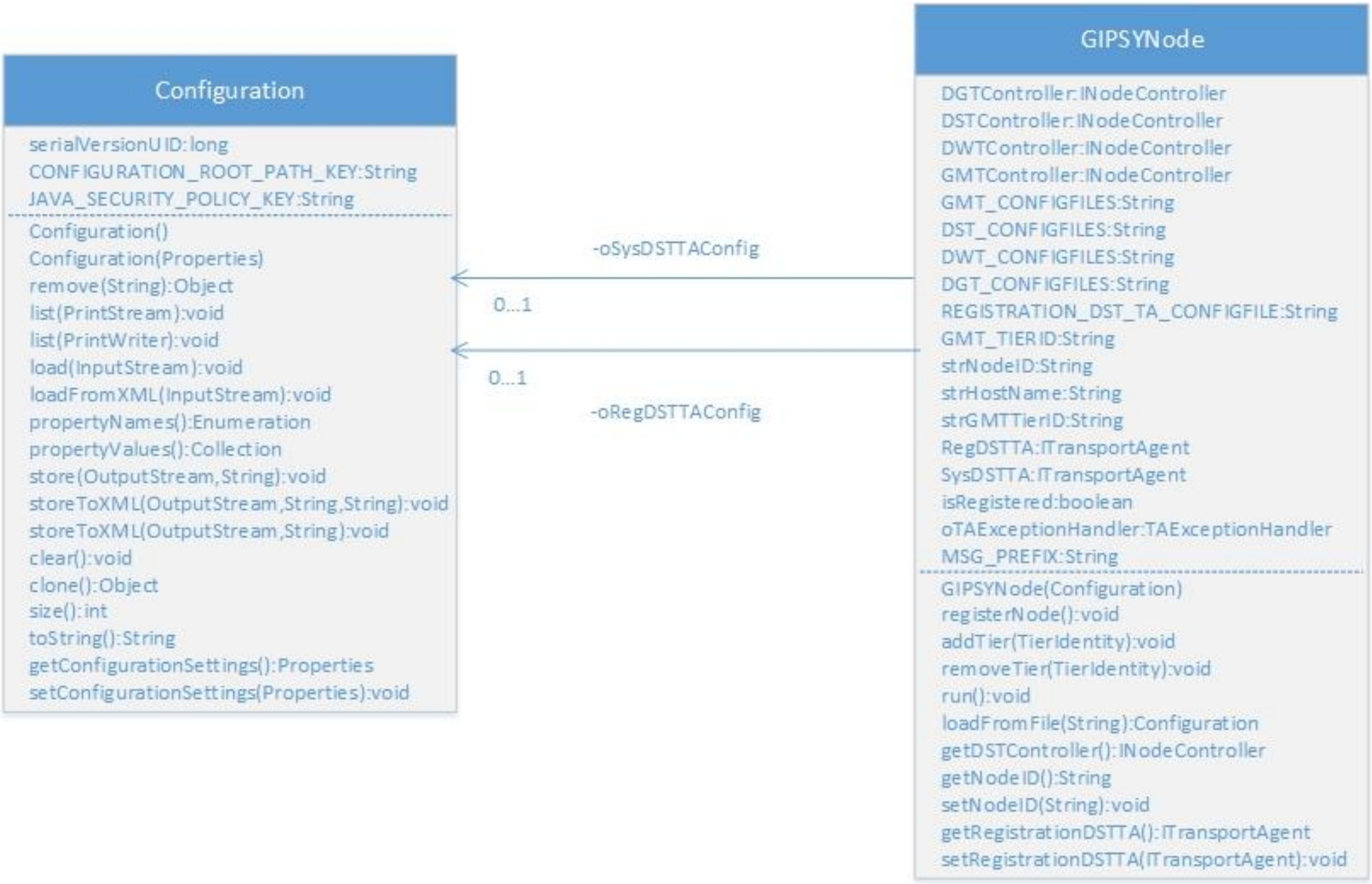

**Figure 56: Prototype Pattern**

**Pattern as seen in Code:**

```
public class Configuration implements Serializable
{
	protected Properties oConfigurationSettings = null;

	private static final long serialVersionUID = 123L;

public Configuration()
	{
		super();
		this.oConfigurationSettings = new Properties();
		this.initializeDefaultSettings();
	}

public synchronized Object clone()
	{
		Configuration oNewConfig = new Configuration();
		oNewConfig.setConfigurationSettings((Properties)
this.oConfigurationSettings.clone());
		return oNewConfig;
	}



public Properties getConfigurationSettings()
	{
		return this.oConfigurationSettings;
	}

	public void setConfigurationSettings(Properties configurationSettings)
	{
		this.oConfigurationSettings = configurationSettings;
	}

}

public class GIPSYNode extends Thread
{

private ITransportAgent oRegDSTTA;

private ITransportAgent oSysDSTTA;

private Configuration oRegDSTTAConfig;

private Configuration oSysDSTTAConfig;

public void registerNode()
	throws MultiTierException
	{

// uses oSysDSTTAConfig here

	 }

public GIPSYNode(Configuration poNodeConfig)
	{

	String strTAConfigFile = poNodeConfig.getProperty(REGISTRATION_DST_TA_CONFIGFILE);
try
		{

if(strTAConfigFile != null)
			{
```

```
                        this.oRegDSTTAConfig = loadFromFile(strTAConfigFile);

                        this.oRegDSTTA = TAFactory.getInstance().createTA(this.oRegDSTTAConfig);
                  }
                  else
                  {

                        this.oRegDSTTA = null;
                  }

public void setRegistrationDSTTA(ITransportAgent poRegistrationDSTTA)
      {
            this.oRegDSTTA = poRegistrationDSTTA;
            this.oRegDSTTAConfig = this.oRegDSTTA.getConfiguration();
      }

public void setSystemDSTTA(ITransportAgent poSystemDSTTA)
      {
            this.oSysDSTTA = poSystemDSTTA;
            this.oSysDSTTAConfig = this.oSysDSTTA.getConfiguration();
      }

}
```

**IV. Singleton Pattern:** Singleton pattern restricts the instantiation of a class and ensures that only one instance exists in the class and provide a global point of access to it [45].

In GIPSY `gipsy.GEE.multitier.DST` package has TAFactory class which has singleton pattern. The UML diagram for the pattern is seen in figure 57. The relevant code of the class is given as follows.

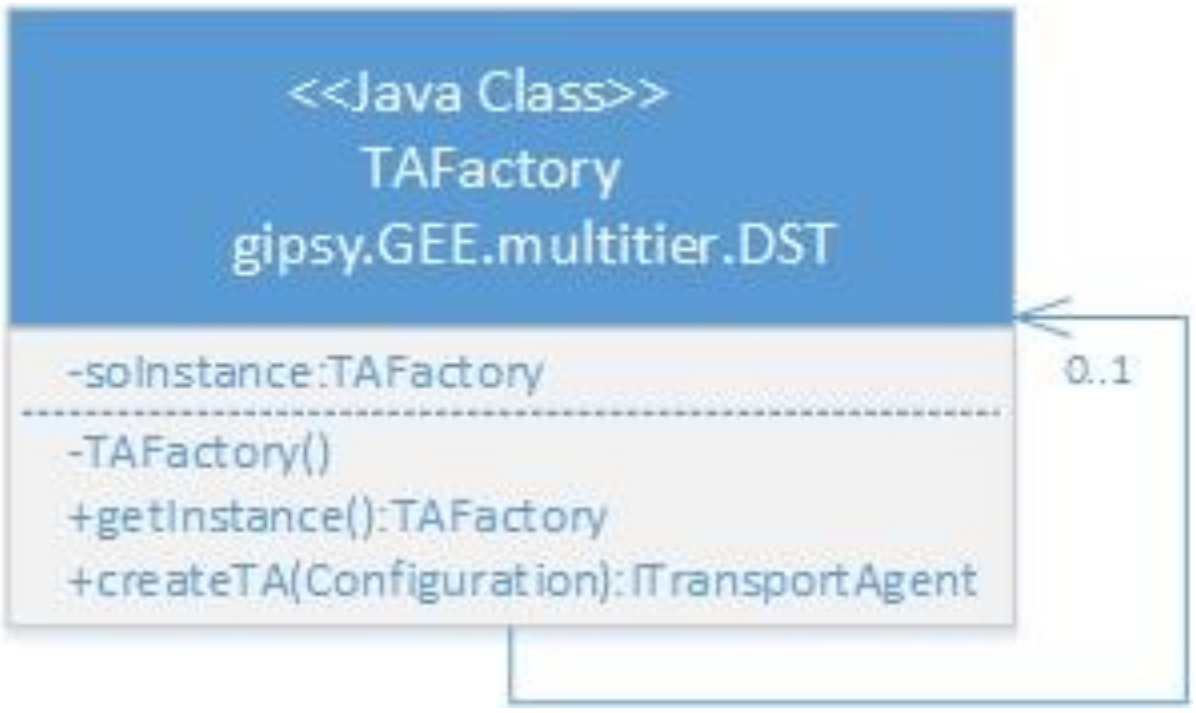

**Figure 57: Singleton Pattern**

**Pattern as seen in the code:**

```
public class TAFactory {

private static TAFactory soInstance = null;

private TAFactory()

public synchronized static TAFactory getInstance() {

if(soInstance == null) {

soInstance = new TAFactory();          }

return soInstance; }
```

### 4.2 Implementation

In this section, the refactoring techniques suggested before have been applied and the resulting changes are discussed in the change logs, patches and diffs.

#### 4.2.1 Refactoring Changesets and Diffs

## I. DMARF

### Case 1: Refactoring for class `NeuralNetwork.java` in `MARF.Classification.NeuralNetwork` package

The figure 58 depicts the diagram of the `NeuralNetwork.java` class after refactoring.

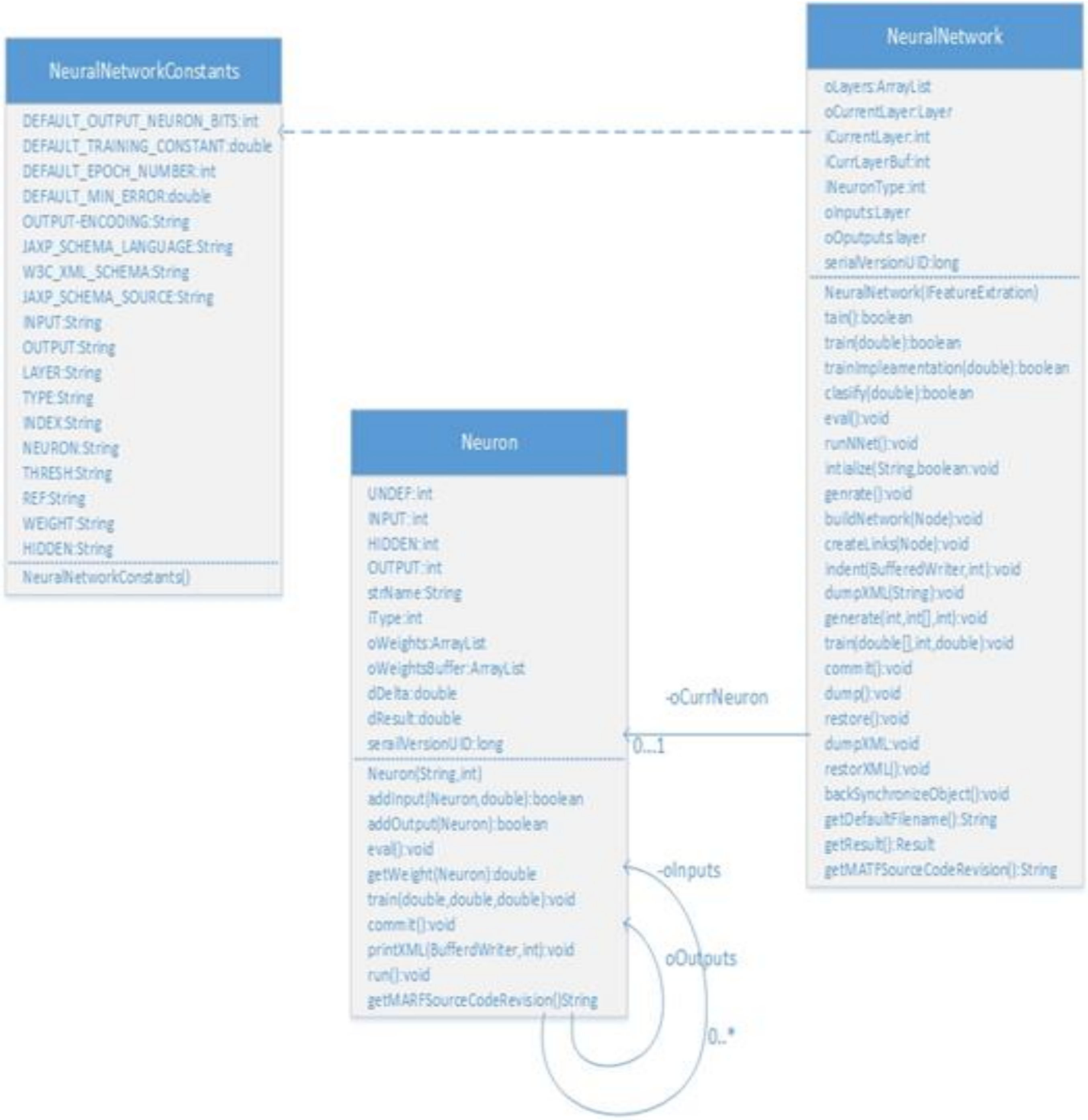

**Figure 58: `NeuralNetwork.java` class after refactoring**

**Change 0/2: Refactor `NeuralNetwork.java` into two classes**

NeuralNetwork class has many Constants. As it is very large class so it is better to separate the constants of that class. It makes the code easy to understand and easy to debug.

In the code there are many literal constants used so those constants are changed to named constants. This will make the code easy to change and easy to understand.

**Change 1/2: Create a new class `NeuralNetworkConstants.java`**

Copy all the declared constants from `NeuralNetwork.class` in this class and also declare new constants for the literal constant used in that class.

Diff of NeuralNetworkConstants class is given as follow.

diff -N src/marf/Classification/NeuralNetwork/NeuralNetworkConstants.java

--- /dev/null 1 Jan 1970 00:00:00 -0000

+++ Src/marf/Classification/NeuralNetwo

rk/NeuralNetworkConstants.java 1 Jan 1970 00:00:00 -0000

@@ -0,0 +1,79 @@

**Change 2/2: Remove the declared constants in `NeuralNetwork.java`**

As Constants of this has been copied to `NeuralNetworkConstants.class`. So all the declared constants in NeuralNetwork.java class are removed. The diff of that change is given as follow.

diff -u -r1.1 NeuralNetwork.java

--- src/marf/Classification/NeuralNetwork/NeuralNetwork.java 13 Aug 2014 02:14:30 -0000 1.1

+++ src/marf/Classification/NeuralNetwork/NeuralNetwork.java 23 Aug 2014 19:14:00 -0000

@@ -52,32 +52,7 @@

@@ -126,30 +101,7 @@

Then the references of the constants and literals in the code are changed. The diff of this change is shown below.

@@ -243,9 +195,9 @@

@@ -445,7 +397,7 @@

@@ -504,7 +456,7 @@

@@ -526,7 +478,7 @@

@@ -546,7 +498,7 @@

@@ -555,7 +507,7 @@

@@ -564,14 +516,14 @@

@@ -583,7 +535,7 @@

@@ -593,7 +545,7 @@

@@ -605,12 +557,12 @@

@@ -668,7 +620,7 @@

@@ -677,14 +629,14 @@

@@ -700,7 +652,7 @@

@@ -711,7 +663,7 @@

@@ -720,7 +672,7 @@

@@ -732,11 +684,11 @@

@@ -788,7 +740,7 @@

@@ -799,7 +751,7 @@

@@ -897,15 +849,15 @@

### Case 2: Refactoring for class `MARF.java` in `marf` package

The figure 59 depicts the diagram of the `MARF.java` class after refactoring.

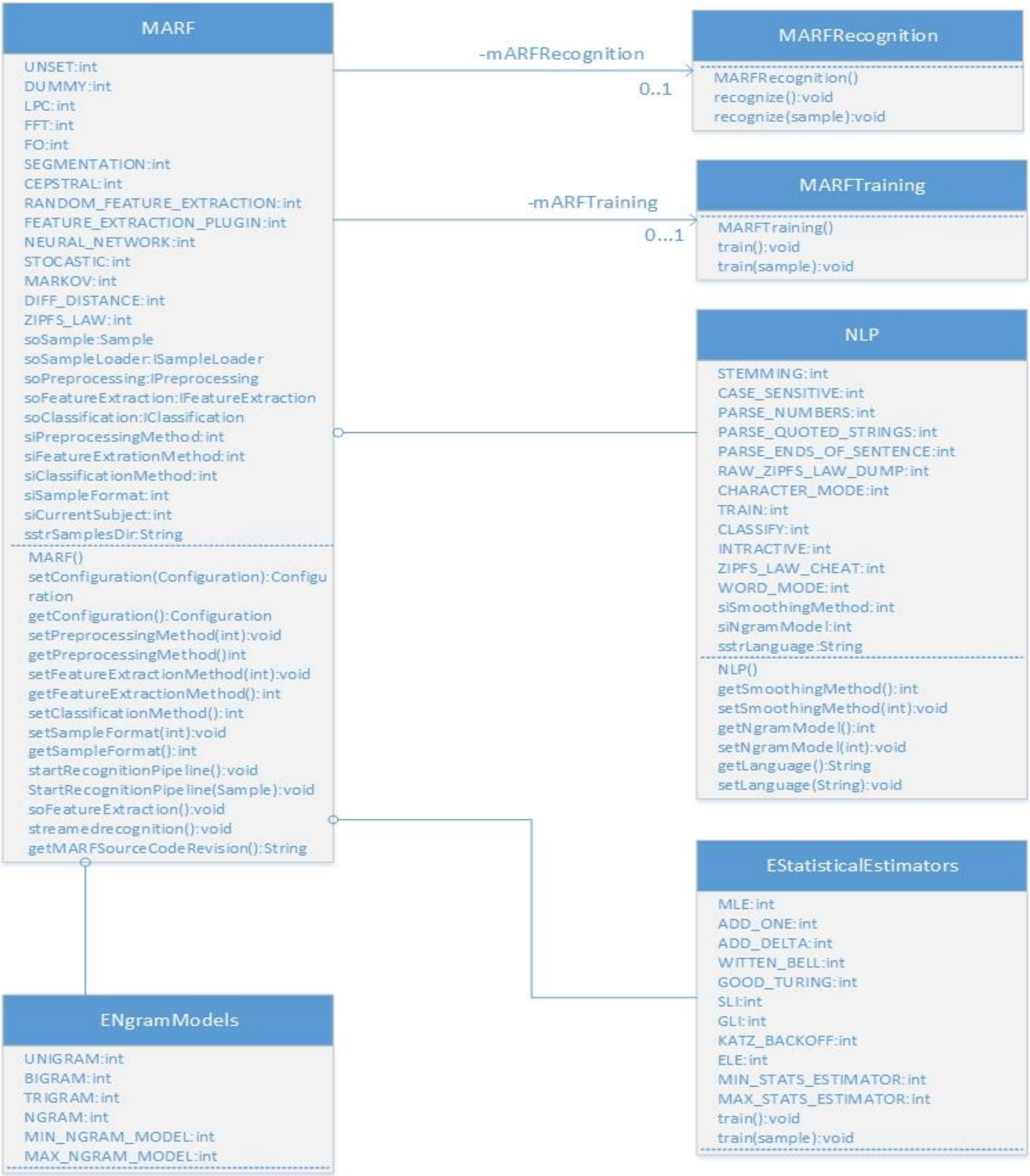

**Figure 59: `MARF.java` class after refactoring**

**Change 0/1: Refactor class `MARF.java` into three different classes**

Original `MARF.java` class is very large because it contains lots of attributes and methods. So, to reduce the complexity of the class extract class refactoring technique has been applied and two new classes has been created. Now new class MARFRecognition.java is responsible for MARF identification mode. New MARFTraining.java class is now responsible for MARF training mode.

The diff of that change is given as follow.

diff -u -r1.1 MARF.java

--- src/marf/MARF.java 13 Aug 2014 02:14:25 -0000 1.1

+++ src/marf/MARF.java 23 Aug 2014 20:08:33 -0000

@@ -47,6 +47,10 @@

@@ -519,7 +523,7 @@

@@ -1302,34 +1306,7 @@

@@ -1341,49 +1318,7 @@

@@ -1402,7 +1337,7 @@

@@ -1426,7 +1361,7 @@

@@ -1479,12 +1414,9 @@

@@ -1501,6 +1433,14 @@

diff -N src/marf/MARFRecognition.java

--- /dev/null 1 Jan 1970 00:00:00 -0000

+++ src/marf/MARFRecognition.java 1 Jan 1970 00:00:00 -0000

@@ -0,0 +1,47 @@

diff -N src/marf/MARFTraining.java

--- /dev/null 1 Jan 1970 00:00:00 -0000

+++ src/marf/MARFTraining.java 1 Jan 1970 00:00:00 -0000

@@ -0,0 +1,51 @@

**Change 1/1: Extract method applied in startRecognitionPipeline() method in `MARF.java` class**

Original startRecognitionPipeline() method in MARF.java class is too big and complex. That's why a new method has been created as private and named soFeatureExtraction() method. It now share responsibilities of original method startRecognitionPipeline(). This new method is created so that original startRecognitionPipeline() method become short and less complex.

The diff of that change is given as follow.

diff -u -r1.1 MARF.java

--- src/marf/MARF.java 13 Aug 2014 02:14:25 -0000 1.1

+++ src/marf/MARF.java 23 Aug 2014 20:08:33 -0000

@@ -47,6 +47,10 @@

@@ -1402,7 +1337,7 @@

## Case 3: Refactoring for class `Resultset.java` in `marf.Storage` package

The figure 60 depicts the diagram of the `ResultSet.java` class after refactoring.

**Change 0/1: Refactor class `ResultSet.java` into four different classes**

Original ResultSet.java class is large because of many methods. So, to reduce the complexity of the class extract class refactoring technique has been applied and three new classes has been created. Now new class ResultSetClosestID.java is responsible for returning the closest ID. New ResultSetID.java class is in charge of retrieving ID of a subject. ResultSetRandom.java class is now responsible for retrieving result of random ID.

The diff of that change is given as follow.

diff -u -r1.1 ResultSet.java

--- src/marf/Storage/ResultSet.java 13 Aug 2014 02:14:27 -0000 1.1

+++ src/marf/Storage/ResultSet.java 23 Aug 2014 17:41:19 -0000

@@ -23,6 +23,10 @@

@@ -104,54 +108,6 @@

@@ -162,25 +118,8 @@

@@ -191,18 +130,7 @@

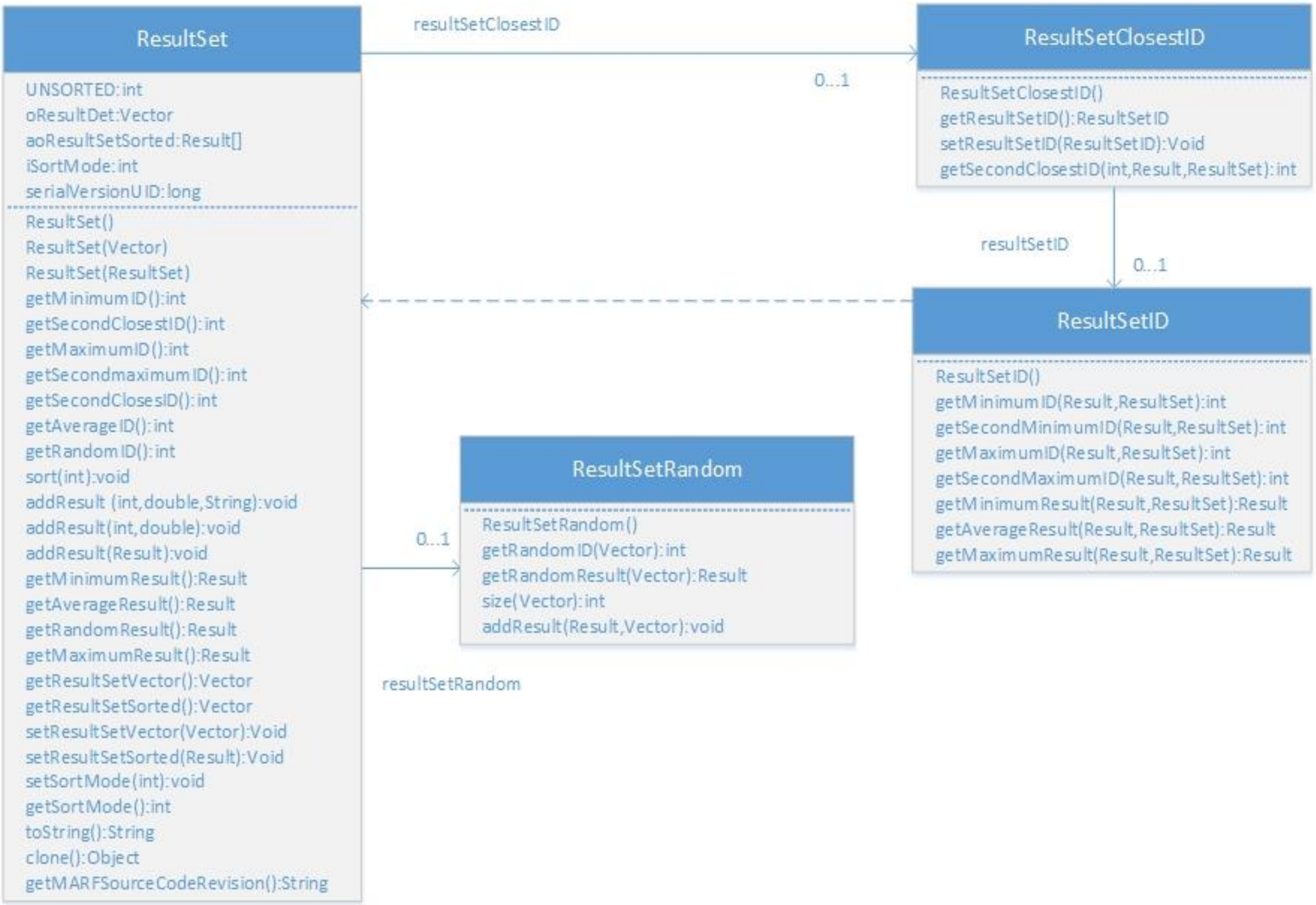

**Figure 60: `ResultSet.java` class after refactoring**

@@ -263,12 +191,7 @@

@@ -278,29 +201,7 @@

@@ -310,18 +211,7 @@

@@ -413,29 +303,7 @@

@@ -446,13 +314,21 @@

diff -N src/marf/Storage/ResultSetClosestID.java

--- /dev/null 1 Jan 1970 00:00:00 -0000

+++ src/marf/Storage/ResultSetClosestID.java 1 Jan 1970 00:00:00 -0000

@@ -0,0 +1,47 @@

diff -N src/marf/Storage/ResultSetID.java

--- /dev/null 1 Jan 1970 00:00:00 -0000

+++ src/marf/Storage/ResultSetID.java 1 Jan 1970 00:00:00 -0000

@@ -0,0 +1,94 @@

diff -N src/marf/Storage/ResultSetRandom.java

--- /dev/null 1 Jan 1970 00:00:00 -0000

+++ src/marf/Storage/ResultSetRandom.java 1 Jan 1970 00:00:00 -0000

@@ -0,0 +1,52 @@

**Change 1/1: Extract method applied in toString() method in `ResultSet.java` class**

One new method has been created and named oSortedBuffer() method. It now share responsibilities of original method toString(). This new method is created so that original toString() method become short and less complex.

The diff of that change is given as follow.

diff -u -r1.1 ResultSet.java

--- src/marf/Storage/ResultSet.java 13 Aug 2014 02:14:27 -0000 1.1

+++ src/marf/Storage/ResultSet.java 23 Aug 2014 17:41:19 -0000

@@ -23,6 +23,10 @@

@@ -413,29 +303,7 @@

## Case 4: Refactoring for class `GrammarCompiler.java` in `marf.nlp.Parsing` package

The figure 61 depicts the diagram of the `GrammerCompiler.java` class after refactoring.

**Change 0/5: Move method fillInTransitionTable() of `GrammarCompiler.java` class to `TransitionTable.java` class**

The method fillInTransitionTable() is more interested in class TransitionTable.java rather than the class GrammarCompiler.java where it belongs. Because it is using an object of TransitionTable.java named soTransitionTable in the method of fillInTansitionTable(). That's why it is more appropriate to move fillInTansitionTable() method to TransitionTable.java class which is in marf.nlp.Parsing package.

The diff of that change is given as follow.

diff -u -r1.1 TransitionTable.java

--- src/marf/nlp/Parsing/TransitionTable.java 13 Aug 2014 02:14:17 -0000 1.1

+++ src/marf/nlp/Parsing/TransitionTable.java 24 Aug 2014 17:45:23 -0000

@@ -6,8 +6,10 @@

@@ -482,6 +484,124 @@

**Change 1/5: Move method getBusted() of `GrammarCompiler.java` class to `Token.java` class in `marf.nlp.Parsing` package**

The method getBusted() is more interested in class Token.java rather than the class GrammarCompiler.java where it belongs. Because it is using an object of Token.java named oToken in the method of getBusted(). That's why it is more appropriate to move getBusted() method to Token.java class which is in marf.nlp.Parsing package.

The diff of that change is given as follow.

diff -u -r1.1 Token.java

--- src/marf/nlp/Parsing/Token.java 13 Aug 2014 02:14:17 -0000 1.1

+++ src/marf/nlp/Parsing/Token.java 24 Aug 2014 17:45:23 -0000

@@ -9,6 +9,7 @@

@@ -238,6 +239,21 @@

**Change 2/5: Move method getGrammarElement() of `GrammarCompiler.java` class to `Grammar.java` class in `marf.nlp.Parsing.GrammarCompiler` package**

The method getGrammarElement() is more interested in class Grammar.java rather than the class GrammarCompiler.java where it belongs. Because it is using an object of Grammar.java named oGrammar in the method of getGrammarElement(). That's why it is more appropriate to move getGrammarElement() method to Grammar.java class which is in marf.nlp.Parsing package.

The diff of that change is given as follow.

diff -u -r1.1 Grammar.java

---
src/marf/nlp/Parsing/GrammarCompiler/Grammar.java 13 Aug 2014 02:14:30 -0000 1.1

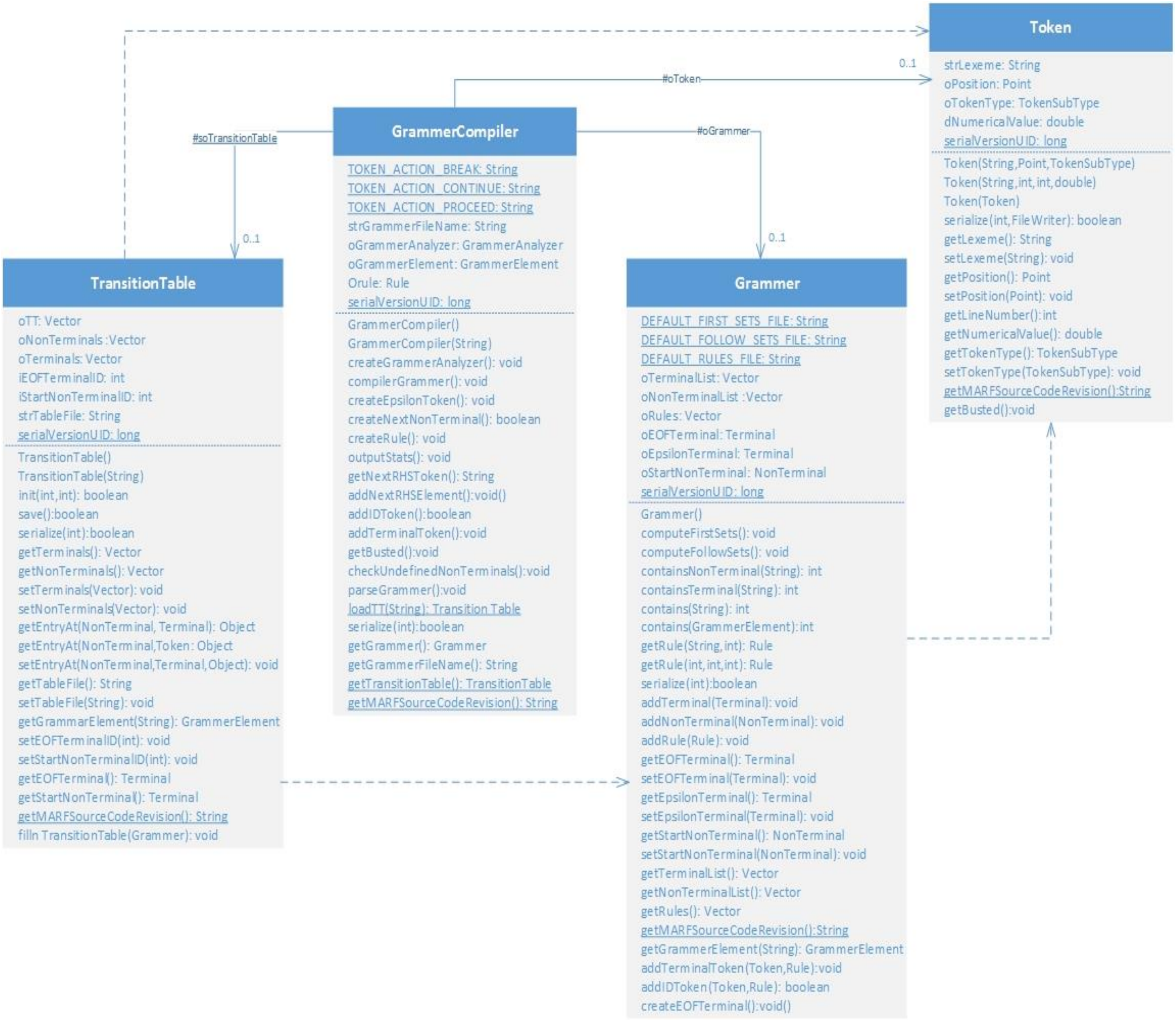

**Figure 61: `GrammerCompiler.java` class after refactoring**

+++ src/marf/nlp/Parsing/GrammarCompiler/Grammar.java 24 Aug 2014 17:45:23 -0000

@@ -2,8 +2,8 @@

@@ -674,6 +674,79 @@

**Change 3/5: Move method addTerminalToken() of `GrammarCompiler.java` class to `Grammar.java` class in `marf.nlp.Parsing.GrammarCompiler` package**

The method addTerminalToken() is more interested in class Grammar.java rather than the class GrammarCompiler.java where it belongs. Because it is using an object of Grammar.java named oGrammar in the method of addTerminalToken(). That's why it is more appropriate to move addTerminalToken() method to Grammar.java class which is in marf.nlp.Parsing package.

The diff of that change is given as follow.

diff -u -r1.1 Grammar.java

--- src/marf/nlp/Parsing/GrammarCompiler/Grammar.java 13 Aug 2014 02:14:30 -0000 1.1

+++ src/marf/nlp/Parsing/GrammarCompiler/Grammar.java 24 Aug 2014 17:45:23 -0000

@@ -2,8 +2,8 @@

@@ -674,6 +674,79 @@

**Change 4/5: Move method addIDToken() of `GrammarCompiler.java` class to `Grammar.java` class in `marf.nlp.Parsing.GrammarCompiler` package**

The method addIDToken() is more interested in class Grammar.java rather than the class GrammarCompiler.java where it belongs. Because it is using an object of Grammar.java named oGrammar in the method of addIDToken(). That's why it is more appropriate to move addIDToken() method to Grammar.java class which is in marf.nlp.Parsing package.

The diff of that change is given as follow.

diff -u -r1.1 Grammar.java

--- src/marf/nlp/Parsing/GrammarCompiler/Grammar.java 13 Aug 2014 02:14:30 -0000 1.1

+++ src/marf/nlp/Parsing/GrammarCompiler/Grammar.java 24 Aug 2014 17:45:23 -0000

@@ -2,8 +2,8 @@

@@ -674,6 +674,79 @@

**Change 5/5: Move method createEOFTerminal() of `GrammarCompiler.java` class to `Grammar.java` class in `marf.nlp.Parsing.GrammarCompiler` package**

The method createEOFTerminal() is more interested in class Grammar.java rather than the class GrammarCompiler.java where it belongs. Because it is using an object of Grammar.java named oGrammar in the method of createEOFTerminal(). That's why it is more appropriate to move createEOFTerminal() method to Grammar.java class which is in marf.nlp.Parsing package.

The diff of that change is given as follow.

diff -u -r1.1 Grammar.java

--- src/marf/nlp/Parsing/GrammarCompiler/Grammar.java 13 Aug 2014 02:14:30 -0000 1.1

+++ src/marf/nlp/Parsing/GrammarCompiler/Grammar.java 24 Aug 2014 17:45:23 -0000

@@ -2,8 +2,8 @@

@@ -674,6 +674,79 @@

## II. GIPSY

### Case 1: Refactoring for classes `DFGCodeGenerator.java` and `DFGTranCodeGenerator.java` in `gipsy.GIPC.DFG.DFGGenerator` package

The figure 62 depicts the diagram of the **`DFGCodeGenerator.java`** class and **`DFGTranCodeGenerator.java`** after refactoring.

**Change 0/1: Refactoring Code Duplication among `DFGCodeGenerator` and `DFGTranCodeGenerator` Classes**

Most of the methods and attributes are same in both of those classes except there are two additional attributes "FunName" and "dimshape" in DFGTranCodeGenerator and implementation of methods drawDFG(), drawDFG(SimpleNode) and generateDFG(Simplenode,int,String) is different.

In DFGTranCodeGenerator class method genTable() and genTable(SimpleNode) have same implementation as methods in generateDimensionIndexTable() and generateDimensionIndexTable(SimpleNode) in class DFGGenerateCode, so in order to make them identical we changed the name of genTable() and genTable(SimpleNode) to generateDimensionIndexTable() and generateDimensionIndexTable(SimpleNode).

In addition to that there was one attribute private int funnum = 1 was unused in both of the classes so we removed that.

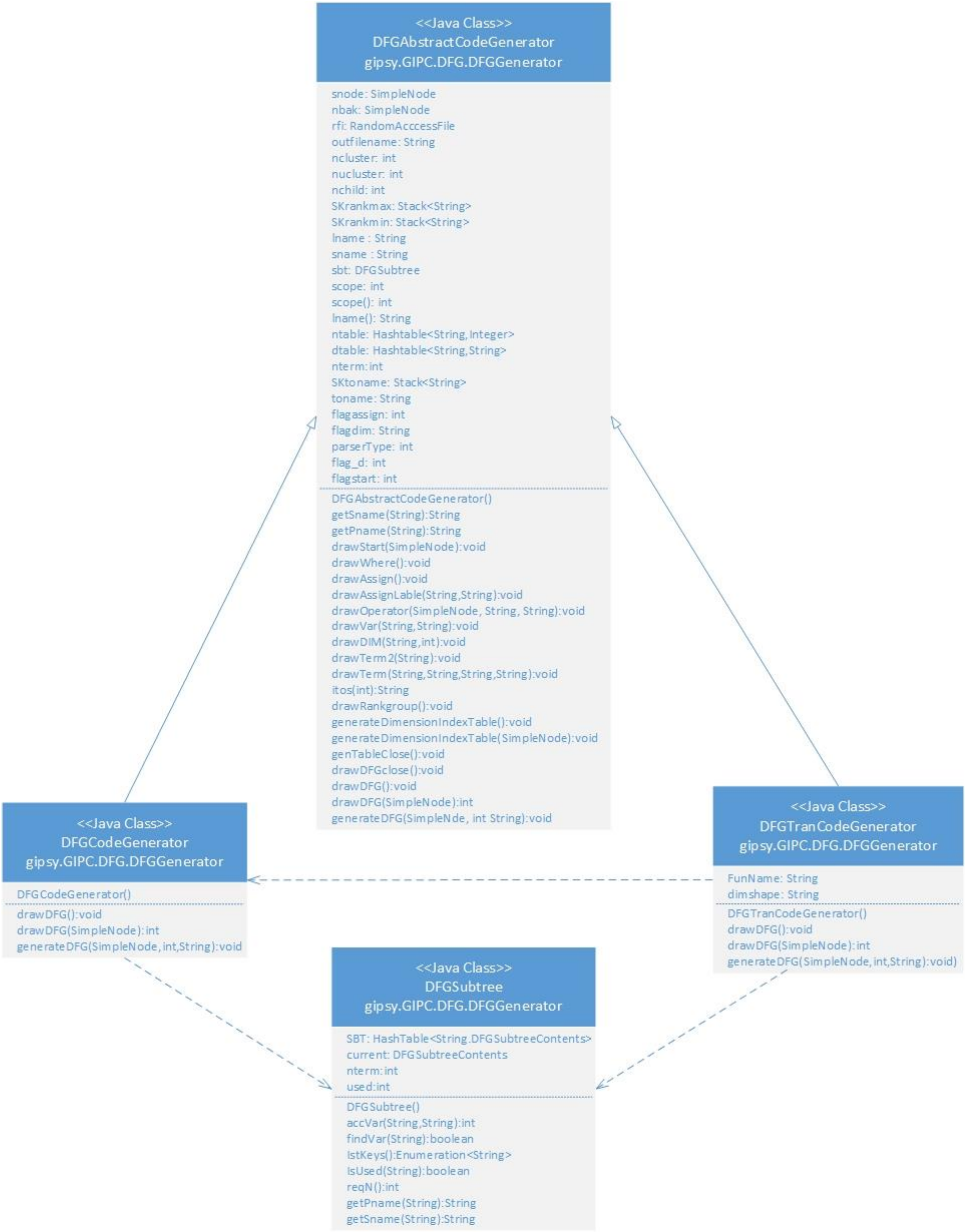

**Figure 62: `DFGCodeGenerator.java` and `DFGTranCodeGenerator.java` class after refactoring**

**Change 1/2: Create new `DFGAbstractCodeGenerator` class.**

Create new abstract class, in which we copy common attributes and common methods of both DFGCodeGenerator and DFGCodeTranGenerator classes. Then we declare the abstract methods, drawDFG(), drawDFG(SimpleNode) and generateDFG(Simplenode,int,String). The diff of DFGAbstractCodeGenerator.java class is as follow.

diff -N src/gipsy/GIPC/DFG/DFGGenerator/DFGAbstractCodeGenerator.java

--- /dev/null 1 Jan 1970 00:00:00 -0000

+++ src/gipsy/GIPC/DFG/DFGGenerator/DFGAbstractCodeGenerator.java 1 Jan 1970 00:00:00 -0000

@@ -0,0 +1,452 @@

**Change2/2: Remove common methods and attributes from `DFGCodeGenerator` and `DFGTranCodeGenerator` class.**

Extend both classes from DFGAbstractCodeGenerator class and remove the common attributes and methods from

DFGCodeGenerator. The diff of that class is given as follow.

diff -u -r1.1 DFGCodeGenerator.java

--- src/gipsy/GIPC/DFG/DFGGenerator/DFGCodeGenerator.java 13 Aug 2014 02:41:24 -0000 1.1

+++ src/gipsy/GIPC/DFG/DFGGenerator/DFGCodeGenerator.java 22 Aug 2014 22:04:01 -0000

@@ -20,316 +20,15 @@

Similarly remove the common attribute and common methods from class DFGCodeTranGenerartor. Following shows the diff of that class.

diff -u -r1.1 DFGTranCodeGenerator.java

--- src/gipsy/GIPC/DFG/DFGGenerator/DFGTranCodeGenerator.java 13 Aug 2014 02:41:24 -0000 1.1

+++ src/gipsy/GIPC/DFG/DFGGenerator/DFGTranCodeGenerator.java 22 Aug 2014 22:04:01 -0000

@@ -17,530 +17,219 @@

In method drawDFG(SimpleNode) change the call of DFGCodeGenerator class according to hierarchy. This is done in both classes.

The diff of that change in DFGCodeGenerator.java class is as follow.

@@ -434,7 +133,7 @@

@@ -536,122 +235,12 @@

Change the access type to protected of those attributes in class DFGAbstractController, that are used by drawDFG(), drawDFG(SimpleNode) and generateDFG(SimpleNode, int, String), in refactored class.

## Case 2: Refactoring for class `GMTWrapper.java` in `gipsy.GEE.multitier.GMT` package

The figure 63 depicts the diagram of the `GMTWrapper.java class` after refactoring.

**Change 0/2: Refactor class `GMTWrapper.java` into two classes**

Original GMTWrapper.java class is too large which contain many attributes and methods. So, to reduce the complexity of the class extract class refactoring technique has been applied and one new class has been created. Now new class GMTWrapperAllocateDeallocateTier.java is responsible for allocating and deallocating tier in GIPSY's multitier architecture.

The diff of that change is given as follow.

diff -u -r1.1 GMTWrapper.java

--- src/gipsy/GEE/multitier/GMT/GMTWrapper.java 13 Aug 2014 02:41:14 -0000 1.1

+++ src/gipsy/GEE/multitier/GMT/GMTWrapper.java 24 Aug 2014 00:31:28 -0000

@@ -45,6 +45,7 @@

@@ -103,7 +104,7 @@

@@ -140,20 +141,7 @@

@@ -172,7 +160,7 @@

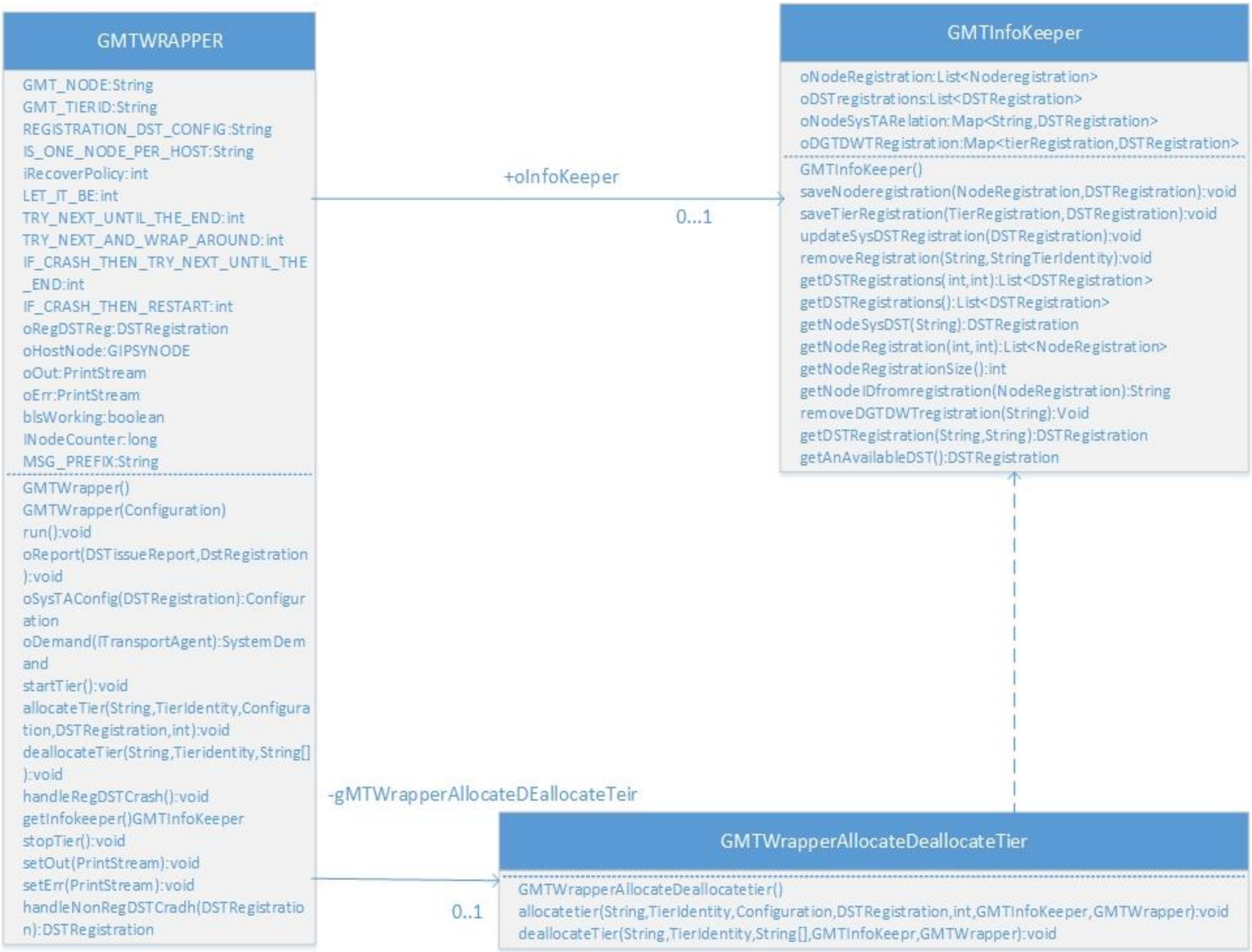

**Figure 63: `GMTWrapper.java` class after refactoring**

@@ -196,18 +184,7 @@

@@ -404,7 +381,7 @@

@@ -449,9 +426,7 @@

@@ -468,6 +443,31 @@

@@ -593,212 +593,24 @@

@@ -866,27 +678,6 @@

@@ -900,10 +691,10 @@

diff -N src/gipsy/GEE/multitier/GMT/GMTWrapperAllocateDeallocateTier.java

--- /dev/null 1 Jan 1970 00:00:00 -0000

+++ src/gipsy/GEE/multitier/GMT/GMTWrapperAllocateDeallocateTier.java 1 Jan 1970 00:00:00 -0000

@@ -0,0 +1,165 @@

**Change 1/2: Extract method applied in run() method of `GMTWrapper.java` class**

Original run() method in GMTWrapper.java class is too big and complex. That's why three new methods have been created as private and named as oDemand() method, oSysTAConfig() method and oReport() method. They now share responsibilities of original method run(). This new method is created so that original run() method become short and less complex.

The diff of that change is given as follow.

diff -u -r1.1 GMTWrapper.java

--- src/gipsy/GEE/multitier/GMT/GMTWrapper.java 13 Aug 2014 02:41:14 -0000 1.1

+++ src/gipsy/GEE/multitier/GMT/GMTWrapper.java 24 Aug 2014 00:31:28 -0000

@@ -45,6 +45,7 @@

@@ -468,6 +443,31 @@

**Change 2/2: Move method getAnAvailableDST() of `GMTWrapper.java` class to `GMTInfoKeeper.java` class in `gipsy.GEE.multitier.GMT` package**

The method getAnAvailableDST() is more interested in class GMTInfoKeeper.java rather than the class GMTWrapper.java where it belongs. Because it is using GMTInfoKeeper.java class's method getDSTRegistrations(). That's why getAnAvailableDST() method is moved to GMTInfoKeeper.java class which is in gipsy.GEE.multitier.GMT package.

The diff of that change is given as follow.

diff -u -r1.1 GMTInfoKeeper.java

--- src/gipsy/GEE/multitier/GMT/GMTInfoKeeper.java 13 Aug 2014 02:41:14 -0000 1.1

+++ src/gipsy/GEE/multitier/GMT/GMTInfoKeeper.java 24 Aug 2014 00:31:28 -0000

@@ -349,4 +349,19 @@

## Case 3: Refactoring for class `GIPC.java` in `gipsy.GIPC` package

The figure 64 depicts the diagram of the `GIPC.java class` after refactoring.

**Change 0/1: Refactor class `GIPC.java` into two classes**

Original GIPC.java class is too large which contain many attributes and methods. So, to reduce the complexity of the class extract class refactoring technique has been applied and one new class has been created. Now new class GIPCSetupConfiguration.java is responsible for configuring the GIPC.

The diff of that change is given as follow.

diff -u -r1.1 GIPC.java

--- src/gipsy/GIPC/GIPC.java 13 Aug 2014 02:41:11 -0000 1.1

+++ src/gipsy/GIPC/GIPC.java 24 Aug 2014 02:49:27 -0000

@@ -55,6 +55,8 @@

@@ -184,11 +186,6 @@

@@ -257,7 +254,7 @@

@@ -275,8 +272,8 @@

@@ -288,124 +285,10 @@

@@ -414,8 +297,8 @@

@@ -428,18 +311,7 @@

@@ -460,7 +332,7 @@

@@ -471,7 +343,7 @@

@@ -479,7 +351,7 @@

@@ -490,7 +362,7 @@

@@ -498,7 +370,7 @@

@@ -507,7 +379,7 @@

@@ -516,7 +388,7 @@

@@ -525,7 +397,7 @@

@@ -679,7 +551,7 @@

@@ -693,12 +565,12 @@

@@ -743,21 +615,13 @@

@@ -783,6 +647,14 @@

@@ -935,7 +807,7 @@

diff -N src/gipsy/GIPC/GIPCSetupConfiguration.java

--- /dev/null 1 Jan 1970 00:00:00 -0000

+++ src/gipsy/GIPC/GIPCSetupConfiguration.java 1 Jan 1970 00:00:00 -0000

@@ -0,0 +1,124 @@

**Change 1/1: Extract method applied in process() method of `GIPC.java` class**

Original process() method in GIPC.java class is too big and complex. That's why a new method has been created as private and named as oGEE() method. It now share responsibilities of original method process(). This new

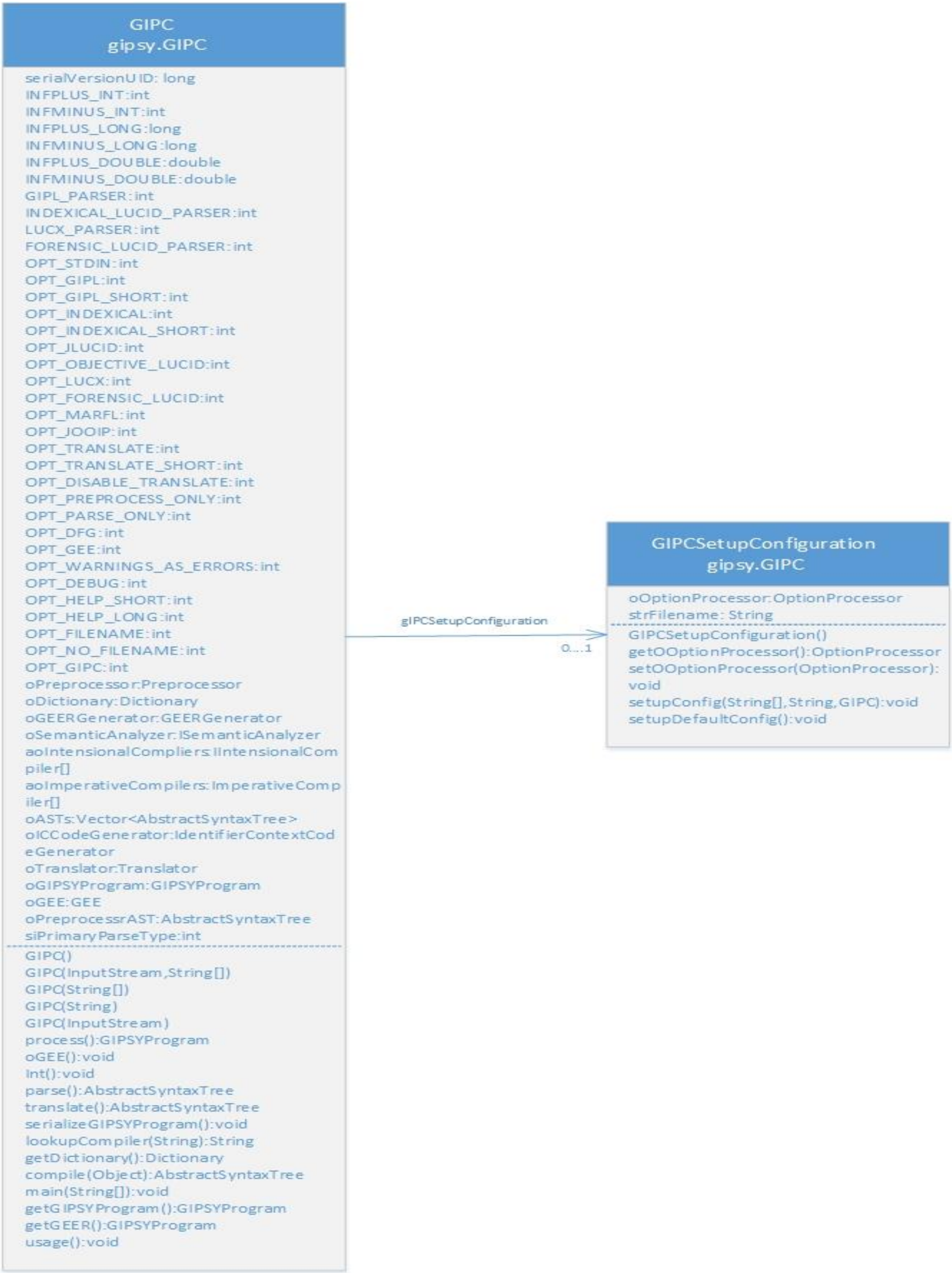

**Figure 64: `GIPC.java` class after refactoring**

method is created so that original process() method become short and less complex.

The diff of that change is given as follow.

diff -u -r1.1 GIPC.java

--- src/gipsy/GIPC/GIPC.java 13 Aug 2014 02:41:11 -0000 1.1

+++ src/gipsy/GIPC/GIPC.java 24 Aug 2014 02:49:27 -0000

@@ -55,6 +55,8 @@

@@ -783,6 +647,14 @@

### Case 4: Refactoring for class `GIPSYGMTOperator.java` in `gipsy.RIPE.editors.RunTimeGraphEditor.ui` package

The figure 65 depicts the diagram of the `GIPSYGMTOperator.java class` after refactoring.

**Change 0/2: Refactor class `GIPSYGMTOperator.java` into four different classes**

Original GIPSYGMTOperator.java class is too large which contain many attributes and methods. So, to reduce the complexity of the class extract class refactoring technique has been applied and four new classes has been created. Now new class GIPSYGMTOperatorInitializeComponents.java is responsible for initializing components. New GIPSYGMTOperatorPanelCreation.java class is in charge of creating panel. GIPSYGMTOperatorState.java class is now responsible for keeping track of operator state. To create panel for GIPSY tiers GIPSYTiersPanelCreation.java class has been created.

The diff of that change is given as follow.

diff -u -r1.1 GIPSYGMTOperator.java

--- src/gipsy/RIPE/editors/RunTimeGraphEditor/ui/GIPSYGMTOperator.java 13 Aug 2014 02:41:14 -0000 1.1

+++ src/gipsy/RIPE/editors/RunTimeGraphEditor/ui/GIPSYGMTOperator.java 23 Aug 2014 00:16:32 -0000

@@ -62,7 +62,13 @@

@@ -72,44 +78,19 @@

@@ -124,125 +105,19 @@

@@ -250,7 +125,7 @@

@@ -259,7 +134,7 @@

@@ -267,48 +142,45 @@

@@ -316,77 +188,10 @@

@@ -410,25 +215,10 @@

@@ -447,129 +237,12 @@

@@ -613,7 +286,7 @@

@@ -657,7 +330,8 @@

@@ -675,9 +349,11 @@

@@ -686,7 +362,7 @@

@@ -716,52 +392,9 @@

@@ -789,36 +422,7 @@

@@ -845,7 +449,7 @@

@@ -916,4 +520,24 @@

diff -N src/gipsy/RIPE/editors/RunTimeGraphEditor/ui/GIPSYGMTOperatorInitializeComponents.java

--- /dev/null 1 Jan 1970 00:00:00 -0000

+++ src/gipsy/RIPE/editors/RunTimeGraphEditor/ui/GIPSYGMTOperatorInitializeComponents.java1 Jan 1970 00:00:00 -0000

@@ -0,0 +1,228 @@

diff -N src/gipsy/RIPE/editors/RunTimeGraphEditor/ui/GIPSYGMTOperatorPanelCreation.java

--- /dev/null 1 Jan 1970 00:00:00 -0000

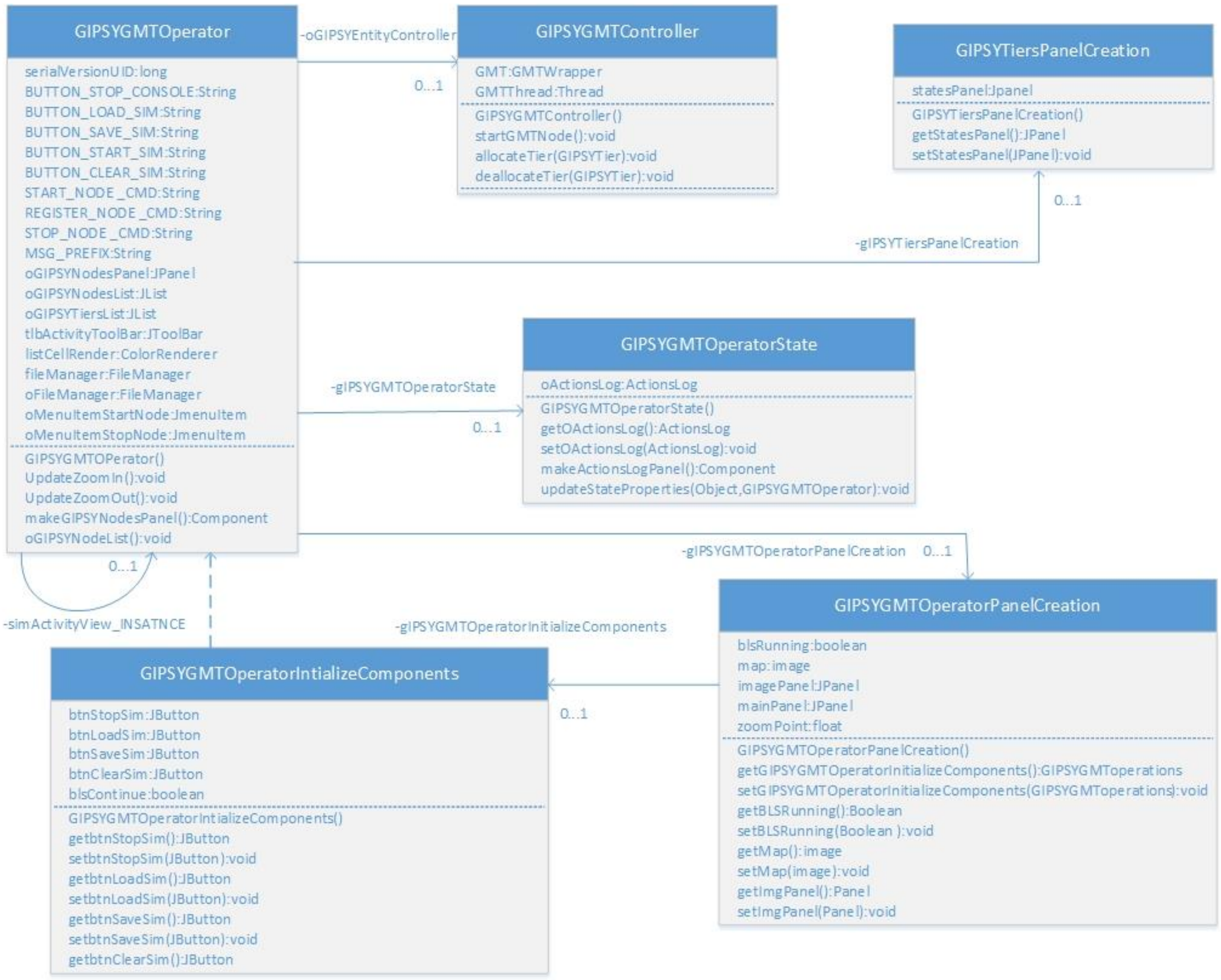

**Figure 65: `GIPSYGMTOperator.java` class after refactoring**

+++ src/gipsy/RIPE/editors/RunTimeGraphEditor/ui/GIPSYGMTOperatorPanelCreation.java 1 Jan 1970 00:00:00 -0000

@@ -0,0 +1,244 @@

diff -N src/gipsy/RIPE/editors/RunTimeGraphEditor/ui/GIPSYGMTOperatorState.java

--- /dev/null 1 Jan 1970 00:00:00 -0000

+++ src/gipsy/RIPE/editors/RunTimeGraphEditor/ui/GIPSYGMTOperatorState.java 1 Jan 1970 00:00:00 -0000

@@ -0,0 +1,67 @@

diff -N src/gipsy/RIPE/editors/RunTimeGraphEditor/ui/GIPSYTiersPanelCreation.java

--- /dev/null 1 Jan 1970 00:00:00 -0000

+++ src/gipsy/RIPE/editors/RunTimeGraphEditor/ui/GIPSYTiersPanelCreation.java 1 Jan 1970 00:00:00 -0000

@@ -0,0 +1,63 @@

**Change 1/2: Extract method applied in makeGIPSYNodesPanel() method in `GIPSYGMTOperator.java` class**

Two new methods has been created. One is oGIPSYNodeList() and another is oNodeListPopUpMenu(). They now share many responsibilities of original method makeGIPSYNodesPanel(). These two new methods are created so that original makeGIPSYNodesPanel() method become short and less complex.

The diff of that change is given as follow.

diff -u -r1.1 GIPSYGMTOperator.java

--- src/gipsy/RIPE/editors/RunTimeGraphEditor/ui/GIPSYGMTOperator.java 13 Aug 2014 02:41:14 -0000 1.1

+++ src/gipsy/RIPE/editors/RunTimeGraphEditor/ui/GIPSYGMTOperator.java 23 Aug 2014 00:16:32 -0000

@@ -62,7 +62,13 @@

@@ -267,48 +142,45 @@

**Change 2/2: Move method startInstance() of `GIPSYGMTOperator.java` class to `GIPSYGMTController.java` class in `gipsy.RIPE.editors.RunTimeGraphEditor.operator` package**

The method startInstance() is more interested in class GIPSYGMTController.java rather than the class GIPSYGMTOperator.java where it belongs. Because it is calling method startGMTNode() and allocateTier() which are in GIPSYGMTController.java class. That's why startInstance() method is moved to GIPSYGMTController.java class which is in gipsy.RIPE.editors.RunTimeGraphEditor.operator package.

The diff of that change is given as follow.

diff -u -r1.1 GIPSYGMTController.java

--- src/gipsy/RIPE/editors/RunTimeGraphEditor/operator/GIPSYGMTController.java 13 Aug 2014 02:41:12 -0000 1.1

+++ src/gipsy/RIPE/editors/RunTimeGraphEditor/operator/GIPSYGMTController.java 23 Aug 2014 00:16:32 -0000

@@ -94,5 +94,18 @@

#### 4.2.1 Test Cases

The following test cases are implemented to check if the behavior of the code has changed or not after refactoring.

##### I. DMARF

**Test Case 1: Test Case for `NeuralNetwork.java class`**

The MARF test case is created in JUnit. The test case is placed in `marf.Classification.NeuralNetwork` package. It verifies all the constant values that are moved to new class NeuralNetworkConstants.java. It also checks the object initialization of new class.

**Test Case 2: Test Case for `MARF.java` class**

The MARF test case is created in JUnit. The test case is placed in `marf` package. It verifies the constant values and methods of MARF.java class to check if the behavior of the class has changed or not.

##### II. GIPSY

**Test Case 1: Test case for `DFGCodeGenerator.java` and `DFGTranCodeGenerator.java` classes**

The Gipsy test case is created in JUnit. The test case is placed in `gipsy.GIPC.DFGGenerator` package. It checks the object initialization of refactored classes `DFGCodeGenerator` and `DFGTranCodeGenerator` and also verify the initialized value of attributes. The different values are also assigned to the attributes and check weather values are changing or not. The behavior of the methods `getPname(String)`, `getSname(String)` are also verified weather they are providing expected value for give String values.

**Test Case 2: Test case for `GIPC.java` class**

The GIPSY test case is created in JUnit. The test case is placed in `gipsy.GIPC` package. It verifies the constant values and methods of GIPC.java class to check if the behavior of the class has changed or not.

## 5. Conclusion

After performing research on the DMARF and GIPSY case studies by studying numerous research papers and comprehensive analysis of the source code, the architectures of the case studies have been understood. The distributed architecture of DMARF allow it to have autonomous properties which help aid in the designing of biometric and security applications. The architectures of both DMARF and GYPSY can be used together to allow research in more robust fields such as Intensional CyberForensics. Eight design patterns have been found by the analysis of the code which help gain a better understanding as to how the different components and the modules of the system interact with each other along with the relationships between them. This analysis helped identify code smells which bring down the quality of the structure of the code. The structure was improved without changing the overall behavior of the system by applying the suggested refactoring techniques. Test cases are used to verify if the behavior of the system was changed or not. The changes made in the refactored code are described in the patch sets and the change logs.

In the process of analysis of the architecture and source code, a better understanding has been gained to perform architectural and structural analysis to improve the overall quality of the code by keeping the internal logic of the code. More research can be done to better know about architectural patterns like the Pipes and Filters pattern which was identified within these case studies.Better architecture will lead to better testing, better maintainability and improved overall quality of the final product.

## TEAM REFERENCES

| Name | Student ID | Paper (From Project PDF) | |
|---|---|---|---|
| | | DMARF | GIPSY |
| Zinia Das | 7099533 | 11 | 29 |
| Mohammad Iftekharul Hoque | 6993451 | 16 | 21 |
| Renuka Milkoori | 7188196 | 14 | 29 |
| Jithin Nair | 7093888 | 13 | 15 |
| Rohan Nayak | 6846696 | 28 | 13 |
| Dhana Shree | 7166440 | 12 | 20 |
| Swamy Yogya Reddy | 7170386 | 27 | 5 |
| Arslan Zaffar | 6406645 | 18 | 7 |

| Name | Design Patterns |
|---|---|
| Zinia Das | Singleton |
| Mohammad Iftekharul Hoque | Strategy |
| Renuka Milkoori | Factory |
| Jithin Nair | Abstract Factory |
| Rohan Nayak | Decorator |
| Dhana Shree | Prototype |
| Swamy Yogya Reddy | Observer |
| Arslan Zaffar | Adapter |

## APPENDIX

### A1. Terminology

| Abbreviations used in this Work | |
|---|---|
| MARF | Modular Audio Recognition Framework |
| DMARF | Distributed Modular Audio Recognition Framework |
| NLP | Natural Language Processing |
| ASSL | Autonomic System Specification Language |
| JDSF | Java Data Security Framework |
| RMI | Remote Method Invocation |
| CORBA | Common Object Request Broker Architecture |
| SNMP | Simple Network Management Protocol |
| IP | Internet Protocol |
| TCP | Transmission Control Protocol |
| MIB | Management Information Base |
| WAL | Write Ahead Logging |
| GIPSY | General Intensional Programming System |
| HOIL | Higher Order Intensional Language |
| DGT | Demand Generator Tier |
| DWT | Demand Worker Tier |
| DST | Demand Store Tier |
| GIM | GIPSY Instance Manager |
| GMT | GIPSY Instance Manager |
| IDS | Intensional Data Dependency Structure |
| ICP | Intensional Communication Procedures |
| GEE | General Eduction Engine |
| GIPC | General Intensional Programming Language Compiler |
| CST | C Sequential Threads |
| IDP | Intensional Demand Propagator |
| IVW | Intensional Value Warehouse |
| RIPE | Run time Interactive Programming Environment |
| RFE | RIPE Functional Executor |
| GIPL | General Intensional Programming Language |
| AGIPSY | Autonomic General Intensional Programming System |
| GEER | General Eduction Engine Resources |
| UML | Unified Modeling Language |

### A.2 Lines of Code Breakdown

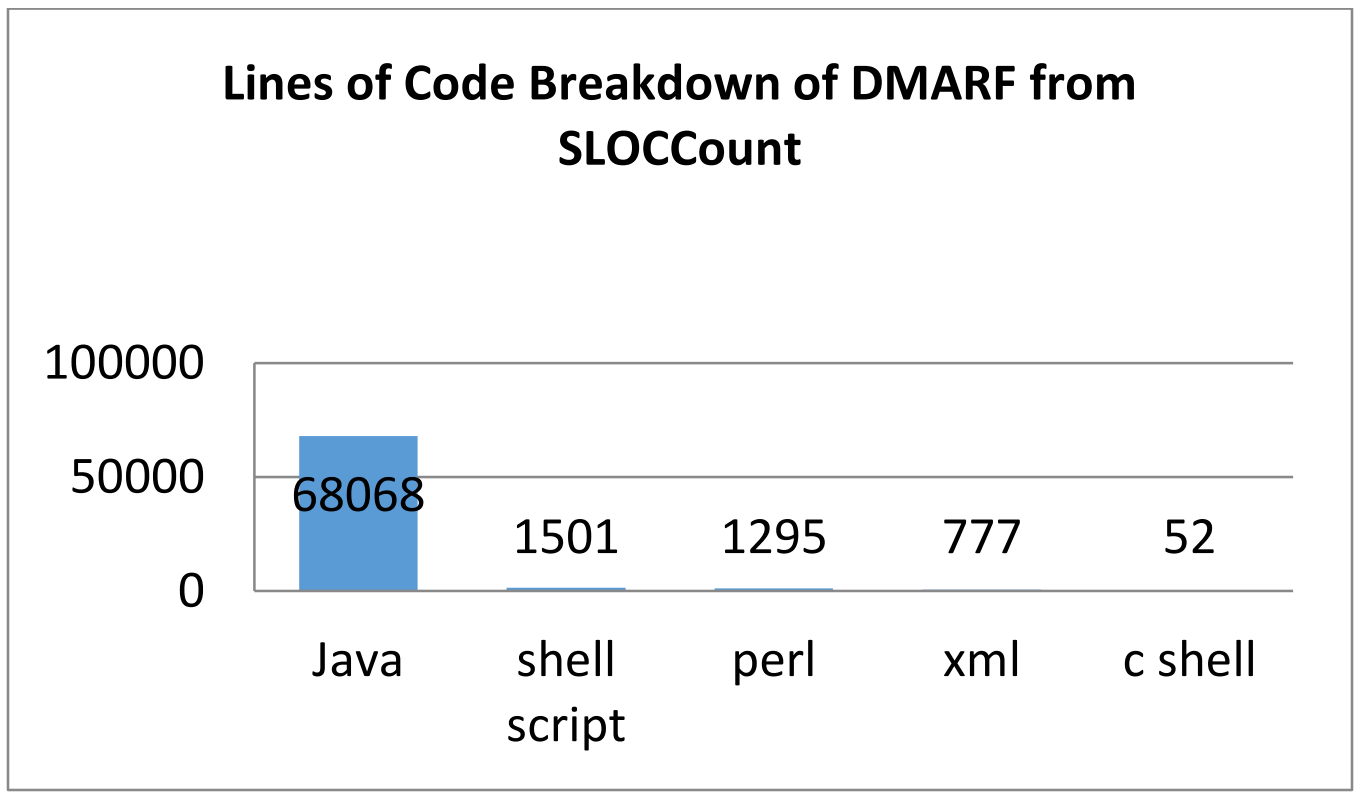

**Figure 66: Breakdown of lines of code DMARF using SLOCCount**

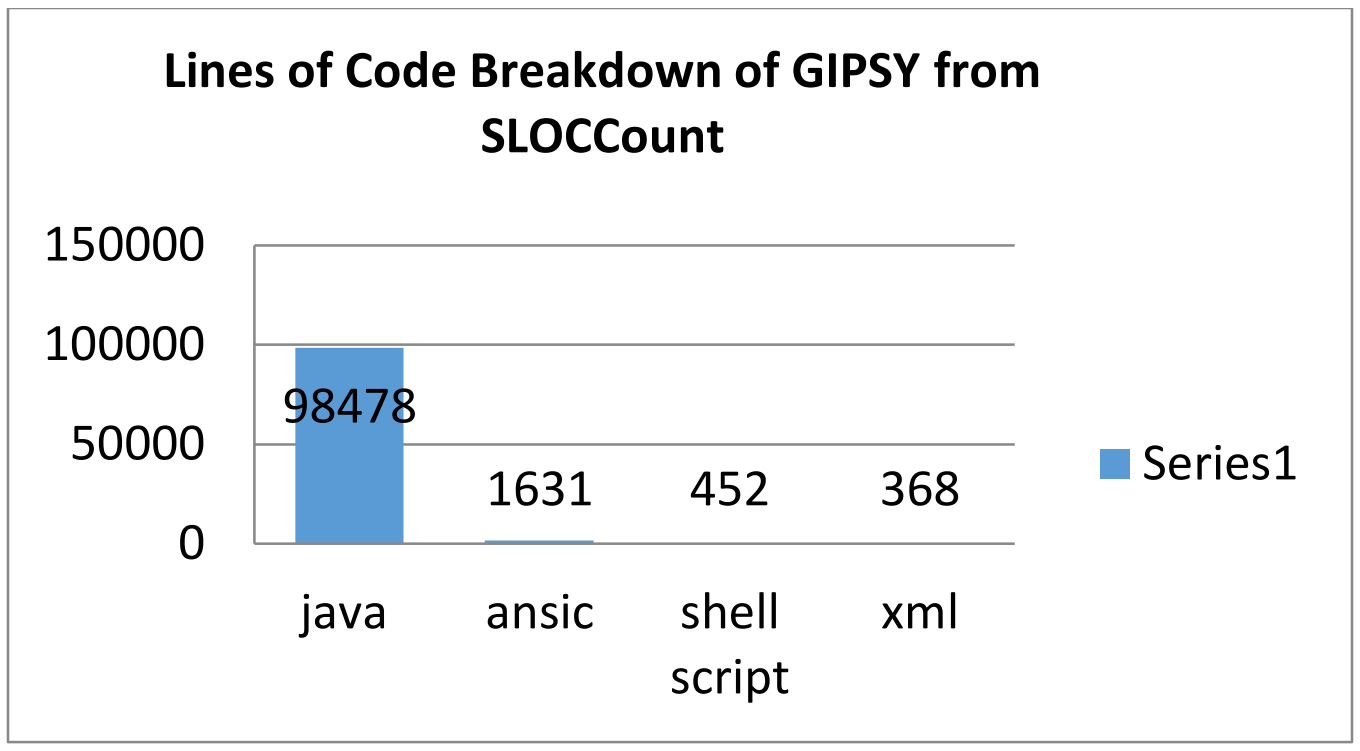

**Figure 67: Breakdown of lines of code GIPSY using SLOCCount**

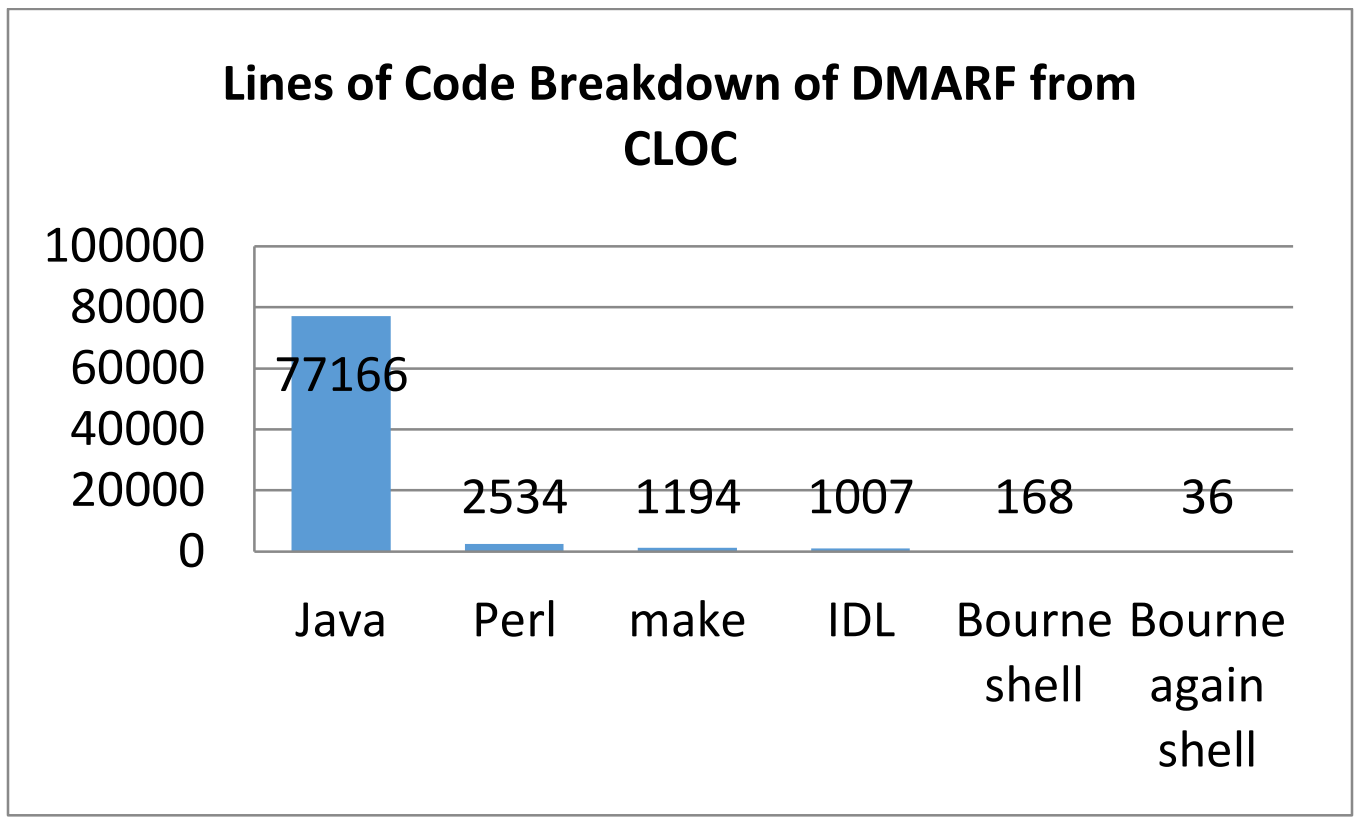

**Figure 68: Breakdown of lines of code DMARF using CLOC**

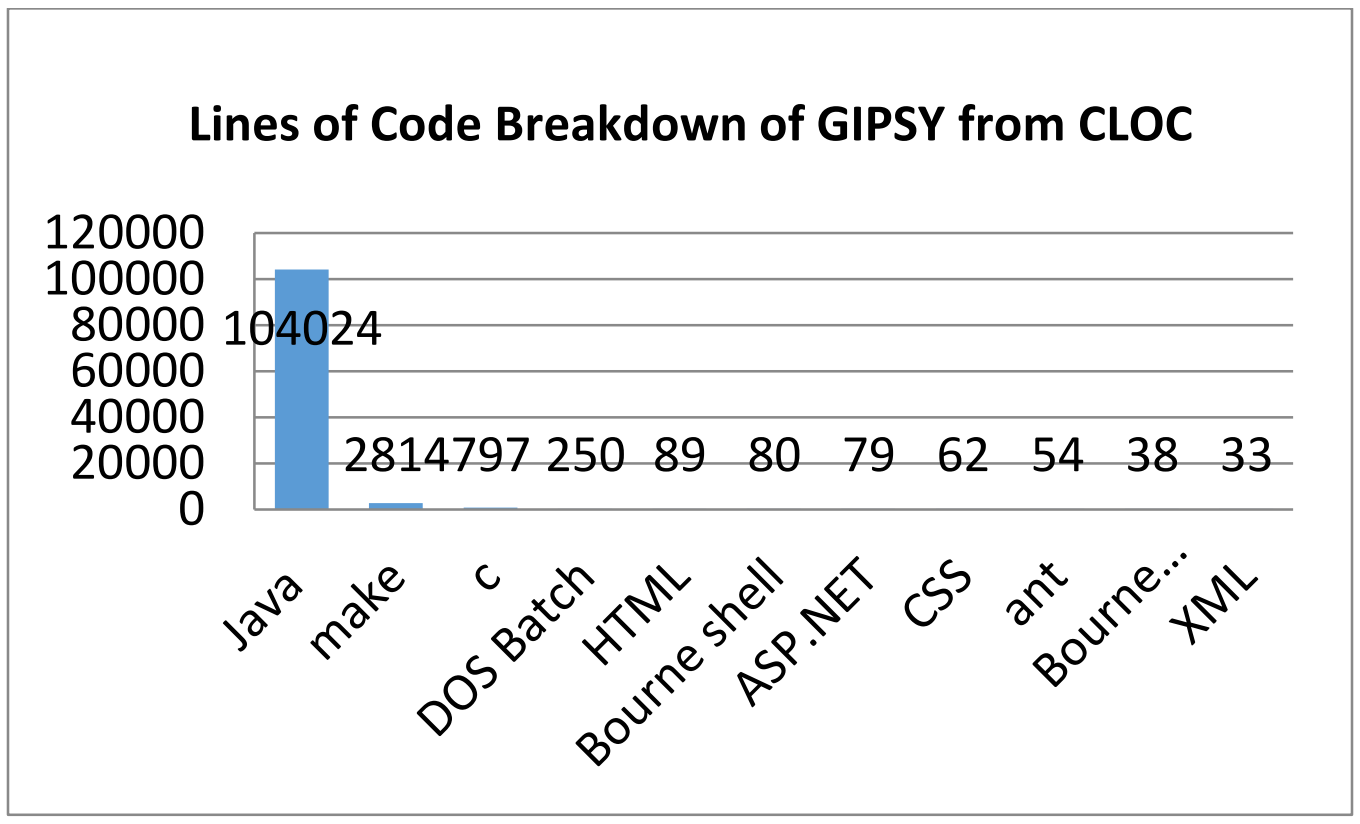

**Figure 69: Breakdown of lines of code GIPSY using CLOC**